\documentclass[prd, reprint, showpacs, nofootinbib, raggedbottom]{revtex4-1}
\usepackage{epsf}
\usepackage{amsmath, mathtools, graphicx, color, latexsym, array, hyperref, amssymb, gensymb}
\usepackage[inline]{enumitem}
\usepackage[retainorgcmds]{IEEEtrantools}
\usepackage[explicit]{titlesec}
\usepackage[caption=false]{subfig}

\allowdisplaybreaks

\newcolumntype{S}{>{\centering\arraybackslash}m{2.1cm}}
\newcolumntype{N}{>{\centering\arraybackslash}m{2.1cm}}
\newcolumntype{B}{>{\centering\arraybackslash}m{2.1cm}}
\newcolumntype{T}{>{\centering\arraybackslash}m{2cm}}
\newcolumntype{d}[1]{D{.}{.}{#1} }
\newcommand*{\figref}[1]{Figure \ref{#1}}
\newcommand*{\tabref}[1]{Table \ref{#1}}
\newcommand*{\secref}[1]{Section \ref{#1}}
\newcommand*{\secsref}[2]{Sections \ref{#1} and \ref{#2}}
\newcommand*{\appendref}[1]{Appendix \ref{#1}}

\renewcommand*{\imath}{\iota}

\newcommand*{\Action}{\mathcal{S}}

\newcommand*{\metric}{g}
\newcommand*{\RicScal}{R}
\newcommand*{\RicTens}{\RicScal}
\newcommand*{\Riemann}{\RicScal}

\newcommand*{\EinsTens}{G}
\newcommand*{\StressEnergy}{T}

\newcommand*{\firstform}{h}
\newcommand*{\secondform}{\chi}
\newcommand*{\firstprojection}[2]{\firstform^{#1}_{\phantom{#1}{#2}}}

\newcommand{\vecb}[1]{\mathbf{#1}}
\newcommand{\Tensb}[1]{\boldsymbol{#1}}

\newcommand*{\FriedScale}{a}
\newcommand*{\CosmoConst}{\Lambda}

\newcommand*{\LattScale}{a}
\newcommand*{\LattTime}{\tau}

\newcommand*{\Cauchy}{\Sigma}
\newcommand*{\Cauchyt}[1]{\Cauchy_{#1}}

\newcommand*{\CWstrutlength}{m}
\newcommand*{\CWstrut}[1]{\CWstrutlength_{#1}}
\newcommand*{\CWdiag}[1]{d_{#1}}
\newcommand*{\TrapArea}[1]{\Area{trap}{#1}}

\newcommand*{\TriArea}[1]{\Area{tri}{#1}}
\newcommand*{\CWdef}{\delta}

\newcommand*{\CWdeficit}[1]{\CWdef_{#1}}
\newcommand*{\TrapDeficit}[1]{\CWdeficit{#1}^{\,\text{trap}}}
\newcommand*{\DeficitA}[1]{\DeficitS{\text{A}}{#1}}
\newcommand*{\DeficitB}[1]{\DeficitS{\text{B}}{#1}}
\newcommand*{\TriDeficit}[1]{\CWdeficit{#1}^{\,\text{tri}}}

\newcommand*{\Num}{N}
\newcommand*{\N}[1]{\Num_{#1}}

\newcommand*{\Nvert}{{\N{0}}}
\newcommand*{\Nedge}{{\N{1}}}
\newcommand*{\Ntri}{{\N{2}}}
\newcommand*{\Ntet}{{\N{3}}}

\newcommand*{\Odt}[1]{O \! \left(dt^{#1} \right)}
\newcommand*{\Odtone}{O \! \left(dt \right)}
\newcommand*{\Oconst}{O \! \left(1 \right)}

\newcommand*{\axisl}{h}
\newcommand*{\axislen}[1]{\axisl_{#1}}

\newcommand*{\axislensq}[1]{\axislen{#1}^2}
\newcommand*{\axislendot}[1]{\dot{\axisl}_{#1}}

\newcommand*{\dihedral}[2]{\theta^{(#1)}_{#2}}

\newcommand*{\RegTime}[1]{t_{#1}}
\newcommand*{\CWstrutl}[1]{\CWstrut{i}^\ell}
\newcommand*{\CWstrutg}[1]{\CWstrut{i}^g}
\newcommand*{\CWstrutdot}[1]{\dot{\CWstrutlength}_{#1}}
\newcommand*{\Area}[2]{A^\text{{#1}}_{#2}}

\newcommand*{\AreaAi}[2]{\Area{A{#1}}{#2}}
\newcommand*{\AreaBi}[2]{\Area{B{#1}}{#2}}

\newcommand*{\DeficitS}[2]{\CWdef^{#1}_{#2}}

\newcommand*{\CWTrapDihe}{\theta}
\newcommand*{\CWTrapdihedral}[1]{\CWTrapDihe_{#1}}

\newcommand*{\dM}{\delta M}

\newcommand*{\CWlength}{l}
\newcommand*{\CWLattl}{\CWlength}
\newcommand*{\CWLattlenZ}[1]{\CWLattl_{#1}}  %Z indicates zeroth order term
\newcommand*{\CWLattlenZdot}[1]{\dot{\CWLattl}_{#1}}
\newcommand*{\CWLattlen}[2]{\CWLattl^{({#1})}_{#2}}
\newcommand*{\CWLattlendot}[2]{\dot{\CWLattl}^{({#1})}_{#2}}
\newcommand*{\CWLattdl}[2]{\delta \CWLattlen{#1}{#2}}
\newcommand*{\CWLattdldot}[2]{\delta \CWLattlendot{#1}{#2}}

\newcommand*{\CWLattdiagp}[1]{\CWdiag{#1}^{AB^\prime}}
\newcommand*{\CWLattdiag}[1]{\CWdiag{#1}^{AE^\prime}}
\newcommand*{\CWLattdiagZ}[1]{\CWdiag{#1}}

\newcommand*{\CWLattstrutv}[2]{\CWstrut{#2}^{#1}}
\newcommand*{\CWLattstrutvdot}[2]{\CWstrutdot{#2}^{#1}}
\newcommand*{\CWLattdstrutv}[2]{\delta \CWLattstrutv{#1}{#2}}

\newcommand*{\CWLattstrutlen}[1]{\CWLattstrutv{EE^\prime}{#1}}
\newcommand*{\CWLattstrutlenp}[1]{\CWLattstrutv{AA^\prime}{#1}}
\newcommand*{\CWLattstrutlendot}[1]{\CWLattstrutvdot{EE^\prime}{#1}}
\newcommand*{\CWLattstrutlenpdot}[1]{\CWLattstrutvdot{AA^\prime}{#1}}

\newcommand*{\CWLattdihedral}[1]{\CWTrapdihedral{#1}}
\newcommand*{\CWLattdihedraldot}[1]{\dot{\CWTrapDihe}_{#1}}
\newcommand*{\CWLattdihedralS}[2]{\CWLattdihedral{#2}^{({#1})}}
\newcommand*{\CWLattddihedralS}[2]{\delta \CWLattdihedral{#1}{#2}}

\newcommand*{\CWLattdihedralPhi}[2]{\phi^{(#1)}_{#2}}
\newcommand*{\CWLattddihedralPhi}[2]{\delta\CWLattdihedralPhi{#1}{#2}}

\newcommand*{\CWLattdeficit}[2]{\CWdeficit{#2}^{(#1)}}
\newcommand*{\CWLattQuadDeficit}[1]{\CWdeficit{#1}^{\,\text{quad}}}
\newcommand*{\CWLattQuadArea}[1]{\Area{quad}{#1}}

\newcommand*{\dpsi}[1]{\delta \psi_{#1}}
\newcommand*{\dpsidot}[1]{\delta \dot{\psi}_{#1}}

\newcommand*{\LinEl}{s}
\newcommand*{\RegLinEl}[1]{\LinEl_{#1}}
\newcommand*{\RegLinEldot}[1]{\dot{\LinEl}_{#1}}

\newcommand*{\distance}[1]{\left\lvert \overrightarrow{#1} \right\rvert}

\renewcommand{\thesection}{\Roman{section}}
\titleformat{\section}[block]
  {\normalfont\bfseries}{\thesection.}{1em}{\centering \MakeUppercase{#1}}
\renewcommand{\thesubsection}{\Alph{subsection}}
\titleformat{\subsection}[block]
  {\normalfont\bfseries}{\thesubsection.}{1em}{\centering {#1}}

\begin{document}

\title{\bf Regge calculus models of closed lattice universes}
\author{Rex G \surname{Liu}}
\email{R.Liu@damtp.cam.ac.uk}
\affiliation{Trinity College, Cambridge CB2 1TQ, UK,}
\affiliation{DAMTP, CMS, Wilberforce Road, Cambridge CB3 0WA, UK.}
\author{Ruth M \surname{Williams}}
\email{R.M.Williams@damtp.cam.ac.uk}
\affiliation{Girton College, Cambridge CB3 0JG, UK,}
\affiliation{DAMTP, CMS, Wilberforce Road, Cambridge CB3 0WA, UK.}

\begin{abstract}
This paper examines the behaviour of closed `lattice universes' wherein masses are distributed in a regular lattice on the Cauchy surfaces of closed vacuum universes.  Such universes are approximated using a form of Regge calculus originally developed by Collins and Williams to model closed FLRW universes.  We consider two types of lattice universes, one where all masses are identical to each other and another where one mass gets perturbed in magnitude.  In the unperturbed universe, we consider the possible arrangements of the masses in the Regge Cauchy surfaces and demonstrate that the model will only be stable if each mass lies within some spherical region of convergence.  We also briefly discuss the existence of Regge models that are dual to the ones we have considered.  We then model a perturbed lattice universe and demonstrate that the model's evolution is well-behaved, with the expansion increasing in magnitude as the perturbation is increased.
\end{abstract}

\pacs{98.80.-k, 04.25.-g}

\maketitle

\section{Introduction}
Modern cosmology is founded upon the so-called \emph{Copernican principle}, which posits that the universe `looks' on average to be the same regardless of where one is in the universe or in which direction one looks.  In other words, every point in the universe is identical; no point is special.  More formally, the Copernican principle states that Cauchy surfaces of the universe can be admitted that are homogeneous and isotropic, and this symmetry can be expressed mathematically by writing the universe's metric in the form
\begin{equation}
d\LinEl^2 = - dt^2 + \FriedScale^2(t) \left[\frac{dr^2}{1- k r^2} + r^2 \big(d\theta^2 + \sin^2\theta\, d\phi^2 \big) \right],
\label{FLRW}
\end{equation}
where $\FriedScale(t)$ is a time-dependent function known as the \emph{scale factor} and $k$ is a curvature constant.  The sign of $k$ determines whether Cauchy surfaces of constant time $t$ will be open, flat, or closed, with $k<0$ being open, $k=0$ being flat, and $k>0$ being closed.  The metric \eqref{FLRW} is known as the \emph{Friedmann-Lema\^itre-Robertson-Walker} (FLRW) metric.  Though the form of this FLRW metric is fixed by consideration of symmetries alone, the function for the scale factor $\FriedScale(t)$ is instead determined by general relativity.  Inserting the metric into the Einstein field equations yields
\begin{IEEEeqnarray}{rCl}
\left( \frac{\dot{\FriedScale}}{\FriedScale} \right)^2 &=& \frac{1}{3} \Big(8\pi\rho + \CosmoConst \Big) - \frac{k}{\FriedScale^2},
\label{Friedmann1}\\
\frac{\ddot{\FriedScale}}{\FriedScale} \hphantom{\Bigg)^2} &=& -\frac{4\pi}{3} \Big(\rho + 3p \Big) + \frac{\CosmoConst}{3},
\label{Friedmann2}
\end{IEEEeqnarray}
where $\rho$ and $p$ are the energy density and pressure of any perfect fluid filling the space, and where $\CosmoConst$ is the cosmological constant.  This pair of differential equations is known as the \emph{Friedmann equations}, and their solution determines $\FriedScale(t)$.

These FLRW models have had great success in explaining much of the universe's behaviour, including most notably the Hubble expansion of the universe, the cosmic microwave background, and baryon acoustic oscillations.  Indeed, the underlying assumption of homogeneity and isotropy appears well-supported by precision measurements showing the CMB to be isotropic to within one part in 100,000 \cite{Smoot-et-al, *Bennett-et-al}.  Yet in spite of this, observations also clearly show that the late, matter-dominated universe is not homogeneous and isotropic except at the coarsest of scales.  Instead, matter is distributed predominantly in clusters and superclusters of galaxies with large voids in between, and the physical effects of such a `lumpy' universe are still not fully understood.

Indeed, there has been intense interest recently over the possible importance of inhomogeneities to observational cosmology.  Perhaps the area of greatest interest concerns the possible effects of inhomogeneities on recent redshift measurements from Type Ia supernovae (SN1a).  When fitted to perfectly homogeneous FLRW models, these measurements have led to the conclusion that the universe's expansion is accelerating \cite{Riess-et-al, *Perlmutter-et-al, *Barris-et-al}, and to account for this acceleration, cosmologists have posited the existence of some exotic matter, known generally as \emph{dark energy}.  However other cosmologists have instead posited that much, if not all, of this acceleration is as an apparent effect, arising from fitting data from an inhomogeneous universe onto a homogeneous model \cite{EllisBuchert, *Wiltshire2007, *Mattsson, *Ellis2011, *CELU}.  It has been argued that this effect actually arises from the non-linear structure of the inhomogeneities and that such a structure could not be adequately modelled simply by perturbing an FLRW universe \cite{KMR, *ClarksonMaartens, *ClarksonUmeh}.  Therefore, as there has so far been no confirmed direct observation of any exotic matter, the relative importance of dark energy compared to inhomogeneities in explaining the supernovae data remains an open question.

For this reason, there has been a resurgence of interest recently in the various approaches to building non-perturbative, inhomogeneous models.  One notable example is the Lema\^itre-Tolman-Bondi (LTB) models.  It has been shown that such models can account for all known cosmological observations, including SN1a redshifts, without requiring any exotic matter, but the observer must sit at the centre of a Hubble-scale under-dense region \cite{MHE, *CS, *Tomita1, *Tomita2, *AAG, *Moffat, *Mansouri, *GH, *CFL, *CGH, *ABNV, *BW, *FLSC}.  In other words, the observer sits at the centre of an isotropic but non-homogeneous universe.

On the other hand, the `Swiss-cheese' models of Einstein and Straus retain an FLRW background but introduce inhomogeneities by replacing co-moving spherical FLRW regions with Schwarzschild or LTB regions \cite{EinsteinStraus, *EinsteinStraus-err, Kantowski, *BiswasNotari, *MKMR, *BTT, *MKM, *CliftonZuntz}.  By an appropriate fitting of these regions into the FLRW background, the resulting space-time will still be an exact solution to the Einstein field equations.  However, because of their FLRW background, these models will still be dynamically identical to FLRW universes.  And although there is no \emph{a priori} reason to believe the optical properties should be identical as well, recent studies have shown that they are, in fact, broadly similar \cite{Kantowski, *BiswasNotari, *MKMR, *BTT, *MKM, *CliftonZuntz}.

In this paper, we shall consider a different class of universes, one that is not in any way based on FLRW universes.  We shall consider the so-called \emph{lattice universes} where the matter content on each Cauchy surface consists of identical point masses arranged into a regular lattice.  Unlike the LTB or Swiss-cheese models, the lattice universe has a truly discrete matter content and is otherwise vacuum throughout, which is more representative of the actual universe's matter distribution.  Because of the regular arrangement of the point masses, these lattice universes still possess a high degree of symmetry, though not as great as that of FLRW universes.

We shall focus on lattices formed by tessellating 3-spaces of constant curvature with identical regular polyhedral cells.  The possible lattices that can be constructed from such a tessellation have been summarised in \appendref{App_Latt}, and we shall refer to these lattices as Coxeter lattices.  To `construct' a lattice universe then, we select one of the Coxeter lattices of \appendref{App_Latt} and distribute a set of point masses in a regular manner on that lattice, such as at the centres of the cells, at the centres of the faces, at the mid-points of the edges, or at the vertices.  Naturally, after the masses have been arranged in this manner, the Cauchy surfaces will no longer be surfaces of constant curvature; however, the metric is still expected to be invariant under the same symmetry transformations that leave the lattice invariant, symmetries which include discrete translation symmetries, discrete rotational symmetries, and reflection symmetries at the cell boundaries.  In other words, the lattice universe should have a metric of the form
\begin{equation}
d\LinEl^2 = -dt^2 + \gamma^{(3)}_{ab}\left(t,\Tensb{x}\right)\,dx^a dx^b,
\end{equation}
where $\Tensb{\gamma}^{(3)}\left(t,\Tensb{x}\right)$ is the 3-dimensional metric for constant $t$ hypersurfaces, and Latin indices $a,b = 1, 2, \text{ or } 3$ denote spatial co-ordinates only; the spatial metric $\Tensb{\gamma}^{(3)}\left(t,\Tensb{x}\right)$ at constant $t$ would possess the lattice symmetries.  Effectively, the Copernican symmetries of FLRW universes have been reduced to just these symmetries.  This paper will focus exclusively on closed lattice universes based on the tetrahedral Coxeter lattices; these lattices consist of 5, 16, or 600 identical, equilateral tetrahedral cells.

There has been a variety of approaches adopted to modelling such universes.  Using exact methods, Wheeler \citep{Wheeler1982} as well as Clifton \emph{et al.} \citep{CRT} have successfully constructed the exact 3-metric $\Tensb{\gamma}^{(3)}\left(t=0,\Tensb{x}\right)$ for the time-symmetric Cauchy surfaces of closed universes.  Korzy{\'n}ski has further generalised this work by examining the case where there is an arbitrary number of masses, not necessarily arranged in a lattice \citep{Korzynski}.  Clifton \emph{et al.} have also examined the dynamics of the closed universes by evolving their initial data along certain highly-symmetric curves \citep{CGRT, *CGR}.  Bruneton and Larena have modelled the dynamics of the flat universe using an exact but perturbative expansion of the metric about Minkowski space-time \citep{BrunetonLarena1}.  Using numerical approaches, Bentivegna and Korzy{\'n}ski have studied the evolution of closed universes from initial data on a hypersurface at time-symmetry \citep{BentivegnaKorzynski2012}, while Yoo \emph{et al.} as well as Bentivegna and Korzy{\'n}ski have studied the dynamics of the flat universe \citep{YATN, *YON, BentivegnaKorzynski2013}.  On the other hand, Lindquist and Wheeler \cite{LW, *LW-err, *Houches} have devised an approximation to the lattice universe, generalised by Clifton and Ferreira \citep{CF, *CF-err, *CFO}, wherein each polyhedral lattice cell gets approximated by a spherical cell with Schwarzschild geometry inside.  This has been applied to study the evolution of the closed, flat, as well as open universes \citep{LW, *LW-err, CF, *CF-err, *CFO, RGL}.

In this paper, we shall consider another approach to modelling the lattice universe; we shall adopt a Regge calculus formalism originally developed by Collins and Williams (CW) to model closed FLRW universes \citep{CW}.  Regge calculus \cite{Regge} is a highly versatile formalism that can in principle approximate any solution of the Einstein field equations using a piece-wise linear manifold; this makes it particularly suitable to studying systems where an exact solution is difficult to obtain.  The Regge manifold is constructed by gluing flat blocks together such that neighbouring blocks share an entire face; as the blocks are flat, the metric inside is the Minkowski metric.  The Regge manifold is generally referred to as a \emph{skeleton}.  The solutions of Regge calculus are generally expected to converge at second order in the skeletal edge-lengths to the corresponding continuum solutions of the Einstein field equations \citep{BrewinGentle}.  Although we shall focus on closed lattice universes in this paper, much of the work is readily generalisable to flat and open universes.

Collins and Williams have constructed their skeleton from a one-parameter family of space-like Cauchy surfaces that foliate the entire skeleton.  Each Cauchy surface triangulates a closed FLRW Cauchy surface with equilateral tetrahedra such that all vertices, edges, and faces in the triangulation are identical.  We note that closed FLRW Cauchy surfaces of constant $t$ can be embedded as 3-spheres in 4-dimensional Euclidean space $\mathbf{E}^4$.  The scale factor $\FriedScale(t)$ can always be re-scaled so that the FLRW curvature constant $k$ becomes unity, in which case, $\FriedScale(t)$ would equal the 3-sphere radius of the embedding.  The embedding is then given by
\begin{equation}
\begin{aligned}
r &= \sin \chi,\\
x^1 &= \FriedScale(t)\cos\chi,\\
x^2 &= \FriedScale(t)\sin\chi\cos\theta,\\
x^3 &= \FriedScale(t)\sin\chi\sin\theta\cos\phi,\\
x^4 &= \FriedScale(t)\sin\chi\sin\theta\sin\phi,
\end{aligned}
\label{FLRW:closedE4}
\end{equation}
for $0 \leq \chi, \theta \leq \pi$ and $0 \leq \phi < 2\pi$.  CW Cauchy surfaces would triangulate such 3-sphere Cauchy surfaces with identical equilateral tetrahedra, and according to Coxeter \cite{Coxeter}, such a triangulation is only possible using 5, 16, and 600 tetrahedra.  In fact, the triangulations correspond to the three closed tetrahedral Coxeter lattices of \appendref{App_Latt}; this makes the CW skeletons particularly apposite for modelling lattice universes, as their Cauchy surfaces naturally provide suitable lattices into which the masses of the lattice universe could be embedded.  \tabref{tab:primary} tabulates the numbers of vertices, edges, triangles, and tetrahedra for each of the possible triangulations.
\begin{table} [htb]
\renewcommand{\arraystretch}{1.2}
\caption{\label{tab:primary}The number of simplices in each of the three triangulations of the 3-sphere with equilateral tetrahedra as well as the number of triangles meeting at any edge.  We introduce $\Ntet$, $\Ntri$, $\Nedge$, and $\Nvert$ to denote the number of tetrahedra, triangles, edges, and vertices in the Cauchy surface.}
\begin{tabular*}{8.6cm}{>{\centering\arraybackslash}m{1.7cm} >{\centering\arraybackslash}m{1.7cm} >{\centering\arraybackslash}m{1.2cm} >{\centering\arraybackslash}m{1.7cm} >{\centering\arraybackslash}m{1.7cm}}
\hline\hline
Tetrahedra $(\Ntet)$ & Triangles $(\Ntri)$ & Edges $(\Nedge)$ & Vertices $(\Nvert)$ & Triangles per edge \\
\hline
5 & 10 & 10 & 5 & 3\\
16 & 32 & 24 & 8 & 4\\
600 & 1200 & 720 & 120 & 5\\
\hline\hline
\end{tabular*}
\end{table}
The CW Cauchy surfaces are then joined together by a series of time-like edges called \emph{struts} connecting each vertex on one surface with its time-evolved image on the next.  Because all vertices on a surface are identical, all struts between any two surfaces are identical as well.  With this construction, the CW Cauchy surface at discrete time parameter $\RegTime{i}$ can be characterised by just two distinct lengths, the tetrahedral edge-length $\CWLattlenZ{}(\RegTime{i})$ and the strut-length $\CWstrut{}(\RegTime{i})$.  Surfaces at different $\RegTime{i}$ are completely identical apart from an overall re-scaling of the $\CWLattlenZ{}(\RegTime{i})$ length-scale.  Therefore, Collins and Williams interpreted $\CWLattlenZ{}(\RegTime{i})$ to be the Regge calculus analogue of $\FriedScale(t)$.  By then solving the Regge action and then taking the continuum time limit $(\RegTime{i+1} - \RegTime{i}) \to 0$, they were able to derive an expression for the length-scale as a function $\CWLattlenZ{}(t)$ of continuum time $t$.  The CW skeletons were first applied to model closed dust-filled FLRW universes and were found to yield very accurate results, with the models with a greater number of tetrahedra yielding higher accuracy \cite{CW, Brewin}.

In this paper, we shall also consider a different type of lattice universe.  We shall consider a closed lattice universe where a single mass gets perturbed in magnitude.  However as a prelude to this investigation, we shall first explore the properties of the completely unperturbed lattice universe.  Thus, this paper is organised as follows.  We shall begin, in the second section, with a brief exposition of Regge calculus with CW skeletons.  In the third section, we then explore the behaviour of the unperturbed lattice universe.  We shall show that the universe's behaviour depends on where the masses are located: the universe becomes unconditionally divergent if the masses are placed at the vertices of the Cauchy surface but is unconditionally convergent if the masses are placed anywhere in a spherical region around the centres of the tetrahedra.  Thereafter, we shall work only with models where the masses are at the tetrahedral centres.  We also make this choice for the following reason: in FLRW universes, a test particle that is co-moving with the universe would be following a geodesic as well, and we suspect this to also be the case in the lattice universe; thus if we want particles in our model also to be both co-moving and following geodesics across the entire Regge space-time, then we must place the particles at the centres of the tetrahedra; at any other location, co-moving particles will not follow global geodesics.  In the fourth section, we shall perturb one of these masses, construct the corresponding perturbed Regge model, and derive the relevant equations governing the model's evolution.  We shall focus exclusively on the 5-tetrahedra model: this model would involve only two sets of perturbed edges whereas the other two models would require many more, with each set having its own independent length.  In the fifth section, we shall consider the application of the initial value equation at the moment of time symmetry to this model: we shall derive certain conditions that the initial conditions of the Regge equations must satisfy in order to be consistent with this equation.  In the final section, we shall examine the behaviour of the model for various perturbations and compare it against that of the unperturbed model; we shall then close with a brief discussion of certain assumptions inherent in our model.

In this paper, we shall use geometric units where $G = c = 1$.

\section{Regge calculus with CW skeletons}
\label{CWRegge}

As mentioned above, Regge calculus approximates any continuous space-time using a piece-wise linear manifold composed of flat blocks glued together at their faces.  Regge calculus customarily uses 4-simplices as the blocks, though as we shall soon see with the CW skeleton, this is not always the case.

Curvature in the skeleton manifests itself as conical singularities concentrated on the sub-faces of co-dimension 2; these sub-faces are known as \emph{hinges}.  Each block will have two sub-faces of co-dimension 1 meeting at any of its hinges; we shall refer to such sub-faces simply as the \emph{faces} of the block.  If a hinge were flat, then the dihedral angles between all faces meeting at the hinge would sum to $2\pi$; any deviation from $2\pi$ provides a measure of the curvature and is known as the \emph{deficit angle}.  The deficit angle $\CWdeficit{i}$ at a hinge labelled $i$ is given by
\begin{equation}
\CWdeficit{i} = 2\pi - \sum_j \dihedral{i}{j},
\label{def_ang}
\end{equation}
where $\dihedral{i}{j}$ is the dihedral angle at the hinge formed by the faces of the block labelled $j$ and the summation is over all blocks meeting at the hinge.

If the skeleton consists solely of 4-simplices, then to completely specify the skeletal geometry, one need only specify the lengths of all edges.  This follows because the internal geometry of any $n$-simplex, including its angles and the areas of its sub-simplices, is completely determined when the lengths of its $C(n+1,2)$ edges are specified.  Thus the skeletal edge-lengths serve as the Regge analogue of the metric; in analogy to how the metric is determined by the Einstein field equations in general relativity, the edge-lengths are determined by the \emph{Regge field equations}, the Regge analogue to the Einstein field equations.

The Regge field equations are obtained by following a variational approach similar to how the Einstein field equations are obtained from the Einstein-Hilbert action.  In general relativity, the Einstein field equations can be derived by varying the Einstein-Hilbert action
\begin{equation}
\Action_{EH} = \displaystyle{\frac{1}{16\pi}\int \RicScal \, \sqrt{-\metric} \; d^4x \; - \mkern-20mu \sum_{i \, \in \, \left\{ \text{particles} \right\} } \mkern-15mu M_i \int d\RegLinEl{i}}
\label{Einstein-Hilbert}
\end{equation}
with respect to the metric tensor $\metric_{\mu\nu}$, where $\RicScal$ is the Ricci scalar, $\metric = \det (\metric_{\mu\nu})$, $M_i$ is the mass of particle $i$, and $d\RegLinEl{i}$ is its line element; the summation is over all particles in the space-time.  When applied to a Regge skeleton, this reduces to the \emph{Regge action} \cite{Regge}
\begin{equation}
\Action_{Regge} = \displaystyle{\frac{1}{8\pi}\sum_{i \,\in\, \left\{ \text{hinges}\right\}} \mkern-9mu A_i\, \CWdeficit{i}} \; - \mkern-26mu \sum_{\substack{i \, \in \, \left\{ \text{particles} \right\} \\ j \, \in \, \left\{ \text{blocks} \right\} }} \mkern-22mu M_i \, \RegLinEl{ij},
\label{ReggeAction}
\end{equation}
where $A_i$ is the area of a hinge in the Regge skeleton, $\CWdeficit{i}$ its corresponding deficit angle, and $\RegLinEl{ij}$ the length of particle $i$'s path through block $j$; the first summation is over all hinges in the skeleton while the second is over all particles and all blocks of the skeleton.  Note that if particle $i$ never passes through block $j$, then $\RegLinEl{ij}$ will accordingly be zero.

Since the skeletal edge-lengths are the Regge analogue of the metric, the Regge action is varied with respect to an edge-length $\ell_j$ to get the Regge field equations
\begin{equation}
0 = \displaystyle{ \mkern-15mu \sum_{i \,\in\, \left\{ \text{hinges}\right\}} \mkern-15mu \CWdeficit{i}} \, \frac{\partial A_i}{\partial \ell_j} \; - \; 8\pi \mkern-25mu \sum_{\substack{i \, \in \, \left\{ \text{particles} \right\} \\ j \, \in \, \left\{ \text{blocks} \right\} }} \mkern-22mu M_i \, \frac{\partial \RegLinEl{ij}}{\partial \ell_j},
\label{ReggeEqn}
\end{equation}
where the variation of the deficit angles has cancelled out owing to the well-known Schl\"afli identity \citep{Regge},
$$
\sum_i A_i \frac{\partial \dihedral{i}{k}}{\partial \ell_j} = 0;
$$
this identity holds for any individual block $k$, with the summation being over all hinges in the block and $\dihedral{i}{k}$ being the block's dihedral angle at hinge $A_i$.\footnote{In the standard formulation of Regge calculus, one actually uses a simplicial manifold where every block is a 4-simplex, and the Schl\"afli identity is usually formulated in terms of simplices rather than arbitrary blocks.  However, any block can always be triangulated into simplices, and one can then apply the simplicial form of the Schl\"afli identity to the triangulated block to obtain the form of the identity we have above, using the chain rule if necessary to satisfy any constraints on the block's geometry.}

\begin{figure}[htb]
{\fontsize{8pt}{9.6pt}\input{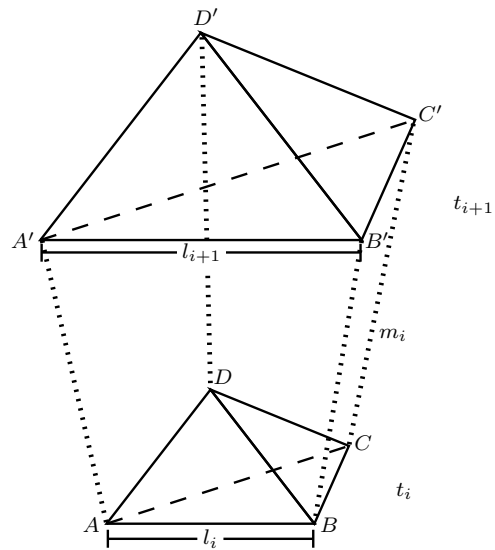}}
\caption{\label{fig:4block}An equilateral tetrahedron of edge-length $\CWLattlenZ{i}$ at time $\RegTime{i}$ evolves to a tetrahedron of edge-length $\CWLattlenZ{i+1}$ at time $\RegTime{i+1}$, tracing out a 4-dimensional world-tube.  The struts are all of equal length.}
\end{figure}

Collins and Williams originally constructed their skeleton not out of 4-simplices but out of 4-blocks; each 4-block corresponds to the truncated world-tube of a tetrahedron as it evolves from one Cauchy surface to the next.  If we denote the Cauchy surface at time $\RegTime{i}$ by $\Cauchyt{i}$ and the surface at time $\RegTime{i+1}$ by $\Cauchyt{i+1}$, then a 4-block between $\Cauchyt{i}$ and $\Cauchyt{i+1}$ consists of a tetrahedron in $\Cauchyt{i}$ with edges of length $\CWLattlenZ{i} = \CWLattlenZ{}(\RegTime{i})$, a tetrahedron in $\Cauchyt{i+1}$ with edges of length $\CWLattlenZ{i+1} = \CWLattlenZ{}(\RegTime{i+1})$, and four equal-length struts connecting each tetrahedral vertex in $\Cauchyt{i}$ to its time-evolved counterpart in $\Cauchyt{i+1}$.  \figref{fig:4block} depicts a typical 4-block;
for simplicity, we shall sometimes refer to the tetrahedron in $\Cauchyt{i+1}$ as the upper tetrahedron and the tetrahedron in $\Cauchyt{i}$ as the lower tetrahedron, as that is how they appear in the figure.

Since we are no longer using 4-simplices as the skeletal building block, specification of the edge-lengths will in general not be sufficient to determine the entire skeletal geometry.  There are two approaches one could take to completing the geometry.  The first is to fully triangulate each 4-block into 4-simplices by introducing extra edges; one would then have to specify the lengths of these new edges.  The second, which was taken by Collins and Williams, is to specify the 4-block's internal geometry.  Regardless of which approach is taken, the 4-block's geometry would be determined by two requirements:
\begin{enumerate*}[label=(\roman*)]
\item that all struts have the same length; and
\item that there be no twist or shear along the 4-block.
\end{enumerate*}
These requirements would manifest themselves either as a specification of the new edge-lengths in the first approach or as a specification of the 4-block's internal geometry in the second.  Brewin has likened the requirements to a choice of lapse and shift function in the ADM formalism.  Indeed, the standard form of the FLRW metric \eqref{FLRW} also implies a certain foliation of FLRW space-time, and Collins and Williams' choice seems closest to the lapse and shift implicit in this foliation.

These constraints imply that the lower tetrahedron would simply expand or contract uniformly about its centre when evolving to the upper tetrahedron.  The geometry can best be understood by introducing a co-ordinate system for the 4-block.  As depicted in \figref{fig:4block}, we label the vertices of the lower tetrahedron by $A$, $B$, $C$, $D$, and their counterparts in the upper tetrahedron by $A^\prime$, $B^\prime$, $C^\prime$, $D^\prime$, respectively.  The lower vertices' co-ordinates are then
\begin{equation}
\renewcommand{\arraystretch}{2}
\begin{aligned}
A &= \displaystyle{\left(-\frac{\CWLattlenZ{i}}{2}, -\frac{\CWLattlenZ{i}}{2\sqrt{3}}, -\frac{\CWLattlenZ{i}}{2\sqrt{6}}, \, \imath \RegTime{i} \right)},\\
B &=\displaystyle{\left(\frac{\CWLattlenZ{i}}{2}, -\frac{\CWLattlenZ{i}}{2\sqrt{3}}, -\frac{\CWLattlenZ{i}}{2\sqrt{6}}, \, \imath \RegTime{i} \right)},\\
C &= \displaystyle{\left(0, \frac{\CWLattlenZ{i}}{\sqrt{3}}, -\frac{\CWLattlenZ{i}}{2\sqrt{6}}, \, \imath \RegTime{i} \right)},\\
D &= \displaystyle{\left(0, 0, \frac{\sqrt{3}\, \CWLattlenZ{i}}{2\sqrt{2}}, \, \imath \RegTime{i} \right)};
\end{aligned}
\label{vertices}
\end{equation}
the co-ordinates of their upper counterparts are given by an analogous expression where each letter becomes primed and each subscript changes from $i$ to $i+1$.  Although a Euclidean metric is being used, the imaginary unit $\imath$ has been introduced to the time co-ordinate so that inner products would effectively yield a signature of $(+,+,+,-)$.  In Collins and Williams' approach, the 4-block's internal geometry is constrained to be that represented by these co-ordinates for given edge-lengths $\CWLattlenZ{i}$, $\CWLattlenZ{i+1}$, and $\CWstrut{i}$, that is, where the tetrahedron simply expands or contracts uniformly about its centre.

\begin{figure}[htb]
{\fontsize{8pt}{9.6pt}\input{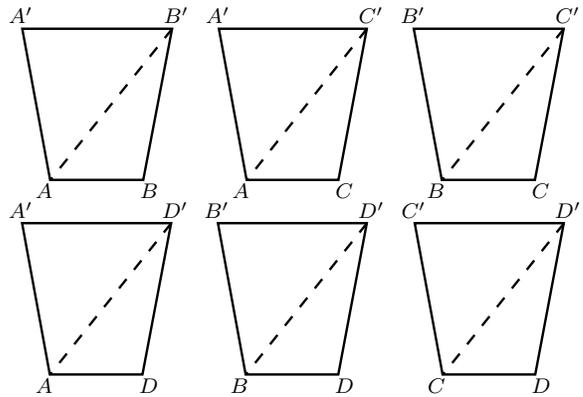}}
\caption{\label{fig:allhinges}The world-sheets generated by the six tetrahedral edges and their triangulation into triangular time-like hinges.}
\end{figure}

If we choose to fully triangulate the skeleton instead, then we shall introduce diagonals of type $AD^\prime$, $BD^\prime$, $CD^\prime$, $AC^\prime$, $BC^\prime$, and $AB^\prime$ in each 4-block; this corresponds to one diagonal above each of the tetrahedral edges, as illustrated in Figure \ref{fig:allhinges}, and these diagonals divide the 4-block into four distinct 4-simplices, $ABCDD^\prime$, $ABCC^\prime D^\prime$, $ABB^\prime C^\prime D^\prime$, and $AA^\prime B^\prime C^\prime D^\prime$.
Each 4-block in the skeleton will get triangulated in this manner, and it can be shown that this will lead to a consistent triangulation of the entire skeleton.  To be consistent with the geometry described by \eqref{vertices}, the diagonals would then be constrained to have a length given by
\begin{equation}
\CWdiag{i}^{\,2} = \frac{1}{3}\CWLattlenZ{i}^{\,2} + \frac{1}{24}\left(3\, \CWLattlenZ{i+1} + \CWLattlenZ{i} \right)^2 - \delta \RegTime{i}^{\,2},
\end{equation}
where $\delta \RegTime{i}$ denotes the quantity $\delta \RegTime{i} := \RegTime{i+1} - \RegTime{i}$.

When it comes to varying the Regge action, the standard approach would be to regard each edge as being independent of all others; the action would then be varied one edge-length at a time, keeping all other lengths constant, and each edge would generally lead to an independent Regge equation.  We shall refer to this approach as \emph{local variation}.  Local variation is always performed on a fully triangulated skeleton so as to avoid ambiguities in the varied skeleton's geometry.  Consider for instance varying the edge $AB$ in quadrilateral $ABA^\prime B^\prime$ from length $\CWLattlenZ{i}^{AB}$ to $\tilde{\CWlength}_i^{AB}$: without any constraints on the quadrilateral's internal geometry, there would be a wide range of possibilities that the varied quadrilateral could be, all possibilities having struts of length $\CWstrut{i}$ and tetrahedral edges of lengths $\tilde{\CWlength}_i^{AB}$ and $\CWLattlenZ{i+1}^{AB}$.  Thus constraints on the varied geometry are needed to determine a unique possibility.  Under local variation, the constraint is imposed by specifying a diagonal's length and then keeping that length constant when another edge is varied.

Collins and Williams however use a different approach to vary the skeleton: they would vary entire sets of identical edges at once; for instance, they would vary all tetrahedral edges on a Cauchy surface at once, or all struts between a pair of consecutive surfaces.  Under this variation, the tetrahedra in each 4-block would remain equilateral, the struts connecting them would continue having equal length, and the 4-block itself would continue having no twist or shear; thus even when varied, all 4-blocks would continue having the same internal geometry as that specified by the 4-block co-ordinate system \eqref{vertices}.  We shall refer to this approach as \emph{global variation}.  In standard general relativity, global variation would be analogous to requiring the metric in the Einstein-Hilbert action to be of FLRW form \eqref{FLRW}, and then varying the action with respect to $\FriedScale(t)$; this effectively imposes Copernican symmetries on the metric prior to varying the Einstein-Hilbert action.  On the other hand, local variation is analogous to the more standard approach where the Einstein-Hilbert action is varied first, yielding the Einstein field equations, and then the metric is set to be of FLRW form.

However we have previously modelled the $\Lambda$-FLRW universe using the CW formalism \cite{RGL-Williams}, and in those models, we showed that locally varying the Regge action for a CW skeleton led to a set of non-physical equations.  When we triangulated the skeleton, we had assumed that all tetrahedral edges in a Cauchy surface and all struts between a pair of surfaces would remain identical.  Thus, when we locally varied the skeleton, we would set the relevant edges to be equal to help simplify the resulting Regge equations; this has the effect of reducing the number of independent Regge equations to just one per distinct set of edges.  We were effectively assuming that had we not set the lengths equal, the complete set of Regge equations would reveal them to be equal anyway.  However, we believe this assumption to be unfounded because we believe the diagonals actually disrupt the symmetry between edges, rendering them no longer identical to each other.  The diagonals are not actually distributed uniformly across the skeleton: some vertices are attached to more diagonals than others; for instance in our 4-block triangulation above, vertex $A$ gets attached to three diagonals while vertex $D$ gets attached to none.  Thus, the struts, which correspond to the vertices' world-lines, would no longer be identical.  For similar reasons, other geometric objects, such as tetrahedral edge-lengths and their associated world-sheets, may no longer be identical either.  Therefore, it no longer makes sense to set edge-lengths to be equal after locally varying the skeleton; rather, each edge would have to be evolved independently of all others using its own Regge evolution equation, and there may potentially be one independent equation for each edge in the Cauchy surface.  We did not encounter any similar problem when we globally varied the Regge action; in that case, all edges of the same type would indeed be truly identical to each other.  Thus drawing on this lesson, in this paper, we shall consider only solutions to the global Regge equation for all models.

However, Brewin \cite{Brewin} has shown that, in certain cases, the global Regge equation can be related to the local one through a chain rule.  In such cases, the local solutions would form a subset of the global ones.  Although we have chosen to consider only global solutions, for the perturbed lattice universes, we shall derive the global solutions by using such a chain rule.  This chain rule relationship between the global and local Regge equations will be further elaborated on when we consider the perturbed universes in the fourth section.

Brewin has also drawn several analogies between the ADM formalism and the CW formalism.  He has likened the tetrahedral edge-lengths to the 3-metric of an ADM foliation and the Regge equations obtained from varying the tetrahedral edges to the ADM evolution equations.  He has also likened the struts and diagonals to the ADM lapse and shift functions, respectively, and the Regge equations obtained from their variation to the ADM Hamiltonian and momentum constraints, respectively.  Thus in this paper, we shall refer to the Regge equations obtained from the tetrahedral edges as evolution equations, from the struts as Hamiltonian constraints, and from the diagonals as momentum constraints.  In our study of the $\Lambda$-FLRW universe, we also found the Hamiltonian constraint to be a first integral of the evolution equation; thus we could study the universe's behaviour from the Hamiltonian constraint alone.  Similarly, it can be shown that for the models of the unperturbed lattice universe, which will be considered below, the Hamiltonian constraints are also first integrals of the evolution equations, though we shall not provide the proof.  We shall assume a similar conclusion holds for models of the perturbed lattice universe.  This conclusion means that we can determine the models' evolution from their constraint equations alone, and this is what we shall do.

Before leaving this section, we wish to make a final comment on using co-ordinate system \eqref{vertices} to calculate geometric quantities in the CW skeleton.  This co-ordinate system greatly facilitates the calculation of any such quantity, but these quantities would get expressed in terms of the time difference $\delta \RegTime{i}$ rather than purely in terms of skeletal edge-lengths.  In Regge calculus, it is the skeletal edge-lengths that are to be varied, so we shall have to convert $\delta \RegTime{i}$ into edge-lengths.  Only the strut-lengths $\CWstrut{i}$ depend on $\delta \RegTime{i}$, since increasing the time separation between a pair of consecutive Cauchy surfaces lengthens $\CWstrut{i}$ but leaves $\CWLattlenZ{i}$ and $\CWLattlenZ{i+1}$ unchanged.  Therefore by using \eqref{vertices} to calculate a strut-length, such as that of $AA^\prime$, we obtain the relation
 \begin{align}
\CWstrut{i}^2 &= \left(\frac{3}{8}\,\CWLattlenZdot{i}^{\,2} - 1 \right) \delta \RegTime{i}^2,\label{strut}\\
\intertext{where we have introduced the notation}
\CWLattlenZdot{i} &:= \frac{\CWLattlenZ{i+1}-\CWLattlenZ{i}}{\RegTime{i+1}-\RegTime{i}}.\nonumber
\end{align}

\section{Regge calculus of closed, regular lattice universes}
We shall now apply the CW formalism to model closed, regular lattice universes.  In this context, the Regge action \eqref{ReggeAction} can be greatly simplified.  First, a non-triangulated CW skeleton consists of only two distinct types of hinges, space-like triangular hinges corresponding to the triangles of the equilateral tetrahedra and time-like trapezoidal hinges generated by the world-sheets of the tetrahedral edges as they evolve from one Cauchy surface to the next.  An example of a triangular hinge would be $ABC$ in the 4-block described by \eqref{vertices}, and an example of a trapezoidal hinge would be $ABA^\prime B^\prime$.  Secondly, in the regular lattice universe, all masses are identical.  Hence, the action \eqref{ReggeAction} can be expressed as
\begin{equation}
8\pi \Action = \mkern-40mu \sum_{i \in \left\{\substack{\text{trapezoidal}\\\text{hinges}}\right\}} \mkern-40mu \TrapArea{i} \TrapDeficit{i} + \mkern-35mu \sum_{i \in \left\{\substack{\text{triangular}\\\text{hinges}}\right\}} \mkern-35mu \TriArea{i} \TriDeficit{i} - 8\pi N_p\, M \sum_{i \in \{ \RegTime{i} \}} \RegLinEl{i},
\label{EqualMassAction}
\end{equation}
where $M$ is the common mass of each particle, $N_p$ the total number of particles in the universe, and $\RegLinEl{i}$ the length of one particle's trajectory between Cauchy surfaces $\Cauchyt{i}$ and $\Cauchyt{i+1}$.  As we shall be varying with respect to the struts alone, the space-like triangular hinges can be ignored.

When this action is varied globally with respect to the struts $\CWstrut{j}$, we obtain the Regge equation
\begin{equation}
0 = \sum_i \frac{\partial \TrapArea{i}}{\partial \CWstrut{j}} \TrapDeficit{i} - 8\pi N_p\, M \sum_{i \in \{ \RegTime{i} \}} \frac{\partial \RegLinEl{i}}{\partial \CWstrut{j}},
\label{EqualMassReggeEqn}
\end{equation}
where the first summation is still over all trapezoidal hinges.

There are several possible ways to arrange the masses into a regular lattice on a Cauchy surface; examples include placing masses at the centres of the tetrahedra, the centres of the triangles, the mid-points of the tetrahedral edges, or the tetrahedral vertices.  Each of these configurations will yield a regular lattice; the new cell boundaries would lie along planes equidistant to pairs of masses that are nearest neighbours to each other, and the masses would consequently lie at the centres of the new cells.  A 2-dimensional analogue has been illustrated in \figref{fig:new_lattice} of \appendref{App_Latt}, where the original lattice, drawn in solid lines, consists of equilateral triangles tessellating flat 2-dimensional space; masses placed at the mid-points of the triangular edges result in a new lattice consisting of rhomboidal cells, drawn in dashed lines.  However as discussed in \appendref{App_Latt}, not all new lattices would correspond to Coxeter lattices; those that do not would have non-regular polytopes as lattice cells and would consequently have reduced lattice symmetries.  However we see that in this way, the CW formalism can allow us to go beyond Coxeter lattices and model other lattice universes, something which would not be possible with the LW formalism because spherical cells, which that formalism uses, can only well-approximate lattice cells that are regular polytopes \cite{RGL}.  Once a particular arrangement of masses has been chosen, then by symmetry, the masses will maintain that arrangement on all subsequent Cauchy surfaces.  The masses are therefore co-moving with respect to the Cauchy surface.

Let us consider masses located in each tetrahedron at the general location of
$$
\Tensb{v}_i = \alpha\, \vecb{A} + \beta\, \vecb{B} + \gamma\, \vecb{C} + \delta\, \vecb{D},
$$
with vectors $\vecb{A}$, $\vecb{B}$, $\vecb{C}$, $\vecb{D}$ denoting the position vectors of the tetrahedron's four vertices relative to the tetrahedral centre, and with constants $\alpha$, $\beta$, $\gamma$, $\delta$ satisfying $0 \leq \alpha, \beta, \gamma, \delta \leq 1$ and $\alpha+\beta+\gamma+\delta = 1$.  Each mass is located at a distance $\lvert \Tensb{v}_i \rvert$ from the tetrahedral centre given by
\begin{equation}
\begin{split}
\Tensb{v}_i\cdot\Tensb{v}_i ={} & \frac{1}{8} \left[3 (\alpha^2 + \beta^2 + \gamma^2 + \delta^2) \right. \\
& \left. \hphantom{\frac{1}{8} [} {}- 2 (\alpha \beta + \alpha \gamma + \alpha \delta + \beta \gamma + \beta \delta + \gamma \delta) \right] \CWLattlenZ{i}^2,
\end{split}
\label{Unperturbed:MassDistance}
\end{equation}
with $\CWLattlenZ{i}$ being the length of a tetrahedral edge; as $\lvert \Tensb{v}_i \rvert / \CWLattlenZ{i}$ is a constant of time, we shall denote this constant by $v$.

We shall work with 4-block co-ordinates given by \eqref{vertices} and their $\Cauchyt{i+1}$ counterparts to determine all geometric quantities; we can consider position vectors $\vecb{A}$, $\vecb{B}$, $\vecb{C}$, $\vecb{D}$ as being identical to the position of vertices $A$, $B$, $C$, $D$ in \eqref{vertices}.  In these co-ordinates, the length $\RegLinEl{i}$ of each mass' line element between $\Cauchyt{i}$ and $\Cauchyt{i+1}$ is given by
$$
\RegLinEl{i}^2 = v^2 \delta \CWLattlenZ{i}^2 - \delta \RegTime{i}^2,
$$
where $\delta \CWLattlenZ{i}$ denotes the difference $\delta \CWLattlenZ{i} := \CWLattlenZ{i+1} - \CWLattlenZ{i}$.  In terms of the strut-length $\CWstrut{i}$, $\RegLinEl{i}^2$ can be expressed as
\begin{equation}
\RegLinEl{i}^2 = \left(v^2 - \frac{3}{8}\right)\delta \CWLattlenZ{i}^2 + \CWstrut{i}^2,
\label{UnperturbedLineElement}
\end{equation}
where we have made use of \eqref{strut} to substitute for $\delta \RegTime{i}^2$.  Then varying $\RegLinEl{i}$ with respect to $\CWstrut{j}$ yields
\begin{equation}
\frac{\partial \RegLinEl{i}}{\partial \CWstrut{j}} = \frac{\CWstrut{i}}{\RegLinEl{i}} \delta_{ij}.
\label{s_m}
\end{equation}

The area of any trapezoidal hinge between $\Cauchyt{i}$ and $\Cauchyt{i+1}$ is
\begin{equation}
\TrapArea{i} = \frac{\imath}{2}(\CWLattlenZ{i+1}+\CWLattlenZ{i})\left[\frac{1}{4}(\CWLattlenZ{i+1}-\CWLattlenZ{i})^2 - \CWstrut{i}^2 \right]^{\frac{1}{2}},
\label{TrapArea}
\end{equation}
and varying this with respect to $\CWstrut{j}$ yields
\begin{equation}
\frac{\partial \TrapArea{i}}{\partial \CWstrut{j}} = -\frac{\imath}{2}\CWstrut{i}(\CWLattlenZ{i+1}+\CWLattlenZ{i})\left[\frac{1}{4}(\CWLattlenZ{i+1}-\CWLattlenZ{i})^2 - \CWstrut{i}^2 \right]^{-\frac{1}{2}} \delta_{ij}.
\label{TrapAreaVaried}
\end{equation}

Because all 4-blocks meeting at a trapezoidal hinge are identical, the hinge's deficit angle simplifies to
\begin{equation}
\TrapDeficit{i} = 2 \pi - n \CWTrapdihedral{i},
\label{TrapDeficit}
\end{equation}
where $n$ is the number of faces meeting at the hinge and $\CWTrapdihedral{i}$ is the dihedral angle between any two adjacent faces.  Since each trapezoidal hinge corresponds to the world-sheet of a tetrahedral edge and each face on this hinge to the world-tube of a triangle at this edge, $n$ is equal to the number of triangles meeting at an edge; this number is given by the last column of \tabref{tab:primary}.

To determine $\CWTrapdihedral{i}$, let us consider the representative hinge $ABA^\prime B^\prime$.  Faces $ABCA^\prime B^\prime C^\prime$ and $ABDA^\prime B^\prime D^\prime$ meet at this hinge and will be separated by a dihedral angle of $\CWTrapdihedral{i}$; thus, we can determine $\CWTrapdihedral{i}$ from the scalar product of the two faces' unit normals.  Let $\Tensb{\hat{n}}_1$ denote the unit normal pointing into $ABCA^\prime B^\prime C^\prime$ and $\Tensb{\hat{n}}_2$ the unit normal out of $ABDA^\prime B^\prime D^\prime$; then in co-ordinate system \eqref{vertices}, they have co-ordinates
\begin{IEEEeqnarray}{rCl}
\renewcommand{\arraystretch}{2.7}
\hat{n}_1^\mu &=& \displaystyle \frac{\left(0, 0, 1, -\imath \frac{1}{2\sqrt{6}}\, \CWLattlenZdot{i} \right)}{\left(1-\frac{1}{24}\, \CWLattlenZdot{i}^{\,2}\right)^{\frac{1}{2}}}\\
\shortintertext{and}
\hat{n}_2^\mu &=&\displaystyle \frac{\left(0, -2\sqrt{2}, 1, \imath \frac{\sqrt{3}}{2\sqrt{2}}\, \CWLattlenZdot{i}\right)}{3\left(1-\frac{1}{24}\, \CWLattlenZdot{i}^{\,2}\right)^{\frac{1}{2}}};
\end{IEEEeqnarray}
and their scalar product leads to the relation
\begin{equation}
\cos \CWTrapdihedral{i} =\frac{1+\frac{1}{8}\, \CWLattlenZdot{i}^{\,2}}{3-\frac{1}{8}\, \CWLattlenZdot{i}^{\,2}}. \label{cosq}
\end{equation}

We now have all the relevant geometric quantities necessary to solve the Regge equation \eqref{EqualMassReggeEqn}.  Thus by substituting \eqref{TrapAreaVaried}, \eqref{TrapDeficit}, \eqref{strut}, and \eqref{s_m} into \eqref{EqualMassReggeEqn}, we are led to the Hamiltonian constraint
\begin{equation}
\CWLattlenZ{i} = 8\pi M\, \frac{N_p}{N_1} \left[ \frac{\frac{1}{8} \CWLattlenZdot{i}^2-1}{v^2\, \CWLattlenZdot{i}^2 - 1} \right]^\frac{1}{2} \frac{1}{2\pi - n\CWLattdihedral{i}},
\label{discrete_equal_M_Regge0}
\end{equation}
where $N_1$ is the total number of tetrahedral edges on a Cauchy surface.  By using \eqref{cosq}, we can express $\CWLattlenZdot{i}$ as a function of the dihedral angle $\CWLattdihedral{i}$ and thereby re-express \eqref{discrete_equal_M_Regge0} as
\begin{equation}
\CWLattlenZ{i} = 8 \pi M\, \frac{N_p}{N_1} \frac{\tan \frac{\CWLattdihedral{i}}{2}}{\left[8 v^2 \tan^2 \frac{\CWLattdihedral{i}}{2} - \frac{1}{2}\left(8v^2-1\right)\right]^\frac{1}{2}} \frac{1}{2\pi - n \CWLattdihedral{i}}.
\label{discrete_equal_M_Regge}
\end{equation}

We shall now determine the continuum time limit of the constraint equation.  When taking this limit, such that as $\delta \RegTime{i} \to 0$,
\begin{IEEEeqnarray*}{rCl}
\CWLattdihedral{i} &\to& \CWLattdihedral{} + \Odtone,\\
\CWLattlenZ{i} &\to& \CWLattlenZ{},\\
\CWLattlenZ{i+1} &\to& \CWLattlenZ{} + \CWLattlenZdot{} dt + \Odt{2},
\end{IEEEeqnarray*}
the various geometric quantities become
\begin{IEEEeqnarray}{rCl}
\CWLattlenZdot{i} &\to& \CWLattlenZdot{} + \Odtone,\\
\CWstrut{i}^2 &\to& \left(\frac{3}{8}\CWLattlenZdot{}^{\,2} - 1 \right)\, dt^2 + \Odt{3},\label{strut-cont-time}\\
\cos \CWTrapdihedral{i} &\to& \cos \CWTrapdihedral{} \approx \frac{1+\frac{1}{8}\CWLattlenZdot{}^{\,2}}{3-\frac{1}{8}\CWLattlenZdot{}^{\,2}} + \Odtone.\label{cosq-cont-time}
\end{IEEEeqnarray}
Relation \eqref{cosq-cont-time} can then be inverted to parametrise $\CWLattlenZdot{}$ in terms of $\CWTrapdihedral{}$, thus yielding
\begin{equation}
\CWLattlenZdot{}^{\,2} = 8\left[1-2 \, \tan^2\left(\frac{1}{2}\, \CWTrapdihedral{}\right)\right].
\label{ldot}
\end{equation}
Finally, the constraint equation \eqref{discrete_equal_M_Regge} becomes
\begin{equation}
\CWLattlenZ{} = 8 \pi M\, \frac{N_p}{N_1} \frac{\tan \frac{\CWLattdihedral{}}{2}}{\left[8 v^2 \tan^2 \frac{\CWLattdihedral{}}{2} - \frac{1}{2}\left(8v^2-1\right)\right]^\frac{1}{2}} \frac{1}{2\pi - n \CWLattdihedral{}}.
\label{cont_l_equal_mass}
\end{equation}
Equations \eqref{ldot} and \eqref{cont_l_equal_mass} provide a parametric description of the universe's evolution in terms of $\CWLattdihedral{}$.

It is often easiest to study a model's evolution by considering the evolution of its Cauchy surface's volume, and we shall do this below.  In the continuum time limit, the CW Cauchy surface volume is given by
\begin{equation}
U_\Ntet(\RegTime{}) = \frac{\Ntet}{6\sqrt{2}} \, \CWLattlenZ{}(\RegTime{})^3,
\label{ParentU}
\end{equation}
where $\Ntet$ is the number of tetrahedra in the surface and is given by the first column of \tabref{tab:primary}; the volume's rate of expansion is given by
\begin{equation}
\dot{U}_\Ntet(\RegTime{}) = \frac{\Ntet}{2\sqrt{2}}\, \CWLattlenZ{}(\RegTime{})^2\, \CWLattlenZdot{}(\RegTime{}).
\label{ParentdUdt}
\end{equation}

Both $\CWLattlenZdot{}$ and the lengths of the struts place constraints on the range of $\CWLattdihedral{}$.  In the continuum time limit, the strut-length is given by relation \eqref{strut-cont-time}, and for the strut to remain time-like, we require that $\CWLattdihedral{} > \frac{\pi}{3}$.  On the other hand, for $\CWLattlenZdot{}^2$ to be positive, we require that $\CWLattdihedral{} \leq 2 \arctan \frac{1}{\sqrt{2}}$.  Thus, $\CWLattdihedral{}$ is constrained to lie in the range
\begin{equation}
\frac{\pi}{3} < \CWLattdihedral{} \leq 2 \arctan \frac{1}{\sqrt{2}}.
\label{qconstraint}
\end{equation}

For the square root in \eqref{cont_l_equal_mass} to be real, we also require that
\begin{equation}
\CWLattdihedral{} > 2 \arctan \left[ \frac{v^2-\frac{1}{8}}{2v^2} \right]^\frac{1}{2};
\end{equation}
thus $\CWLattlenZ{}$ diverges if $\CWLattdihedral{} = 2 \arctan \left[ \frac{v^2-\frac{1}{8}}{2v^2} \right]^\frac{1}{2}$.  We can compare this with the range of $\CWLattdihedral{}$, as given by \eqref{qconstraint}, to see under what conditions will divergence occur.  If $2 \arctan \left[ \frac{v^2-\frac{1}{8}}{2v^2} \right]^\frac{1}{2} \geq \frac{\pi}{3}$, then $\CWLattlenZ{}$ will definitely diverge; this happens if
$$
v^2 \geq \frac{3}{8}.
$$
Equality would correspond to placing the masses at the tetrahedral vertices.  However the lower bound of $\CWLattdihedral{} > \frac{\pi}{3}$ came from requiring that the struts be time-like, not from any direct constraints on $\CWLattlenZ{}$ itself.  Thus, even if the model satisfies $2 \arctan \left[ \frac{v^2-\frac{1}{8}}{2v^2} \right]^\frac{1}{2} < \frac{\pi}{3}$, we may still see the beginnings of a divergence in $\CWLattlenZ{}$ that is abruptly cut off by the $\CWLattdihedral{} = \frac{\pi}{3}$ bound.  Therefore, for $\CWLattlenZ{}$ to be unconditionally convergent, we require the more stringent constraint that 
$$
0 < - \frac{1}{2} (8v^2-1),
$$
which comes from requiring the square-root in \eqref{cont_l_equal_mass} to be real and non-zero, even when $\CWLattdihedral{} = 0$; thus $v^2$ must satisfy
$$
v^2 < \frac{1}{8}.
$$
When $v^2 = 1/8$, the masses are located at the mid-points of the tetrahedral edges.  Therefore, the universe will only be unconditionally convergent if the masses are placed within a spherical region in the centre of the tetrahedron with a boundary that just touches the mid-points of the tetrahedral edges.  Such a region would include the centres of the triangles and of the tetrahedra itself.

Interestingly, Collins and Williams \cite{CW} found a similar result in their study of closed dust-filled universes.  They were considering different ways to measure the `time' of the universe by using the proper time $\tau$ of test particles located at different positions in the tetrahedron.  We note that $\tau$ is actually identical to the continuum time limit of $\RegLinEl{i}$, that is, the continuum time limit of the square-root of \eqref{UnperturbedLineElement}.  They found that $dU_\Ntet / d\tau$, where $U_\Ntet$ is the volume of the universe as given by \eqref{ParentU}, would diverge if the test particle were outside the same spherical region of convergence as the one we obtained.  We suspect this region of convergence may be a generic feature of any model based on CW skeletons.

Although masses situated at or near vertices will cause the resulting model to diverge, there may be a way around this problem.  Each closed Coxeter lattice admits a dual Coxeter lattice centred on the original lattice's vertices, as noted in \appendref{App_Latt}, although the dual lattice may not necessarily be a tetrahedral lattice.  Each model thus admits a dual model using Cauchy surfaces based on the dual lattice, and we can always extend the CW formalism to perform Regge calculus with these non-tetrahedral models.  Where masses would have been at the vertices in the original model, when translated to the dual model, they would now be at the centres of the dual cells.  As a result, the particles would now be both co-moving with respect to the Cauchy surface and following geodesics globally across the entire Regge space-time; this follows because even with non-tetrahedral cells, each cell would simply expand or contract uniformly about its centre as the Cauchy surface evolves, so a particle at the cell centre would simply propagate along the temporal direction only.  We additionally conjecture that in non-tetrahedral models, there would also be an analogous spherical region of convergence centred at the cell centres regarding the placement of massive particles.  However, it remains to be seen whether the spherical regions of both the original and dual models would completely cover the entire Cauchy surface such that for any configuration of the particles, there is always a model for which the configuration would be well-behaved.  As noted in \appendref{App_Latt}, the 5-tetrahedra lattice is dual to itself, so in this case, the region of convergence in the dual model is exactly identical to the region of convergence in the original model, and if masses are positioned at the vertices of the original model, then when they are translated to the dual model, the resulting model will behave in exactly the same way as the original model with the masses at its cell centres.

In the foregoing discussion, we have only considered universes consisting of a single set of regularly distributed masses.  We can also consider the more general case where we have not one but several different sets of masses, each set being regularly distributed and having equal magnitude, though the magnitude can differ between sets.  If we work in the original lattice, then each set would be arranged in a way that preserves the lattice symmetries, since, as mentioned above, the sets would lie at such sites as the lattice vertices or the mid-points of the edges.  Therefore, even if the magnitudes of the masses differ between sets, because the lattice symmetries remain preserved, the edges in particular would all remain identical to each other; hence, the lattice would still have only one length-scale $\CWLattlenZ{}$.  A similar argument applies when the masses are translated to the dual model.  Therefore for such models with multiple sets of masses, it can be shown that the continuum time equation for $\CWLattlenZ{}$ becomes
\begin{equation}
\CWLattlenZ{} = 8 \pi \sum_i M_i\, \frac{N_i}{N_1} \frac{\tan \frac{\CWLattdihedral{}}{2}}{\left[8 v_i^2 \tan^2 \frac{\CWLattdihedral{}}{2} - \frac{1}{2}\left(8v_i^2-1\right)\right]^\frac{1}{2}} \frac{1}{2\pi - n \CWLattdihedral{}},
\end{equation}
where $i$ labels the set, the summation is over all sets, $N_i$ is the number of particles in set $i$, $M_i$ is the magnitude of the masses in set $i$, and $v_i$ denotes the parameter $v$ for set $i$.  It is clear that this equation has the same convergence conditions as those for the single-set universe.  We note though that there are certain set combinations that would not lead to well-behaved models, even when translated to dual models.  One example would be a 5-tetrahedra model where one set lies at the tetrahedral centres and another at the vertices; one of the two sets would always lie outside the region of convergence, regardless of whether the original model or the dual was being considered.

\begin{figure*}[htb]
\subfloat[\hspace{-0.7cm}\label{fig:EqualMassGraphs-tets}]{\fontsize{8pt}{9.6pt}\input{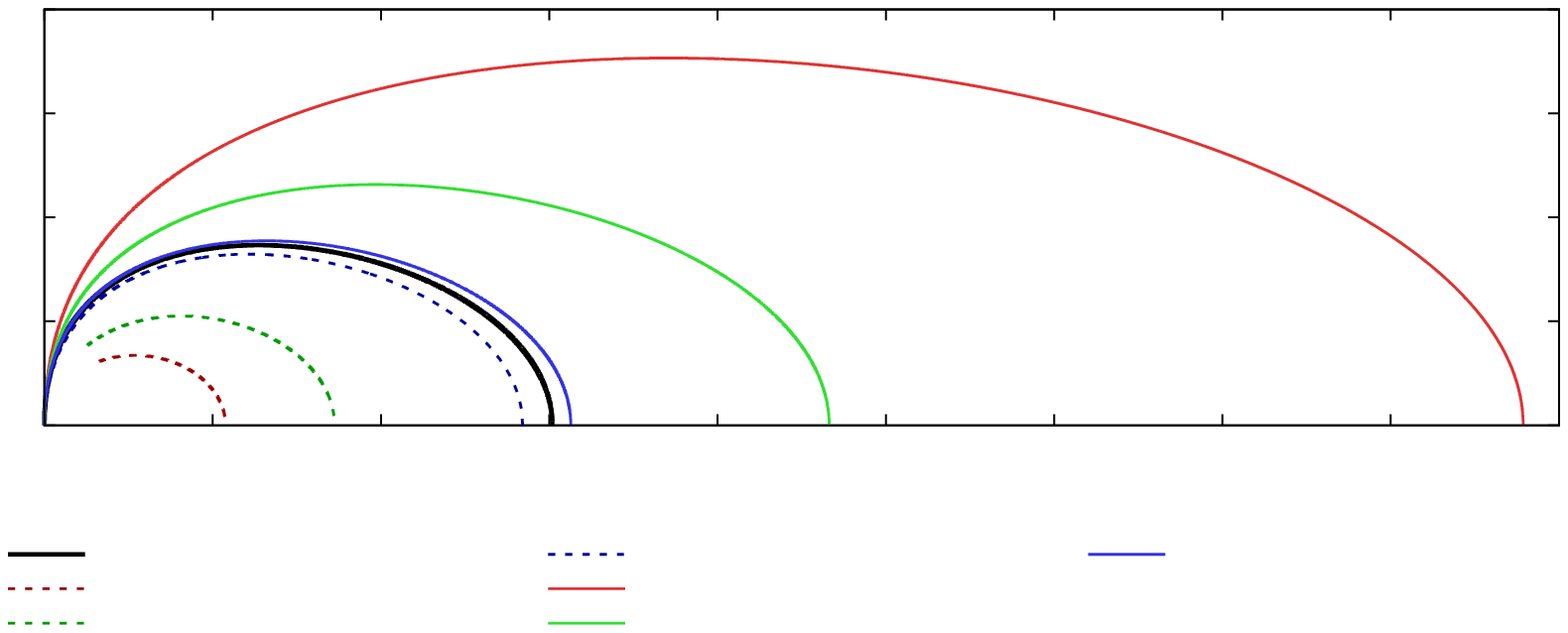}}\\
\subfloat[\hspace{-0.7cm}\label{fig:EqualMassGraphs-edges}]{\fontsize{8pt}{9.6pt}\input{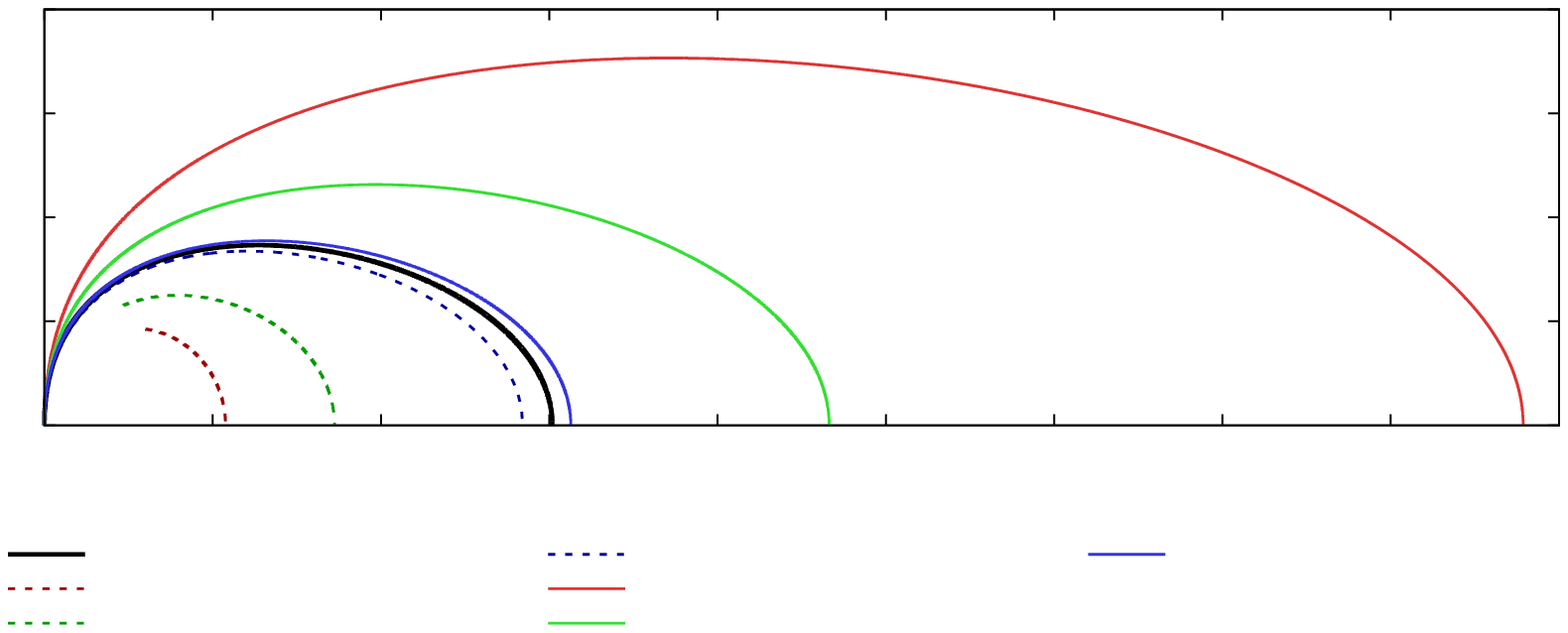}}
\caption{\label{fig:UnperturbedGraphs}The expansion rate of the universe's volume $dU/dt$ versus the volume $U$ itself for the dust-filled FLRW universe, the three different Regge models, and the three equivalent LW models.  In the Regge models, the masses have been positioned (a) at the centres of the tetrahedra and (b) at the mid-points of the tetrahedral edges.  The Regge universe volume is given by the sum of the volumes of the Cauchy surface's constituent tetrahedra, while the LW universe volume is given by the volume of a 3-sphere with a radius equal to the lattice universe scale factor $\LattScale(\LattTime)$.  The universe's total mass is the same across all universes.}
\end{figure*}
Returning to the single-set universes, to illustrate the behaviour of our various Regge models, we have plotted $dU/dt$ against $U$ in \figref{fig:UnperturbedGraphs}, where $U$ is the universe's volume as given by \eqref{ParentU} for the Regge models.  For comparison, we have also plotted the corresponding graphs for the dust-filled closed FLRW model and for the equivalent LW models in \cite{RGL}.  Since the FLRW Cauchy surfaces are 3-spheres, the FLRW universe's volume is given by
\begin{equation}
U_\text{FLRW}(t) = 2\pi^2 \FriedScale(t)^3,
\label{FLRW:closedU}
\end{equation}
and its rate of expansion by 
\begin{equation}
\dot{U}_\text{FLRW}(t) = 6\pi^2 \FriedScale(t)^2 \, \dot{\FriedScale}(t);
\label{FLRW:closeddUdt}
\end{equation}
for a dust-filled FLRW universe, the scale factor $\FriedScale(t)$ is given parametrically by
\begin{align}
\FriedScale &= \displaystyle \frac{\FriedScale_0}{2}(1 - \cos \eta),\\
t &= \displaystyle \frac{\FriedScale_0}{2}(\eta - \sin \eta),
\end{align}
where 
\begin{equation}
\FriedScale_0 = \frac{8\pi\rho_0}{3},
\label{FLRW_a0}
\end{equation}
and where $\rho_0$ is the energy density when $a=1$.  We have taken the LW universe's `volume' to be the volume of a 3-sphere as well, though with the radius set equal to the LW lattice universe scale factor $\LattScale(\LattTime)$ instead;\footnote{We do not imply anything physical by this LW 3-sphere; it has merely been defined to facilitate comparison of length-scales between LW and Regge models.  We could instead have chosen to graph $\dot{\LattScale}(\LattTime)$ against $\LattScale(\LattTime)$ for the LW universes and $\CWLattlenZdot{}/Z$ against $\CWLattlenZ{}/Z$, where $Z$ is some suitably chosen constant that converts the Regge edge-lengths into an embedding 3-sphere radius; possible embeddings were explored at great length in \cite{RGL-Williams}.  The comparison of the models would effectively be equivalent.} that is, the universe's volume and expansion rate are given, respectively, by 
\begin{align}
U_\text{LW}(\LattTime) ={} & 2\pi^2 \LattScale(\LattTime)^3,\label{closedU_LW}\\
\dot{U}_\text{LW}(\LattTime) ={} & 6\pi^2 \LattScale(\LattTime)^2 \, \dot{\LattScale}(\LattTime),
\label{closed_dUdtau_LW}
\end{align}
and the LW scale factor $\LattScale(\LattTime)$ is given parametrically by
\begin{align}
\LattScale ={} & \displaystyle \frac{r_{\scriptscriptstyle E_b}}{2}\sqrt{\frac{r_{\scriptscriptstyle E_b}}{2m}} \, (1 - \cos\eta),\\
\LattTime ={} & \displaystyle \frac{r_{\scriptscriptstyle E_b}}{2}\sqrt{\frac{r_{\scriptscriptstyle E_b}}{2m}}(\eta - \sin \eta),
\end{align}
where
\begin{align*}
r_{\scriptscriptstyle E_b} ={} & \frac{2m}{1-E_b},\\
E_b ={} & \cos^2 \psi_{\Ntet},
\end{align*}
and where the angle $\psi_{\Ntet}$ is given by
$$
\frac{1}{\Ntet} = \frac{2\psi_\Ntet - \sin 2\psi_\Ntet}{2\pi},
$$
with $\Ntet$ being the number of tetrahedra in the model, as listed in the first column of \tabref{tab:primary}.  Two classes of Regge models are shown in \figref{fig:UnperturbedGraphs}: on the one extreme, \figref{fig:EqualMassGraphs-tets} shows the behaviour when particles are positioned at the centres of the tetrahedra, while on the other, \figref{fig:EqualMassGraphs-edges} shows the behaviour when they are at the mid-points of the tetrahedral edges, right on the boundary of the region of unconditional convergence.

There are several features to note from these graphs.  The evolution of all lattice models are well-behaved and closed.  Moreover, the behaviour of the approximations converge to that of the FLRW universe as the number of particles is increased, which is to be expected because the matter content then becomes more dust-like.  However, the Regge models and the LW models converge from opposite directions.  It is difficult to say \emph{a priori} whether the LW approximation or the Regge approximation is more representative of the actual lattice universe: on the one hand, the average matter density is reduced as the number of particles is reduced, and this would weaken the gravitational interaction between particles, but on the other, the mass of each particle gets increased, and this would strengthen the interaction.  However, comparison with exact initial value data on the time-symmetric hypersurface seems to favour the LW models \cite{CRT}: using various measures for the lattice scale factor $\LattScale_{latt}$ on such a hypersurface, Clifton \emph{et al.}\ have shown that the ratio $\LattScale_{latt} / \FriedScale_{\scriptscriptstyle FLRW}$ would always approach unity from above as the number of lattice masses increases.

We note that there is a degree of ambiguity in choosing a common measure with which to compare the two approximations.  It was demonstrated in \cite{RGL-Williams} that CW Cauchy surfaces can be embedded into 3-spheres in $\mathbf{E}^4$; the radius of the embedding 3-sphere could serve as a measure of the CW Cauchy surface's scale factor, but it was shown that there were multiple ways to define the embedding and hence the radius, although each possible radius would simply be a constant re-scaling of the tetrahedral edge-length $\CWLattlenZ{}(t)$.  Each radius implies a different 3-sphere volume, which results in volume graphs of different magnitudes, although regardless of which scaling is chosen, the graphs would still remain closed and well-behaved.  Thus, it is within the realm of possibility that the models are actually equivalent and that the apparent difference is due to an inappropriate choice of scaling.

However, we suspect that the principal factor behind the two approximations' discrepancy is the finite resolution of the lattice cell's geometry in the Regge models.  We have constructed Regge models where the entire geometry of each cell has been reduced to one quantity, that of the tetrahedral edge-length; the cell's interior geometry consists uniformly of Minkowski space-time throughout, with no hinges or extra edges to provide additional geometric information, and such a model is clearly very coarse-grained.  In contrast, LW models approximate each cell's interior geometry with Schwarzschild space-time, and, although not perfect, this approximation should still be more accurate than completely uniform Minkowski space-time.  One could perhaps fine-grain the Regge models by further triangulating each lattice cell into smaller tetrahedra; this would certainly introduce more edges and hence more geometric information into each cell, although the smaller tetrahedra would no longer be identical nor equilateral.  One algorithm for subdividing these tetrahedra has been developed by Brewin \cite{Brewin}.  Alternatively, one might be able to construct a finer-grained 5-cell universe, for instance, by distributing five massive particles in a regular manner in a 600-tetrahedra Cauchy surface; it remains to be seen whether such a regular distribution of the masses can be obtained.  Though assuming it is possible, then each cell would now be associated with several edge-lengths rather than just one, and this would also provide greater geometric information about each lattice cell.  Nevertheless, we see that even these coarse-grained models can still reveal much qualitative information about the lattice universe's behaviour, most notably its stability and its closed evolution; thus we shall continue to use them to study the lattice universe.

Our first method for increasing the number of edge-lengths per lattice universe effectively `spaces-out' the masses on the CW Cauchy surface so that each lattice universe cell would consist of multiple Cauchy surface cells.  It may actually be easier to do this in Regge models of flat and open lattice universes.  The CW formalism should readily be extendible to Cauchy surfaces based on the flat and open Coxeter lattices described in \appendref{App_Latt}.  Because such Cauchy surfaces would extend indefinitely outwards in all directions, it would be far easier to `space-out' the masses: for instance, in Cauchy surfaces based on the flat cubic Coxeter lattice, one could easily place a particle at the centre of every $n^\text{th}$ cube for some suitably large $n$, and each lattice universe cell would then consist of $n^3$ Cauchy surface cubes.  In extending the CW formalism to these other Coxeter lattices, it would also be interesting to see what analogous regions of convergence and dual models these new models would admit.  For the moment though, we leave such investigations to future work.

We shall henceforth work with universes where the masses are located at the centres of the tetrahedra.  This would place the masses right in the middle of the region of convergence.  Furthermore, as mentioned above, this is the only position in the tetrahedra where co-moving masses would also be following geodesics of the Regge space-time, which is closest to what happens in an FLRW universe.

\section{Perturbation of a single mass}
We shall now construct a universe where the magnitude of a single central mass gets perturbed from $M$ to $M^\prime$.  This would induce a commensurate perturbation in the surrounding geometry, specifically in the lengths of the edges surrounding the perturbed mass.  In general, there will be several sets of edges, each with its own independent length, and we must determine what those sets are.  We begin with the edges in the tetrahedron enclosing the perturbed mass.  The Cauchy surface will remain symmetric about this single perturbed mass; therefore the enclosing tetrahedron will still remain equilateral but with perturbed edge-lengths $\CWLattlen{1}{i}$.  We shall refer to this tetrahedron as a Type I tetrahedron.  Next, we consider the four tetrahedra that share a face with the Type I tetrahedron.  We shall refer to these as Type II tetrahedra.  These tetrahedra will no longer be equilateral: they will each have an equilateral base in the face shared with the Type I tetrahedron, but they will also have three other edges identical to each other but different from the $\CWLattlen{1}{i}$ edges.  In the 5-tetrahedra model, these two sets of edges are the only ones present, and we denote the length of the second set by $\CWLattlen{0}{i}$.  In the 16 and 600-tetrahredra models, we must keep going; we must next consider the new sets of tetrahedra neighbouring the ones already considered; each additional set may introduce a new set of edges; and each additional set of edges will generally have a length different from the lengths of all other edges so far considered.  Thus, there will be far more than just two sets of edges in these models.  For example, the 16-tetrahedra model has four distinct edge-lengths.  One might consider simplifying the model by constraining all edges to have a common length $\CWLattlen{0}{i}$ apart from the edges $\CWLattlen{1}{i}$ of the equilateral tetrahedron enclosing the perturbed mass; that is, one might attempt to localise all perturbation in geometry to just this equilateral tetrahedron.  We attempted this but found that the resulting model was self-inconsistent.  Therefore, we cannot study all three models in a general manner but must construct and examine each one individually.  We have chosen to focus only on the 5-tetrahedra model as it is the most tractable, involving only two distinct edge-lengths.

We shall once again work with a co-ordinate system for the tetrahedra's 4-blocks, though each tetrahedral type requires its own system.  Since the Type I tetrahedron is equilateral, we can use the same co-ordinates for its 4-block as those in \eqref{vertices} and its $\Cauchyt{i+1}$ counterpart, although we must replace $\CWLattlenZ{i}$ and $\CWLattlenZ{i+1}$ with $\CWLattlen{1}{i}$ and $\CWLattlen{1}{i+1}$, respectively; we shall continue to label the lower tetrahedron's vertices by $A$, $B$, $C$, $D$ and their upper counterparts by $A^\prime$, $B^\prime$, $C^\prime$, $D^\prime$.

For a Type II tetrahedron's 4-block, we shall label the vertices of the lower tetrahedron by $A$, $B$, $C$, $E$, with $E$ denoting the tetrahedron's apex, and we shall denote their counterparts in the upper tetrahedron by $A^\prime$, $B^\prime$, $C^\prime$, $E^\prime$; vertices $A$, $B$, $C$ are shared with the Type I tetrahedron in $\Cauchyt{i}$ and $A^\prime$, $B^\prime$, $C^\prime$ with the Type I tetrahedron in $\Cauchyt{i+1}$.  We assign to the lower tetrahedron the co-ordinates
\begin{equation}
\renewcommand{\arraystretch}{2}
\begin{aligned}
A &= \displaystyle{\left(-\frac{\CWLattlen{1}{i}}{2}, -\frac{\CWLattlen{1}{i}}{2\sqrt{3}}, 0, \, 0 \right)},\\
B &=\displaystyle{\left(\frac{\CWLattlen{1}{i}}{2}, -\frac{\CWLattlen{1}{i}}{2\sqrt{3}}, 0, \, 0 \right)},\\
C &= \displaystyle{\left(0, \frac{\CWLattlen{1}{i}}{\sqrt{3}}, 0, \, 0 \right)},\\
E &= \displaystyle{\left(0, 0, \axislen{i}, 0 \right)},
\end{aligned}
\label{SLvertices}
\end{equation}
where we have introduced the symbol
$$\axislen{i} := \sqrt{\left(\CWLattlen{0}{i}\right)^2 - \frac{1}{3}\left(\CWLattlen{1}{i}\right)^2}$$
to denote the length of the tetrahedron's central axis.

By symmetry, the base $A^\prime B^\prime C^\prime$ of the upper tetrahedron should simply be a uniform expansion or contraction of the lower tetrahedron's base; there may also be a shift $\delta z_i$ in the spatial direction orthogonal to the base, but that is all.  This shift would be determined by the lengths of the struts.  Thus, the upper base vertices $A^\prime$, $B^\prime$, $C^\prime$ should have the co-ordinates
\begin{IEEEeqnarray*}{rCl}
A^\prime &=& \displaystyle{\left(-\frac{\CWLattlen{1}{i+1}}{2}, -\frac{\CWLattlen{1}{i+1}}{2\sqrt{3}}, \delta z_i, \, \imath \delta \RegTime{i}^{(1)} \right)},\\
B^\prime &=& \displaystyle{\left(\frac{\CWLattlen{1}{i+1}}{2}, -\frac{\CWLattlen{1}{i+1}}{2\sqrt{3}}, \delta z_i, \, \imath \delta \RegTime{i}^{(1)} \right)},\\
C^\prime &=& \displaystyle{\left(0, \frac{\CWLattlen{1}{i+1}}{\sqrt{3}}, \delta z_i, \, \imath \delta \RegTime{i}^{(1)} \right)},
\end{IEEEeqnarray*}
where $\delta t_i^{(1)}$ is the temporal shift of the upper base with respect to the lower one.  Both $\delta z_i$ and $\delta \RegTime{i}^{(1)}$ can be deduced from the fact that
\begin{enumerate*}[label=(\roman*)]
\item struts $AA^\prime$, $BB^\prime$, $CC^\prime$ are shared with the Type I tetrahedron and must therefore have the same length in both co-ordinate systems; and
\item diagonals such as $AB^\prime$, $AC^\prime$, $BC^\prime$ are shared with the Type I tetrahedron and must therefore have the same length in both co-ordinate systems.
\end{enumerate*}
From these two constraints, it can be deduced that
$\delta \RegTime{i}^{(1)} = \delta \RegTime{i} = \RegTime{i+1}-\RegTime{i}$ and that $\delta z_i = \pm \frac{1}{2\sqrt{6}} \delta \CWLattlen{1}{i}$; but in the limit where $M^\prime \to M$, we must recover the unperturbed model, so we take $\delta z_i$ to be $\delta z_i = - \frac{1}{2\sqrt{6}}\, \delta \CWLattlen{1}{i}$.  

The only constraint on apex $E^\prime$ is that it lie a distance $\CWLattlen{0}{i}$ from each of $A^\prime$, $B^\prime$, and $C^\prime$.  This is equivalent to the two constraints that $E^\prime$ lie at distance $\axislen{i+1}$ from the base's centre and that the central axis connecting $E^\prime$ to the base's centre lie orthogonally to the base.  As base $A^\prime B^\prime C^\prime$ defines a 2-dimensional plane in a (3+1)-dimensional Minkowski space-time, the subspace orthogonal to it would be a (1+1)-dimensional plane, and the tetrahedron's central axis can be oriented along any direction in this plane.  Combined with the first constraint, the second constraint implies that $E^\prime$ will lie on a hyperbola in this (1+1)-dimensional plane; exactly where on the hyperbola it lies depends on the length of strut $EE^\prime$, which need not be identical to the length of $AA^\prime$.  Thus the axis of the upper tetrahedron may in general be Lorentz-boosted relative to that of the lower tetrahedron.

Therefore, the upper vertices' co-ordinates are given most generally by
\begin{equation}
\renewcommand{\arraystretch}{2}
\begin{aligned}
A^\prime &= \displaystyle{\left(-\frac{\CWLattlen{1}{i+1}}{2}, -\frac{\CWLattlen{1}{i+1}}{2\sqrt{3}}, - \frac{\delta \CWLattlen{1}{i}}{2\sqrt{6}}, \, \imath \delta \RegTime{i} \right)},\\
B^\prime &= \displaystyle{\left(\frac{\CWLattlen{1}{i+1}}{2}, -\frac{\CWLattlen{1}{i+1}}{2\sqrt{3}}, - \frac{\delta \CWLattlen{1}{i}}{2\sqrt{6}}, \, \imath \delta \RegTime{i} \right)},\\
C^\prime &= \displaystyle{\left(0, \frac{\CWLattlen{1}{i+1}}{\sqrt{3}}, - \frac{\delta \CWLattlen{1}{i}}{2\sqrt{6}}, \, \imath \delta \RegTime{i} \right)},\\
E^\prime &= \displaystyle{\left(0, 0, \axislen{i+1} \cosh \psi_i - \frac{\delta \CWLattlen{1}{i}}{2\sqrt{6}}, \, \imath \delta \RegTime{i} + \imath \axislen{i+1} \sinh \psi_i\right)},
\end{aligned}
\label{SUvertices}
\end{equation}
where $\psi_i$ is the relative boost between the upper and lower tetrahedra's axes.

However, as with the unperturbed model, we shall require that all 4-blocks in the perturbed model have no twist or shear as well.  This implies that the world-sheet generated by each tetrahedral edge between $\Cauchyt{i}$ and $\Cauchyt{i+1}$ must be flat; in other words, the four vectors parallel to the four sides of this world-sheet must be co-planar.  When this requirement is imposed on any world-sheet involving $E^\prime$, such as $AEA^\prime E^\prime$, we obtain the further constraint that
\begin{equation}
0 = A_\psi - B_\psi \cosh \psi_i + C_\psi \sinh \psi_i,\\
\end{equation}
where
\begin{equation}
\begin{aligned}
A_\psi ={} & \axislen{i} \, \CWLattlen{1}{i+1},\\
B_\psi ={} & \axislen{i+1} \, \CWLattlen{1}{i},\\
C_\psi ={} & \axislen{i+1} \CWLattlendot{1}{i} \, \left[ \axislen{i+1} - \frac{\CWLattlen{1}{i}}{2\sqrt{6}} \right].
\end{aligned}
\label{psi_params}
\end{equation}
This constraint effectively determines $\psi_i$, and solving it leads to a quadratic equation with two solutions.  However, if we take the limit where $\CWLattlen{0}{i} \to \CWLattlen{1}{i}$ and $\delta z_i \to -\frac{1}{2\sqrt{6}} \, \delta \CWLattlen{1}{i}$, we expect to recover the 4-block of an equilateral tetrahedron; in this limit, $\psi_i$ should vanish, and this only happens for one of the solutions.  Therefore, we require that
\begin{equation}
\psi_i = \ln \left[ \frac{-A_\psi + \sqrt{ A_\psi^2 + C_\psi^2 - B_\psi^2 } }{C_\psi - B_\psi} \,\right]. \label{psi_relation}
\end{equation}
We note that because of the non-zero boost parameter $\psi_i$, the time-like hinges of this skeleton will be quadrilateral but not necessarily trapezoidal.  Indeed, the lengths of the struts would not, in general, be identical: the length of $EE^\prime$ would be different.  Therefore, we shall denote the areas and deficit angles of quadrilateral hinges by $\CWLattQuadArea{i}$ and $\CWLattQuadDeficit{i}$ instead.  We also note that in the continuum time limit where $\delta \RegTime{i} \to 0$, $\psi_i$ must become $\dot{\psi}\, dt$ to leading order in $dt$; this is because the relative boost between the lower and upper tetrahedra must become infinitesimally small as the separation between Cauchy surfaces tends to zero, and therefore, the zeroth order term of $\psi_i$ must be zero.

As mentioned earlier, Brewin has likened a choice of strut-lengths in the CW formalism to a choice of lapse function in the ADM formalism and the choice of having no twist or shear in the 4-blocks to a choice of shift function.  In the ADM formalism, both the lapse and shift functions can be freely chosen independently of each other throughout the Cauchy surface.  Yet in this CW skeleton, our choice of $\psi_i$ implies that if we freely specify the common length of struts $AA^\prime$, $BB^\prime$, $CC^\prime$, then the length of $EE^\prime$ would be completely determined; we do not have any freedom to specify its length separately.  We therefore see that the choice of shift function here has constrained the freedom to choose the lapse function to just the freedom to choose a single strut-length.  Thus the CW formalism is not completely analogous to the ADM formalism, at least for global models.

The global Regge action for the perturbed lattice model can now be expressed as
\begin{equation}
8\pi \Action = \mkern-40mu \sum_{i \in \left\{\substack{\text{quadrilateral}\\\text{hinges}}\right\}} \mkern-40mu \CWLattQuadArea{i} \CWLattQuadDeficit{i} + \mkern-35mu \sum_{i \in \left\{\substack{\text{triangular}\\\text{hinges}}\right\}} \mkern-35mu \TriArea{i} \TriDeficit{i} -8 \pi  \mkern-30mu \sum_{\substack{i \, \in \, \left\{ \text{particles} \right\} \\ j \, \in \, \left\{ \text{4-blocks} \right\} }} \mkern-30mu M_i \, \RegLinEl{ij},
\label{PerturbedAction}
\end{equation}
where $\RegLinEl{ij}$ is the path-length for the mass of magnitude $M_i$ in the 4-block labelled $j$; when varied with respect to the struts $\CWstrut{k}$, it yields the Regge equation
\begin{equation}
0 = \sum_i \frac{\partial \CWLattQuadArea{i}}{\partial \CWstrut{k}} \CWLattQuadDeficit{i} - 8\pi \mkern-26mu \sum_{\substack{i \, \in \, \left\{ \text{particles} \right\} \\ j \, \in \, \left\{ \text{4-blocks} \right\} }} \mkern-22mu M_i \, \frac{\partial \RegLinEl{ij}}{\partial \CWstrut{k}},
\label{PerturbedReggeEqn}
\end{equation}
where the first summation is still over all quadrilateral hinges.  

\subsection{Global and local Regge equations}
\label{GlobLocSec}
As mentioned in \secref{CWRegge}, the global Regge equations can, in certain cases, be related to the local equations through a chain rule, and we have chosen to use such a chain rule relationship to obtain the global Regge equations for the perturbed lattice universe model.  We shall therefore elaborate now on this relationship.

As discussed above, local variation requires that each 4-block of the skeleton be fully triangulated.  We have already described in \secref{CWRegge} how the triangulation of each 4-block is to be carried out.  To understand how the triangulation works more globally, we label the five vertices of the 5-tetrahedra CW Cauchy surface $\Cauchyt{i}$ by $A$, $B$, $C$, $D$, $E$, with $E$ corresponding to the common apex of all non-equilateral tetrahedra; each vertex in $\Cauchyt{i+1}$ is then labelled by the same letter as its counterpart in $\Cauchyt{i}$ but with a prime.  Diagonals are chosen so that the letter of a diagonal's vertex in $\Cauchyt{i}$ alphabetically precedes the letter of the diagonal's vertex in $\Cauchyt{i+1}$; thus, it is actually the relative ordering of the labels rather than the labels themselves that matter.  In the 5-tetrahedra model, this triangulation generates the 10 diagonals of $AB^\prime$, $AC^\prime$, $AD^\prime$, $AE^\prime$, $BC^\prime$, $BD^\prime$, $BE^\prime$, $CD^\prime$, $CE^\prime$, and $DE^\prime$.  This algorithm for choosing the diagonals ensures that each 4-block gets properly triangulated into 4-simplices; and indeed, the triangulation is consistent with the 4-block triangulation described earlier.  

To relate the local and global Regge equations, it is necessary that the global and local Regge actions be identical.  In our model, the Regge action will consist of two components, one corresponding to 
$$\displaystyle{\frac{1}{8\pi}\sum_{i \,\in\, \left\{ \text{hinges}\right\}} \mkern-9mu A_i\, \CWdeficit{i}}$$
and another to
$$\sum_{\substack{i \, \in \, \left\{ \text{particles} \right\} \\ j \, \in \, \left\{ \text{blocks} \right\} }} \mkern-22mu M_i \, \RegLinEl{ij}.$$
We shall show that each of these components are identical in the two actions and hence that the two actions are themselves identical for the model we are considering.

We begin with the first component.  In the global action, there are two types of hinges: the quadrilateral time-like hinges generated by the world-sheets of the tetrahedral edges and the triangular space-like hinges corresponding to the triangles of the tetrahedra in a Cauchy surface.  In the 4-block described by \eqref{SLvertices} and \eqref{SUvertices}, an example of the quadrilateral hinge would be $ABA^\prime B^\prime$, and an example of the triangular hinge would be $ABC$.  In the local action, there are three types of hinges.  The diagonals split each quadrilateral hinge into a pair of time-like triangular hinges, and these pairs of time-like triangular hinges correspond to the first type; an example pair would be $ABB^\prime$ and $A A^\prime B^\prime$.  The second type are same triangular space-like hinges as in the global action.  The third type correspond to isosceles triangular hinges formed by two diagonals and one tetrahedral edge, an example being $ABC^\prime$.  These hinges have no analogue in the global action; however, they do not depend on the struts either, and as we shall be varying the action with respect to the struts only, they will have no contribution to the Regge equations; hence we shall ignore these hinges.  Since each quadrilateral hinge is required to be planar, whether triangulated or not, the area of each quadrilateral hinge will equal the sum of the areas of its constituent time-like triangular hinges.  Naturally, the space-like triangular hinges will have the same areas in the global and the local actions.  Thus the hinge areas appearing in the two actions are identical, and if the corresponding deficit angles are identical, then the first component is identical in the two actions.

After edge-lengths of the same type have been set equal, the deficit angles $\DeficitA{i}$ and $\DeficitB{i}$ of any triangular time-like hinge do become identical to the deficit angle $\CWLattQuadDeficit{i}$ of the original quadrilateral hinge; that is, $\DeficitA{i} = \DeficitB{i} = \CWLattQuadDeficit{i}$.  Setting the edge-lengths equal makes the two triangular hinges be co-planar both with each other and with the original quadrilateral hinge; thus the 4-blocks meeting at the triangular hinges would be flat; the unit normals of the triangulated faces would be identical to the unit normals of the original faces; and the dihedral angles between the triangulated faces would be identical to the dihedral angles between the original faces.  Since the number of faces meeting at the triangulated hinge is the same as the number of faces at the original hinge, the deficit angles for the triangulated hinge and the original hinge are identical.  %Additionally, since the dihedral angles are unchanged, then in the continuum time limit, $\CWlendot{}$ would still be given by \eqref{ldot}.

The deficit angle of the space-like triangular hinges are also identical in the original and the triangulated skeletons; this follows simply because the unit normals to the faces meeting at the hinge will not change because of the triangulation.  Thus we can conclude that the first component of the local and global actions are identical.

As for the second component, clearly the path through the 4-block should not depend on whether the 4-block has been triangulated or not.  Hence the global and local form of this component should also be identical; and therefore the local and global actions are indeed completely identical.

Since the two actions are effectively identical, we can express the global Regge equation as a linear combination of local equations through a chain rule
\begin{equation}
0 = \frac{\partial \Action}{\partial \CWstrut{i}} = \sum_j \frac{\partial \Action}{\partial \CWstrutl{j}}\frac{\partial \CWstrutl{j}}{\partial \CWstrut{i}} + \sum_j \frac{\partial \Action}{\partial \CWLattdiagZ{j}}\frac{\partial \CWLattdiagZ{j}}{\partial \CWstrut{i}},
\label{chain_rule}
\end{equation}
where $\partial \Action / \partial \CWstrutl{j}$ is a local variation with respect to a single strut $\CWstrutl{j}$, with superscript $\ell$ indicating that these struts are to be considered as local rather than global struts, and where $\partial \Action / \partial \CWLattdiagZ{j}$ is a local variation with respect to a single diagonal $\CWLattdiagZ{j}$; the first summation is over all local struts and the second is over all diagonals.

We shall examine the contribution of each summation in \eqref{chain_rule} to the global Regge equation.  First, however, we note that as we are varying locally with respect to the struts, there are two distinct sets of struts in our skeleton relevant to local variation, one corresponding to the struts of the Type I tetrahedron, and the other to the $EE^\prime$ strut of the Type II tetrahedron.  The first set of struts all have the same length, which we shall denote by $\CWLattstrutlenp{i}$ since the set includes strut $AA^\prime$.  The length is given by \eqref{strut} as well, but with $\CWLattlenZdot{i}$ replaced by $\CWLattlendot{1}{i}$, that is, by
\begin{equation}
\left(\CWLattstrutlenp{i}\right)^2 = \left[\frac{3}{8} \left(\CWLattlendot{1}{i}\right)^2 - 1 \right] \delta \RegTime{i}^2.
\label{strutlenp}
\end{equation}
The length of strut $EE^\prime$ will be denoted by $\CWLattstrutlen{i}$ and is given by
\begin{equation}
\begin{split}
\left(\CWLattstrutlen{i}\right)^2 ={} & \left[\axislen{i+1}\cosh\psi_i - \axislen{i} - \frac{1}{2\sqrt{6}} \delta \CWLattlen{1}{i} \right]^2 \\
& {}- \left[ \vphantom{\frac{1}{2\sqrt{6}}} \delta \RegTime{i} + \axislen{i+1}\sinh\psi_i \right]^2.
\end{split}
\label{strutlen}
\end{equation}

We now examine the contribution of the first summation in \eqref{chain_rule}.  We shall always choose the global strut-length $\CWstrut{i}$ to equal either $\CWLattstrutlen{i}$ or $\CWLattstrutlenp{i}$; without loss of generality, let us assume then that $\CWstrut{i} = \CWLattstrutlenp{i}$.  Then it trivially follows that $\partial \CWLattstrutlenp{j} / \partial \CWstrut{i} = \delta_{ij}$, which is clearly $\Oconst$ to leading order in $dt$ when the continuum time limit is taken.  If we substitute for $\delta \RegTime{i}$ in \eqref{strutlen} using \eqref{strutlenp}, it can also be shown that the leading order of $\partial \CWLattstrutlen{j} / \partial \CWstrut{i}$ will be at least $\Oconst$ as well.  We shall be explicitly calculating $\partial \Action / \partial \CWstrutl{j}$ below, and at the end of the calculation, we shall see that the leading order of  $\partial \Action / \partial \CWstrutl{j}$ is also $\Oconst$.  Therefore unless the second summation in \eqref{chain_rule} has negative leading order, the first summation will definitely contribute to the leading order of the global Regge equation.

We now turn to the second summation in \eqref{chain_rule}.  Our skeleton has two diagonals, one triangulating quadrilateral hinges generated by $\CWLattlen{0}{i}$ edges, and the other triangulating quadrilateral hinges generated by $\CWLattlen{1}{i}$ edges.  The first set of hinges are the ones that involve vertices $E$ and $E^\prime$ in the Type II tetrahedra's 4-blocks, and an example of a diagonal on such a hinge would be $AE^\prime$; these diagonals have length $\CWLattdiag{i}$ given by
\begin{equation}
\begin{split}
\left(\CWLattdiag{i}\right)^2 ={} & \frac{2}{3} \left(\CWLattlen{1}{i}\right)^2 - \left(\CWLattlen{0}{i}\right)^2 + \left(\CWLattstrutlen{i}\right)^2 \\
& {} + \axislen{i} \left[2\axislen{i+1}\cosh\psi_i - \frac{1}{\sqrt{6}}\delta \CWLattlen{1}{i} \right].
\end{split}
\label{diag}
\end{equation}
An example of a diagonal on the second set of hinges would be $AB^\prime$, and these diagonals have length $\CWLattdiagp{i}$ given by
\begin{equation}
\left(\CWLattdiagp{i}\right)^2 = \CWLattlen{1}{i+1}\, \CWLattlen{1}{i} + \left(\CWLattstrutlenp{i}\right)^2.
\label{diagp}
\end{equation}
It can be shown that
$$
\frac{\partial \CWLattdiagp{i}}{\partial \CWstrut{i}} = \frac{\partial \CWLattdiagp{i}}{\partial \CWLattstrutlenp{i}} = \frac{\CWLattstrutlenp{i}}{\CWLattdiagp{i}}.
$$
In the continuum time limit, where $\CWLattlen{1}{i} \to \CWLattlen{1}{}(t)$, we have that $\CWLattstrutlenp{i} / \CWLattdiagp{i} \to \CWLattstrutlenp{i} / \, \CWLattlen{1}{}$, which is an $\Odtone$ term.  Thus, $\partial \CWLattdiagp{i} / \partial \CWstrut{i}$ raises the leading order of the second summation in \eqref{chain_rule} by one.  It can also be shown that in the continuum time limit, the leading order term of $\partial \CWLattdiag{j} / \partial \CWstrut{i}$ is at least $\Odtone$ as well.  Therefore $\partial \CWLattdiagZ{j} / \partial \CWstrut{i}$ will be at least $\Odtone$ to leading order for all diagonals.

So unless the leading order of $\partial \Action / \partial \CWLattdiagZ{j}$ is negative, the leading order of the second summation in \eqref{chain_rule} will be at least $\Odtone$.  Naturally, verifying that the leading order of $\partial \Action / \partial \CWLattdiagZ{j}$ is not negative requires a direct calculation of $\partial \Action / \partial \CWLattdiagZ{j}$.  However, such a calculation is beyond the scope of this paper.  Instead, we shall assume the conclusion.  First, if $\partial \Action / \partial \CWLattdiagZ{j}$ had negative leading order, then the corresponding Regge equation would diverge in the continuum time limit; thus our model would break down.  We are assuming this is not the case.  Secondly, we have found many similarities between the Regge model we are studying here and the parent Regge models of the $\Lambda$-FLRW universe studied in \cite{RGL-Williams}.  In that paper, we similarly considered the relationship between global and local Regge equations through an essentially identical chain rule to \eqref{chain_rule}.  We found the leading order of $\partial \CWLattdiagZ{j} / \partial \CWstrut{i}$ to be $\Odtone$ as well for all diagonals.  We also found the leading order of $\partial \CWstrutl{j} / \partial \CWstrut{i}$ to be $\Oconst$; this followed trivially because all struts between pairs of consecutive Cauchy surfaces had equal length, so $\partial \CWstrutl{j} / \partial \CWstrut{i}$ would be unity for struts between the same pair of surfaces.  Finally, we found the leading order of $\partial \Action / \partial \CWstrutl{j}$ to be $\Oconst$ as well.  Since the two models have identical leading orders for three of the partial derivatives appearing \eqref{chain_rule} and its analogue in \cite{RGL-Williams}, we suspect they would have identical leading order for the final partial derivative, $\partial \Action / \partial \CWLattdiagZ{j}$.  In \cite{RGL-Williams}, the order of this term was $\Odtone$, so we suspect it would be the same here.

If this is true, then the second summation in \eqref{chain_rule} would have a higher leading order than the first summation and would therefore not contribute.  As a result, the solutions to $0=\partial \Action / \partial \CWstrutl{j}$ would by themselves satisfy the global Regge equation $0=\partial \Action / \partial \CWstrut{i}$, much like the situation with the $\Lambda$-FLRW Regge models in \cite{RGL-Williams}.

\subsection{Particle trajectories} \label{ParticleTrajSec}
The final term of the Regge action \eqref{PerturbedAction} determines the effect of the masses on the behaviour of the universe.  This term depends on $\RegLinEl{ij}$, the length of the trajectory followed by the mass $M_i$ through the 4-block labelled $j$; thus to fully specify our Regge model, we must specify what trajectory the masses will follow.

Ideally, we should like our masses to follow geodesics throughout the entire universe and also be co-moving with respect to the Cauchy surfaces, as we expect this to be the situation in the continuum universe.  As mentioned earlier on, this is the situation in the perfectly smooth FLRW universe, where test particles co-moving with respect to constant-$t$ Cauchy surfaces are also following geodesics of the space-time.  In the lattice universe, the point-masses should similarly be co-moving with respect to the universe's Cauchy surfaces so as to preserve the lattice symmetries -- we do not expect the gravitational interactions between a symmetric distribution of masses to give rise to an asymmetric motion of the masses.  Yet these particles should also be following geodesics as well.  However we have seen that in the unperturbed Regge model, it is not possible for the particle to be simultaneously co-moving and following geodesics across the entire space-time unless the particles are positioned at the centres of the equilateral tetrahedra.  Even in CW approximations of the perfectly homogeneous and isotropic FLRW space-times, co-moving test particles will not follow geodesics globally either unless the particles are at the centres of the tetrahedra.  Consider, for example, test particles co-moving with respect to the centres of the triangles.  These centres themselves trace out piecewise linear trajectories as the underlying CW skeleton is piecewise linear.  Therefore, test particles co-moving with these points will follow the same piecewise linear trajectories, deflecting every time they cross from one Cauchy surface into the next.  Only trajectories traced out by the centres of the tetrahedra will have no deflection, since the tetrahedra expand or contract uniformly about their centres as they evolve, and this thereby leaves the tetrahedral centres spatially fixed with time.  We note though that each linear segment of a piecewise linear trajectory will still be a local geodesic within the 4-block it traverses because straight-line segments are always geodesics according to the Minkowski metric.  If we take the continuum space-time limit of a CW approximation to an FLRW space-time, we expect to recover the continuum FLRW space-time itself in which any co-moving particle will indeed follow geodesics as well.  The reason only the tetrahedral centres follow global geodesics is because Cauchy surfaces of the CW skeleton are not perfectly isotropic and homogeneous; thus, not all points on the Cauchy surface have been `created equal'.  If co-moving particles do not follow geodesics in CW approximations of the perfectly homogeneous and isotropic FLRW space-times, there is even less reason for them to follow geodesics in approximations of the lattice universe, both perturbed and unperturbed.  Thus we shall only require the point-masses of the perturbed Regge lattice universe to be co-moving with respect to the Cauchy surfaces.  But in the continuum space-time limit, we do hope that these co-moving particles will indeed follow geodesics as well.

As mentioned previously, we have chosen to work with a lattice universe where the masses would be co-moving with the centres of the tetrahedra when the universe is unperturbed.  For the particle in the equilateral Type I tetrahedron, the trajectory is straightforward: by symmetry, the particle should remain,  for its entire trajectory, at the tetrahedron's centre.  For particles in Type II tetrahedra, it is less clear where the particles should be positioned because the tetrahedra are no longer equilateral.  The only clear symmetry here is in the equilateral base.  We can therefore say that a particle should remain above the centre of the equilateral base for the entirety of its trajectory.  The issue lies in fixing the particle's position above the base.  We know that the vertices themselves should be co-moving with respect to the Cauchy surface, so we shall use them as reference points to express the trajectory of the co-moving particle.  The particle's position $\Tensb{p}_i$ on Cauchy surface $\Cauchyt{i}$ can be expressed as
\begin{equation}
\Tensb{p}_i = \alpha \left( \vecb{A} + \vecb{B} + \vecb{C} \right) + \beta\, \vecb{E},
\label{comoving_particle}
\end{equation}
where $\vecb{A}, \vecb{B}, \vecb{C}, \vecb{E}$ are the position vectors of vertices $A, B, C, E$, respectively, and where $\alpha$ and $\beta$ are yet to be determined constants.  For the particle to be inside the tetrahedron, $\alpha$ and $\beta$ must be non-negative and satisfy the constraint $3\, \alpha + \beta = 1$.  For the particle to be co-moving, we require that its position $\Tensb{p}_{i+1}$ on Cauchy surface $\Cauchyt{i+1}$ be given by \eqref{comoving_particle} as well, but with vectors $\vecb{A}, \vecb{B}, \vecb{C}, \vecb{E}$ replaced by $\vecb{A}^\prime, \vecb{B}^\prime, \vecb{C}^\prime, \vecb{E}^\prime$, respectively.  In the 4-block between $\Cauchyt{i}$ and $\Cauchyt{i+1}$, the particle then propagates in a straight line from $\Tensb{p}_i$ to $\Tensb{p}_{i+1}$, and such a trajectory is considered co-moving with respect to the Cauchy surfaces.

There is one situation where there is clearly a unique choice for $\alpha$ and $\beta$.  Should $\CWLattlen{1}{i}$ become equal to $\CWLattlen{0}{i}$ at any moment, then the corresponding tetrahedron will be equilateral; we would then require the particle to lie at the tetrahedron's centre, which means $\alpha$ and $\beta$ must be $\frac{1}{4}$ at this moment.  Yet based on our definition of co-moving trajectories, $\alpha$ and $\beta$ must be constant over the particle's entire trajectory.  Therefore, $\alpha$ and $\beta$ must be $\frac{1}{4}$ over the particle's entire trajectory.  However, we shall take $\alpha$ and $\beta$ to be $\frac{1}{4}$ for all particles, regardless of whether their tetrahedra become equilateral or not.  Such a choice would place the particles at the tetrahedra's centroids, which would in some sense generalise our requirement that the particles be at the tetrahedra's centres.

\subsection{Geometric quantities for the Regge equation}
We now turn to deriving the geometric quantities relevant for the Regge equation.  From \eqref{PerturbedReggeEqn}, it is clear that we need to derive three types of quantities: the varied areas of the time-like hinges, $\partial \Area{}{i} / \partial \CWstrut{k}$; the corresponding deficit angles $\CWdeficit{i}$, or equivalently, the dihedral angle $\CWLattdihedral{i}$ between neighbouring faces; and the varied lengths of the particles' trajectories across 4-blocks, $\partial \RegLinEl{ij} / \partial \CWstrut{k}$.

We shall be taking the continuum time limit of these quantities so that we can express the Regge equation in its continuum time form.  Thus, we shall be needing the continuum time form of the lengths and the boost parameters; these are given by
\begin{IEEEeqnarray*}{rCl}
\CWLattlen{0}{i} &\to& \CWLattlen{0}{}(t),\\
\CWLattlen{0}{i+1} &\to& \CWLattlen{0}{} + \CWLattlendot{0}{} \, dt + \Odt{2},\\
\CWLattlen{1}{i} &\to& \CWLattlen{1}{}(t),\\
\CWLattlen{1}{i+1} &\to& \CWLattlen{1}{} + \CWLattlendot{1}{} \, dt + \Odt{2},\\
\psi_i &\to& \dot{\psi}\, dt  + \Odt{2},\\
\CWLattdiag{i} &\to& \CWLattlen{0}{} + \Odtone,\\
\CWLattdiagp{i} &\to& \CWLattlen{1}{} + \Odtone,\\
\CWLattstrutlen{i} &\to& \CWLattstrutlendot{} dt + \Odt{2},\\
\CWLattstrutlenp{i} &\to& \CWLattstrutlenpdot{} dt + \Odt{2},
\end{IEEEeqnarray*}
where
\begin{IEEEeqnarray}{rCl}
\axislendot{} &:=& \frac{\CWLattlen{0}{} \CWLattlendot{0}{} - \frac{1}{3} \CWLattlen{1}{} \CWLattlendot{1}{}}{\axislen{}},\\
\CWLattstrutlenpdot{} &:=& \left[\frac{3}{8} \left(\CWLattlendot{1}{}\right)^2 - 1 \right]^\frac{1}{2},\\
\CWLattstrutlendot{} &:=& \left[ \left( \frac{1}{2\sqrt{6}}\CWLattlendot{1}{} - \axislendot{} \right)^2 - \left( \axislen{} \dot{\psi} + 1\right)^2 \right]^\frac{1}{2},\\
\dot{\psi} &:=& \frac{\axislendot{} \CWLattlen{1}{} - \axislen{} \CWLattlendot{1}{}}{\axislen{} \CWLattlendot{1}{} \left[ \axislen{} - \frac{\CWLattlen{1}{}}{2 \sqrt{6}} \right]}, \label{psidot}
\end{IEEEeqnarray}
and where $\axislen{} = \sqrt{\left(\CWLattlen{0}{}\right)^2-\frac{1}{3}\left(\CWLattlen{1}{}\right)^2}$ denotes the continuum time limit of $\axislen{i}$.

\begin{figure}[bth]
\input{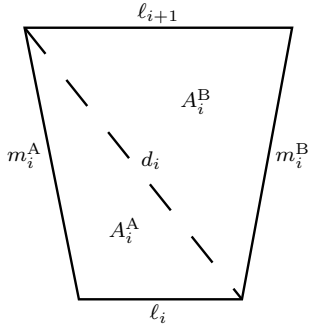}
\caption{\label{fig:hingefig}A diagonal $\CWdiag{i}$ divides a time-like quadrilateral hinge into a lower and upper triangular hinge, labelled A and B, respectively.  The hinge has been represented with the generic tetrahedral edge-lengths of $\ell_i$ and $\ell_{i+1}$; these would equal $\CWLattlen{0}{i}$ and $\CWLattlen{0}{i+1}$, respectively, or $\CWLattlen{1}{i}$ and $\CWLattlen{1}{i+1}$ according to the type of hinge being considered.}
\end{figure}

As mentioned previously, our skeleton has two types of time-like hinges corresponding to the world-sheets of $\CWLattlen{0}{i}$ and $\CWLattlen{1}{i}$ edges.  We shall refer to the quadrilateral hinge generated by $\CWLattlen{1}{i}$ as a Type I hinge and its triangular components as hinges A1 and B1, counterparts to triangles A and B, respectively, in \figref{fig:hingefig}.  
We can use Heron of Alexandria's formula to express these triangular hinges' areas in terms of their edge-lengths.  Hinge A1 has area
\begin{equation}
\begin{split}
\AreaAi{1}{i} ={} & \frac{1}{4}\left[2 \left[ \left(\CWLattlen{1}{i}\right)^2 \left(\CWLattdiagp{i}\right)^2 + \left(\CWLattlen{1}{i}\right)^2 \left(\CWLattstrutlenp{i}\right)^2 \right. \right. \\
& \left. \hphantom{\frac{1}{4} [2 [} {} + \left(\CWLattdiagp{i}\right)^2 \left(\CWLattstrutlenp{i}\right)^2 \right]\\
& \left. \hphantom{\frac{1}{4} [} {} - \left[ \left(\CWLattlen{1}{i}\right)^4 + \left(\CWLattdiagp{i}\right)^4 + \left(\CWLattstrutlenp{i}\right)^4  \right] \right]^\frac{1}{2},
\end{split}
\end{equation}
and, when varied with respect to $\CWLattstrutlenp{i}$, yields
\begin{equation}
\frac{\partial \AreaAi{1}{i}}{\partial \CWLattstrutlenp{i}} = \frac{\CWLattstrutlenp{i}}{8 \, \AreaAi{1}{i}} \left[ \left(\CWLattlen{1}{i}\right)^2 + \left( \CWLattdiagp{i} \right)^2 - \left( \CWLattstrutlenp{i} \right)^2 \right]^\frac{1}{2}.
\end{equation}
Hinge B1 has area
\begin{equation}
\begin{split}
\AreaBi{1}{i} ={} & \frac{1}{4}\left[2 \left[ \left(\CWLattlen{1}{i+1}\right)^2 \left(\CWLattdiagp{i}\right)^2 + \left(\CWLattlen{1}{i+1}\right)^2 \left(\CWLattstrutlenp{i}\right)^2 \right. \right. \\
& \left. \hphantom{\frac{1}{4} [2 [} {} + \left(\CWLattdiagp{i}\right)^2 \left(\CWLattstrutlenp{i}\right)^2 \right] \\
& \left. \hphantom{\frac{1}{4} [} {} - \left[ \left(\CWLattlen{1}{i+1}\right)^4 + \left(\CWLattdiagp{i}\right)^4 + \left(\CWLattstrutlenp{i}\right)^4  \right] \right]^\frac{1}{2},
\end{split}
\end{equation}
and, when varied with respect to $\CWLattstrutlenp{i}$, yields
\begin{equation}
\frac{\partial \AreaBi{1}{i}}{\partial \CWLattstrutlenp{i}} = \frac{\CWLattstrutlenp{i}}{8 \, \AreaBi{1}{i}} \left[ \left(\CWLattlen{1}{i+1}\right)^2 + \left( \CWLattdiagp{i} \right)^2 - \left( \CWLattstrutlenp{i} \right)^2 \right]^\frac{1}{2}.
\end{equation}
It can be shown that in the continuum time limit, the varied areas become
\begin{equation}
\frac{\partial \AreaAi{1}{}}{\partial \CWLattstrutlenp{}} = \frac{\partial \AreaBi{1}{}}{\partial \CWLattstrutlenp{}} = \frac{\CWLattlen{1}{}}{2}\, \CWLattstrutlenpdot{} \left[\frac{1}{8}\left(\CWLattlendot{1}{}\right)^2 - 1 \right]^{-\frac{1}{2}};
\label{VarAreaA1}
\end{equation}
that is, $\partial \AreaAi{1}{} / \partial \CWLattstrutlenp{}$ and $\partial \AreaBi{1}{} / \partial \CWLattstrutlenp{}$ are identical to at least leading order.

We shall refer to the quadrilateral hinge generated by $\CWLattlen{0}{i}$ as a Type II hinge and its triangular components as hinges A2 and B2, also counterparts to triangles A and B, respectively, in \figref{fig:allhinges}.  Hinge A2 has area
\begin{equation}
\begin{split}
\AreaAi{2}{i} ={} & \frac{1}{4}\left[2 \left[ \left(\CWLattlen{0}{i}\right)^2 \left(\CWLattdiag{i}\right)^2 + \left(\CWLattlen{0}{i}\right)^2 \left(\CWLattstrutlen{i}\right)^2 \right. \right.\\
& \left. \hphantom{\frac{1}{4} [2 [} {} + \left(\CWLattdiag{i}\right)^2 \left(\CWLattstrutlen{i}\right)^2 \right] \\
& \left. \hphantom{\frac{1}{4} [} {} - \left[ \left(\CWLattlen{0}{i}\right)^4 + \left(\CWLattdiag{i}\right)^4 + \left(\CWLattstrutlen{i}\right)^4  \right] \right]^\frac{1}{2},
\end{split}
\end{equation}
and, when varied with respect to $\CWLattstrutlen{i}$, yields
\begin{equation}
\frac{\partial \AreaAi{2}{i}}{\partial \CWLattstrutlen{i}} = \frac{\CWLattstrutlen{i}}{8 \, \AreaAi{2}{i}} \left[ \left(\CWLattlen{0}{i}\right)^2 + \left( \CWLattdiag{i} \right)^2 - \left( \CWLattstrutlen{i} \right)^2 \right]^\frac{1}{2}.
\end{equation}
Hinge B2 has area
\begin{equation}
\begin{aligned}
\AreaBi{2}{i} ={} & \frac{1}{4}\left[2 \left[ \left(\CWLattdiag{i}\right)^2 \left(\CWLattlen{0}{i+1}\right)^2 + \left(\CWLattstrutlenp{i}\right)^2 \left(\CWLattlen{0}{i+1}\right)^2 \right. \right.\\
& \left. \hphantom{\frac{1}{4} [2 [} {} + \left(\CWLattdiag{i}\right)^2 \left(\CWLattstrutlenp{i}\right)^2 \right] \\
& \left. \hphantom{\frac{1}{4} [} {} - \left[ \left(\CWLattlen{0}{i+1}\right)^4 + \left(\CWLattdiag{i}\right)^4 + \left(\CWLattstrutlenp{i}\right)^4  \right] \right]^\frac{1}{2},
\end{aligned}
\end{equation}
and, when varied with respect to $\CWLattstrutlenp{i}$, yields
\begin{equation}
\frac{\partial \AreaBi{2}{i}}{\partial \CWLattstrutlenp{i}} = \frac{\CWLattstrutlenp{i+1}}{8 \, \AreaBi{2}{i}} \left[ \left(\CWLattlen{0}{i+1}\right)^2 + \left( \CWLattdiag{i} \right)^2 - \left( \CWLattstrutlenp{i} \right)^2 \right]^\frac{1}{2}.
\end{equation}
It can be shown that in the continuum time limit, the varied areas become
\begin{widetext}
\begin{align}
\begin{split}
\frac{\partial \AreaAi{2}{i}}{\partial \CWLattstrutlen{i}} \to {} & \frac{\partial \AreaAi{2}{}}{\partial \CWLattstrutlen{}}\\
= {} & \frac{\CWLattlen{0}{}}{2} \CWLattstrutlendot{} \left[ \frac{1}{3}\left(\frac{\CWLattlen{1}{}}{\CWLattlen{0}{}}\right)^2 \left( \frac{\CWLattlendot{1}{}}{2\sqrt{6}} - \axislendot{} \right)^2 - \left( \axislen{} \dot{\psi} + 1\right)^2 \right]^{-\frac{1}{2}} \mkern-20mu + \Odtone
\end{split}\label{VarAreaA2}\\
\intertext{and}
\begin{split}
\frac{\partial \AreaBi{2}{i}}{\partial \CWLattstrutlenp{i}} \to {} & \frac{\partial \AreaBi{2}{}}{\partial \CWLattstrutlenp{}}\\
= {} & \frac{\CWLattlen{0}{}}{2} \CWLattstrutlenpdot{} \left[ \frac{1}{3}\left(\frac{\CWLattlen{1}{}}{\CWLattlen{0}{}}\right)^2 \left[\left(\frac{\CWLattlendot{1}{}}{2\sqrt{6}} - \axislendot{} \right)^2 - \left(\CWLattlendot{0}{}\right)^2\right] - \left(\frac{\CWLattlendot{1}{}}{2\sqrt{6}} - \axislendot{} + \frac{\axislen{}}{\CWLattlen{0}{}}\CWLattlendot{0}{} \right)^2 + \left(\CWLattstrutlenpdot{}\right)^2 \right]^{-\frac{1}{2}} \mkern-20mu + \Odtone.
\end{split}\label{VarAreaB2}
\end{align}
\end{widetext}

Since the Cauchy surface is no longer composed solely of equilateral tetrahedra, we must use \eqref{def_ang} to calculate the hinges' deficit angles.  Each 4-block will contribute one dihedral angle to each of its hinges.  The Type I tetrahedron's 4-block will contribute the same dihedral angle to all of its hinges: the tetrahedron is equilateral, so all of its associated A1 hinges are geometrically identical, as are all of its B1 hinges; moreover, when all edges are constrained to be identical, each pair of A1 and B1 hinges become co-planar, as mentioned previously, and the unit normals to the two faces meeting at A1 become identical to the unit normals to the two faces meeting at B1; thus all dihedral angles become identical, and this angle $\CWLattdihedralS{0}{i}$ is given by \eqref{cosq} but with $\CWLattlenZdot{i}$ replaced by $\CWLattlendot{1}{i}$.  In the continuum time limit, we have that $\CWLattdihedralS{0}{i} \to \CWLattdihedralS{0}{}$, and $\CWLattdihedralS{0}{}$ is then given by \eqref{cosq-cont-time} but with $\CWLattlenZdot{}$ replaced by $\CWLattlendot{1}{}$; that is, $\CWLattdihedralS{0}{}$ is given by
\begin{equation}
\cos \CWLattdihedralS{0}{} = \frac{1+\frac{1}{8}\left(\CWLattlendot{1}{}\right)^2}{3-\frac{1}{8}\left(\CWLattlendot{1}{}\right)^2} + \Odtone.
\label{theta0}
\end{equation}

The 4-block of a Type II tetrahedron has four distinct types of hinges: a pair of A1 and B1 hinges for each $\CWLattlen{1}{i}$ edge and a pair of A2 and B2 hinges for each $\CWLattlen{0}{i}$ edge.  We found that all hinges of the same type had the same dihedral angle, even though they are not entirely identical because of the way the Regge skeleton is triangulated.  Additionally, we found that A1 hinges and B1 hinges had the same dihedral angle to leading order in the continuum time limit.  Therefore in the continuum time limit, the 4-block of a Type II tetrahedron will contribute three distinct dihedral angles, which we shall denote $\CWLattdihedralS{1}{}$, $\CWLattdihedralS{2}{}$, and $\CWLattdihedralS{3}{}$ for A1, A2, and B2, respectively.  Each dihedral angle has been calculated by taking the scalar product of the unit normals to the two faces meeting at the corresponding hinge.

One example of an A1 hinge is $ABB^\prime$, which is shared with the Type I tetrahedron's 4-block.  The two faces meeting at this hinge, $ABB^\prime C^\prime$ and $ABB^\prime E^\prime$, are separated by a dihedral angle of $\CWLattdihedralS{1}{i}$.  To leading order in the continuum time limit, this angle is given by
\begin{widetext}
\begin{equation}
\begin{split}
\cos \CWLattdihedralS{1}{} ={} & \left[\left(\frac{1}{8}\left(\CWLattlendot{1}{}\right)^2-3\right)\CWLattlen{1}{} - \frac{1}{2}\sqrt{\frac{3}{2}} \left(\CWLattlendot{1}{}\right)^2\axislen{}\right]\\ 
& {} \times \left[ \left(\frac{1}{8}\left(\CWLattlendot{1}{}\right)^2-3\right) \left[ \left( 9 - \frac{7}{8} \left(\CWLattlendot{1}{} \right)^2 \right) \left( \CWLattlen{1}{} \right)^2 + 3\left( \left( \CWLattlendot{1}{} \right)^2 - 12 \right)\left(\CWLattlen{0}{}\right)^2 - \sqrt{\frac{3}{2}}\,\axislen{}\left(\CWLattlendot{1}{}\right)^2\CWLattlen{1}{} \right] \right]^{-\frac{1}{2}} \mkern-20mu + \Odtone.
\end{split}
\label{theta1}
\end{equation}
\end{widetext}
The B1 counterpart hinge is $AA^\prime B^\prime$, and as mentioned previously, we found the two faces meeting on this hinge, $AA^\prime B^\prime C^\prime$ and $AA^\prime B^\prime E^\prime$, to be separated by the same dihedral angle as well at lowest order in $dt$.

An example of an A2 hinge is $AEE^\prime$.  The two faces meeting at this hinge, $ABEE^\prime$ and $ACEE^\prime$, are separated by a dihedral angle of $\CWLattdihedralS{2}{i}$.  This angle is given by
\begin{widetext}
$$
\cos \CWLattdihedralS{2}{i} = \frac{1}{2} \frac{ \left[ \left(\CWLattlen{0}{i}\right)^2 - \frac{1}{2} \left(\CWLattlen{1}{i}\right)^2 \right]\left(\CWLattstrutlen{i}\right)^2 - \axislensq{i} \left[ \axislen{i+1}\cosh \psi_i - \axislen{i} - \frac{\delta \CWLattlen{1}{i}}{2\sqrt{6}} \right]^2 } { \left[ \left(\CWLattlen{0}{i}\right)^2 - \frac{1}{4} \left(\CWLattlen{1}{i}\right)^2 \right]\left(\CWLattstrutlen{i}\right)^2 - \axislensq{i} \left[ \axislen{i+1}\cosh \psi_i - \axislen{i} - \frac{\delta \CWLattlen{1}{i}}{2\sqrt{6}} \right]^2 },
$$
and in the continuum time limit, this expression becomes
\begin{equation}
\cos \CWLattdihedralS{2}{} = \frac{1}{2} \frac{ \left[ \left(\CWLattlen{0}{}\right)^2 - \frac{1}{2} \left(\CWLattlen{1}{}\right)^2 \right]\left(\CWLattstrutlendot{}\right)^2 - \axislensq{} \left[ \axislendot{} - \frac{\CWLattlendot{1}{}}{2\sqrt{6}} \right]^2 }{ \left[ \left(\CWLattlen{0}{}\right)^2 - \frac{1}{4} \left(\CWLattlen{1}{}\right)^2 \right]\left(\CWLattstrutlendot{}\right)^2 - \axislensq{} \left[ \axislendot{} - \frac{\CWLattlendot{1}{}}{2\sqrt{6}} \right]^2 } + \Odtone.
\label{theta2}
\end{equation}
\end{widetext}

Finally, the B2 counterpart to the above A2 hinge is $AA^\prime E^\prime$.  The two faces meeting here, $AA^\prime B^\prime E^\prime$ and $AA^\prime C^\prime E^\prime$, are separated by a dihedral angle of $\CWLattdihedralS{3}{i}$.  This angle is given by
\begin{widetext}
$$
\cos \CWLattdihedralS{3}{i} = \frac{ 2\left[ \frac{1}{6}\,\CWLattlen{1}{i+1}\,\CWLattlendot{1}{i} - \axislen{i+1} \left( \frac{\CWLattlendot{1}{i}}{2\sqrt{6}}\,\cosh \psi_i + \sinh \psi_i \right) \right]^2 - \left[\left(\CWLattlen{1}{i+1}\right)^2-2 \left(\CWLattlen{0}{i+1}\right)^2 \right] \left[ \frac{1}{8} \left(\CWLattlendot{1}{i}\right)^2 + 1 \right] }{\left[\left(\CWLattlen{1}{i+1}\right)^2-4 \left(\CWLattlen{0}{i+1}\right)^2 \right] \left[ \frac{1}{8} \left(\CWLattlendot{1}{i}\right)^2 - 1 \right] + 4 \left[ \frac{1}{12}\, \CWLattlen{1}{i+1}\, \CWLattlendot{1}{i} +  \axislen{i+1} \left( \frac{\CWLattlendot{1}{i}}{2\sqrt{6}}\,\cosh \psi_i + \sinh \psi_i \right) \right]^2},
$$
and in the continuum time limit, this expression becomes
\begin{equation}
\cos \CWLattdihedralS{3}{} = \frac{2 \left[ \frac{1}{6}\,\CWLattlen{1}{} - \frac{1}{2\sqrt{6}} \, \axislen{} \right]^2 \left(\CWLattlendot{1}{}\right)^2 - \left[\left(\CWLattlen{1}{}\right)^2- 2\, \left(\CWLattlen{0}{}\right)^2 \right] \left[ \frac{1}{8} \left(\CWLattlendot{1}{}\right)^2 + 1 \right]}{\left[\left(\CWLattlen{1}{}\right)^2-4\, \left(\CWLattlen{0}{}\right)^2 \right] \left[ \frac{1}{8} \left(\CWLattlendot{1}{}\right)^2 - 1 \right] + 4 \left[ \frac{1}{12}\, \CWLattlen{1}{} + \frac{1}{2\sqrt{6}} \, \axislen{} \right]^2 \left(\CWLattlendot{1}{}\right)^2} + \Odtone.
\label{theta3}
\end{equation}
\end{widetext}

The last geometric quantity we require is the variation $\partial s_{ij} / \partial m_k$ of the particle's path-length $\RegLinEl{ij}$ through a 4-block with respect to each strut-length $\CWstrut{k}$.    Between a pair of Cauchy surfaces $\Cauchyt{i}$ and $\Cauchyt{i+1}$, there are only two distinct types of path-lengths: one corresponds to the path of the unperturbed masses, and we shall denote this path-length by $\RegLinEl{i}$; the other corresponds to the path of the perturbed mass, and we shall denote this path-length by $\RegLinEl{i}^\prime$.  It can be shown that varying $\RegLinEl{i}$ with respect to each of the relevant struts and then taking the continuum time limit yields
\begin{equation}
\frac{\partial \RegLinEl{}}{\partial \CWLattstrutv{A}{}} = \frac{\partial \RegLinEl{}}{\partial \CWLattstrutv{B}{}} = \frac{\partial \RegLinEl{}}{\partial \CWLattstrutv{C}{}} = \frac{1}{4}\, \frac{\CWLattstrutlenpdot{}}{\RegLinEldot{0}} \left( \frac{1}{4}\, \axislen{} \dot{\psi} + 1 \right) + \Odtone \label{Type2ComovingTrajAA}
\end{equation}
and
\begin{equation}
\frac{\partial \RegLinEl{}}{\partial \CWLattstrutv{E}{}} = \frac{1}{4} \frac{\CWLattstrutlendot{}}{\RegLinEldot{0}} \frac{\frac{1}{4} \axislen{} \dot{\psi} + 1}{\axislen{} \dot{\psi} + 1} + \Odtone, \label{Type2ComovingTrajEE}
\end{equation}
where $\CWLattstrutv{X}{}$ denotes the length, in the continuum time limit, of the strut attached to lower vertex $X$, and where $\RegLinEldot{0}$ denotes the quantity
\begin{equation}
\RegLinEldot{0} := \left[ \left(\frac{1}{4}\, \axislendot{} - \frac{\CWLattlendot{1}{}}{2\sqrt{6}}\right)^2 - \left( \frac{1}{4} \axislen{} \dot{\psi} + 1 \right)^2 \right]^\frac{1}{2}.
\end{equation}
This latter quantity is related to the continuum time limit of $\RegLinEl{i}$, where $\RegLinEl{i} \to \RegLinEl{}$ as $\delta \RegTime{i} \to 0$, through the Taylor expansion
\begin{equation}
\RegLinEl{} \approx \RegLinEldot{0} \, dt + \Odt{2}; \label{Taylor_s}
\end{equation}
the expansion has no zeroth order term because as $\delta \RegTime{i} \to 0$, the separation between Cauchy surfaces becomes infinitesimal, and therefore the particle's path-length from one surface to the next would become infinitesimal as well.  It can also be shown that varying $\RegLinEl{i}^\prime$ with respect to each of the struts and then taking the continuum time limit yields
\begin{equation}
\frac{\partial \RegLinEl{}^\prime}{\partial \CWLattstrutv{A}{}} = \frac{\partial \RegLinEl{}^\prime}{\partial \CWLattstrutv{B}{}} = \frac{\partial \RegLinEl{}^\prime}{\partial \CWLattstrutv{C}{}} = \frac{\partial \RegLinEl{}^\prime}{\partial \CWLattstrutv{D}{}} = -\frac{\imath}{4}\, \CWLattstrutlenpdot{}.
\label{Type1TrajCont}
\end{equation}
The derivation of these results has been explained at length in \appendref{PathVariation}.

\subsection{Solving the Regge equations}
Having now determined all relevant geometric quantities, we can now substitute them into \eqref{PerturbedReggeEqn} to obtain the corresponding Regge equations.   As we have two distinct types of struts, locally varying the Regge action will lead to two distinct equations,
\begin{IEEEeqnarray}{rCl}
\sum_i \frac{\partial \Area{}{i}}{\partial \CWLattstrutlen{k}} \, \CWdeficit{i} &=& \sum_{i,j} 8\pi M_i \frac{\partial \RegLinEl{ij}}{\partial \CWLattstrutlen{k}}, \label{Regge1}\\
\sum_i \frac{\partial \Area{}{i}}{\partial \CWLattstrutlenp{k}} \, \CWdeficit{i} &=& \sum_{i,j} 8\pi M_i \frac{\partial \RegLinEl{ij}}{\partial \CWLattstrutlenp{k}}. \label{Regge2}
\end{IEEEeqnarray}
We can directly obtain the continuum time limit of these equations by substituting in the continuum-time form of the geometric quantities, and this is how we shall proceed.

We begin with the first equation.  Between each pair of consecutive Cauchy surfaces, there is only one strut with length $\CWLattstrutlen{k}$, namely the strut $EE^\prime$ in the Type II tetrahedron's 4-block.  Thus the only relevant geometric quantities are those involving strut $EE^\prime$ and vertex E.  We begin by working out the left-hand side of \eqref{Regge1}, starting with the quantity $\frac{\partial \Area{}{i}}{\partial \CWLattstrutlen{k}}$.  The only edges meeting at vertex $E$ are $AE$-type edges, all of which have length $\CWLattlen{0}{i}$.  Thus the only hinges meeting at strut $EE^\prime$ are hinges like $AEE^\prime$, what we have called A2 hinges above, and for each of these hinges, $\frac{\partial \Area{}{i}}{\partial \CWLattstrutlen{k}}$ is given by \eqref{VarAreaA2} in the continuum time limit.  Next, we consider the corresponding deficit angle.  In the 5-tetrahedra model, three faces meet at each hinge, and hence three dihedral angles contribute to the hinge's deficit angle.  The only dihedral angle at A2 hinges is $\CWLattdihedralS{2}{}$, which is given by \eqref{theta2} in the continuum time limit; thus the deficit angle is $\CWdeficit{i} = 2\pi - 3\, \CWLattdihedralS{2}{}$.  We finally perform the summation on the left-hand side of \eqref{Regge1}.  Because $\frac{\partial \Area{}{i}}{\partial \CWLattstrutlen{k}} \, \CWdeficit{i}$ is identical for all hinges meeting at $EE^\prime$, performing the summation is equivalent to multiplying this term by the number of hinges at $EE^\prime$.  As there are four edges meeting at vertex $E$, there can only be four hinges.

Next, we consider the right-hand side of \eqref{Regge1}.  There are four tetrahedra meeting at vertex $E$, and each carry an unperturbed mass $M$; thus $M_i = M$ for all $i$.  Since all four masses and all four tetrahedra are identical, the quantity $\frac{\partial \RegLinEl{ij}}{\partial \CWLattstrutlen{k}}$ will be identical for all $i$ as well; when index $j=k$, its continuum time form is given by \eqref{Type2ComovingTrajEE}; otherwise it is zero.  Because $M_i \frac{\partial \RegLinEl{ij}}{\partial \CWLattstrutlen{k}}$ is identical for all $i$, performing the summation on the right-hand side of \eqref{Regge1} is equivalent to multiplying $M \frac{\partial \RegLinEl{ij}}{\partial \CWLattstrutlen{k}}$ by four.

Finally we substitute everything into \eqref{Regge1} to obtain the constraint equation
\begin{equation}
\begin{split}
& \CWLattlen{0}{} \left(2\pi - 3\, \CWLattdihedralS{2}{}\right)\\
& \qquad {} = 4\pi M \frac{\frac{1}{4} \axislen{} \dot{\psi} + 1}{\RegLinEldot{0}} \left[ \frac{1}{3}\left(\frac{\CWLattlen{1}{}}{\CWLattlen{0}{}}\right)^{\mkern-5mu 2} \mkern-5mu \left( \frac{\frac{\CWLattlendot{1}{}}{2\sqrt{6}} - \axislendot{}}{\axislen{} \dot{\psi} + 1} \right)^{\mkern-5mu 2} \mkern-5mu - 1 \right]^{\frac{1}{2}} \mkern-5mu .
\end{split}
\label{Regge1_Comoving}
\end{equation}

We next consider the second Regge equation \eqref{Regge2}.  Struts of length $\CWLattstrutlenp{}$ are connected to the four vertices labelled $A$, $B$, $C$, $D$ in the Type I tetrahedron.  As we shall see, varying any of the four associated struts will lead to the same Regge equation.  We begin with the left-hand side of \eqref{Regge2} as well.  Each of the four vertices has four tetrahedral edges attached to it: three edges are attached to the three other vertices in the tetrahedron and have length $\CWLattlen{1}{}$; the fourth is attached to vertex $E$ of a Type II tetrahedron and has length $\CWLattlen{0}{}$.   Each of the four edges generates a pair of time-like triangular hinges, one of which is attached to the vertex's strut.  The length-$\CWLattlen{0}{}$ edge contributes hinges like $AA^\prime E^\prime$ in a Type II tetrahedron's 4-block, what we have called B2 hinges above, and for such hinges, $\frac{\partial \Area{}{i}}{\partial \CWLattstrutlenp{k}}$ is given by \eqref{VarAreaB2}.  The other three hinges correspond to either A1 or B1 hinges, and in either case, $\frac{\partial \Area{}{i}}{\partial \CWLattstrutlenp{k}}$ is given by \eqref{VarAreaA1}.  Once again, the deficit angles at each hinge involve three dihedral angles.  At a B2 hinge, the only relevant dihedral angle is $\CWLattdihedralS{3}{}$, which is given by \eqref{theta3}.  So a B2 hinge has a deficit angle of $\CWdeficit{i} = 2\pi - 3\, \CWLattdihedralS{3}{}$.  At A1 and B1 hinges, the relevant dihedral angles are $\CWLattdihedralS{0}{}$ and $\CWLattdihedralS{1}{}$.  In general, each dihedral angle at a hinge comes from a 4-block meeting at the hinge: each 4-block has two faces meeting at the hinge, and the dihedral angle contributed by the 4-block would be the dihedral angle between these two faces.  Additionally, each 4-block is generated by a tetrahedron attached to the edge that generates the hinge.  So in the case of A1 and B1 hinges, each length-$\CWLattlen{1}{}$ edge is always attached to the Type I tetrahedron and to two Type II tetrahedra; the 4-block of the Type I tetrahedron contributes a single $\CWLattdihedralS{0}{}$ to the hinge's deficit angle; the two Type II tetrahedra each contribute a $\CWLattdihedralS{1}{}$ angle; so the deficit angle of both A1 and B1 hinges is $\CWdeficit{i} = 2\pi - \CWLattdihedralS{0}{} - 2\, \CWLattdihedralS{1}{}$.  Combining the contributions from all four hinges, we can express the left-hand side of \eqref{Regge2} as
\begin{widetext}
\begin{equation*}
\begin{split}
\sum_i \frac{\partial \Area{}{i}}{\partial \CWLattstrutlenp{k}} \CWdeficit{i} = {} & \frac{3}{2}\, \CWLattlen{1}{} \, \CWLattstrutlenpdot{} \left[ \frac{1}{8}\left(\CWLattlendot{1}{}\right)^2 - 1 \right]^{-\frac{1}{2}} \left( 2\pi - \CWLattdihedralS{0}{} - 2\, \CWLattdihedralS{1}{} \right)\\
& {} + \frac{\CWLattlen{0}{}}{2} \CWLattstrutlenpdot{} \mkern-5mu \left[ \frac{1}{3}\left(\frac{\CWLattlen{1}{}}{\CWLattlen{0}{}}\right)^2 \mkern-2.5mu \left[ \! \left(\frac{\CWLattlendot{1}{}}{2\sqrt{6}} - \axislendot{} \right)^2 \mkern-10mu - \! \left(\CWLattlendot{0}{}\right)^2\right] \mkern-2.5mu - \mkern-2.5mu \left(\frac{\CWLattlendot{1}{}}{2\sqrt{6}} - \axislendot{} + \frac{\axislen{}}{\CWLattlen{0}{}}\CWLattlendot{0}{} \right)^2 \mkern-10mu + \! \left(\CWLattstrutlenpdot{}\right)^2 \right]^{-\frac{1}{2}} \mkern-10mu \left( 2\pi - 3\, \CWLattdihedralS{3}{} \right) \! .
\end{split}
\end{equation*}
\end{widetext}

We now move on to the right-hand side of \eqref{Regge2}.  Varying the strut-length will affect the trajectory length $\RegLinEl{ij}$ of four neighbouring masses.  One of these will be the perturbed mass of magnitude $M^\prime$ while the other three will be masses of magnitude $M$.  Thus the right-hand side will be
$$
\sum_{ij} 8\pi M_i \frac{\partial \RegLinEl{ij}}{\partial \CWLattstrutlenp{k}} = 8\pi \left[ M^\prime \frac{\partial \RegLinEl{}^\prime}{\partial \CWLattstrutv{A}{}} + 3\, M \frac{\partial \RegLinEl{}}{\partial \CWLattstrutv{A}{}} \right],
$$
with the quantity $\frac{\partial \RegLinEl{}^\prime}{\partial \CWLattstrutv{A}{}}$ given by \eqref{Type1TrajCont} and $\frac{\partial \RegLinEl{}}{\partial \CWLattstrutv{A}{}}$ by \eqref{Type2ComovingTrajAA}.

Finally substituting everything into \eqref{Regge2}, we obtain
\begin{widetext}
\begin{equation}
\begin{split}
4\pi \left[ M^\prime + \imath \, 3\, M \frac{\frac{1}{4} \axislen{} \dot{\psi} + 1}{\RegLinEldot{0}} \right] ={} & 3\, \CWLattlen{1}{} \left[ 1 - \frac{1}{8}\left(\CWLattlendot{1}{}\right)^2 \right]^{-\frac{1}{2}}\left( 2\pi - \CWLattdihedralS{0}{} - 2\, \CWLattdihedralS{1}{} \right)\\
& {} + \CWLattlen{0}{} \left[ \left(\frac{\CWLattlendot{1}{}}{2\sqrt{6}} - \axislendot{} + \frac{\axislen{}}{\CWLattlen{0}{}}\CWLattlendot{0}{} \right)^2 - \left(\CWLattstrutlenpdot{}\right)^2 \right.\\
& \left. \hphantom{{} + \CWLattlen{0}{} [} {} - \frac{1}{3}\left(\frac{\CWLattlen{1}{}}{\CWLattlen{0}{}}\right)^2\left[\left(\frac{\CWLattlendot{1}{}}{2\sqrt{6}} - \axislendot{} \right)^2 - \left(\CWLattlendot{0}{}\right)^2\right] \right]^{-\frac{1}{2}} \left( 2\pi - 3\, \CWLattdihedralS{3}{} \right),
\end{split}
\label{Regge2_Comoving}
\end{equation}
\end{widetext}
where the quantity $\imath / \RegLinEldot{0}$ appearing on the left-hand side would actually be real; this follows because if the path-length $\RegLinEl{}$ of the unperturbed particles is time-like, then $\RegLinEl{}$ would have to be imaginary, and by virtue of its relationship to $\RegLinEldot{0}$ through Taylor expansion \eqref{Taylor_s}, $\RegLinEldot{0}$ would have to be imaginary as well.  We note that both equations \eqref{Regge1_Comoving} and \eqref{Regge2_Comoving} came from the $\Oconst$ term of the Regge equations $0 = \frac{\partial \Action}{\partial \CWLattstrutlen{}}$ and $0 = \frac{\partial \Action}{\partial \CWLattstrutlenp{}}$; thus, the Regge equations are $\Oconst$ to leading order, as claimed in \secref{GlobLocSec}, when we were relating the local and global Regge equations through chain rule \eqref{chain_rule}.

These Regge equations however involve both $\CWLattlen{0}{}$ and $\CWLattlen{1}{}$ in a non-linear manner, which makes solving for them difficult.  We shall therefore linearise these and all subsequent equations by performing a perturbative expansion up to first order in $\dM := M^\prime - M$.  Under this expansion, we must have that
\begin{align*}
\CWLattlen{0}{} &\approx \CWLattlenZ{} + \CWLattdl{0}{},\\
\CWLattlendot{0}{} &\approx \CWLattlenZdot{} + \CWLattdldot{0}{},\\
\CWLattlen{1}{} &\approx \CWLattlenZ{} + \CWLattdl{1}{},\\
\CWLattlendot{1}{} &\approx \CWLattlenZdot{} + \CWLattdldot{1}{},\\
\psi &\approx \dpsi{},\\
\dot{\psi} &\approx \dpsidot{},\\
\CWLattdihedralS{i}{} &\approx \CWLattdihedral{} + \CWLattddihedralS{i}{} \qquad \text{for } i=0,1,2,3,
\end{align*}
as the zeroth order terms must match the corresponding quantities for the unperturbed model.  The zeroth order angle $\CWLattdihedral{}$ is given by relation \eqref{cosq-cont-time}.

It can then be shown that the zeroth order terms for both Regge equations yield
\begin{equation}
\CWLattlenZ{} = \frac{4\pi M}{2\pi - 3\, \CWLattdihedral{}}\left(1 - \frac{1}{8} \CWLattlenZdot{}^2\right)^\frac{1}{2},
\label{ZerothRegge}
\end{equation}
which is equivalent to the unperturbed Regge equation \eqref{cont_l_equal_mass} for the 5-tetrahedra model with the masses at the tetrahedral centres: for the 5-tetrahedra model, we would have $N_p = 5$, $N=10$, and $n=3$; and since the masses are at the tetrahedral centres, we would have $v^2 = 0$, because recall that $v$ is the ratio between $\lvert \Tensb{v}_i \rvert$, the distance of a mass to its tetrahedron's centre as given by \eqref{Unperturbed:MassDistance}, and $\CWLattlenZ{i}$, the tetrahedral edge-length.  By using \eqref{ldot} to substitute for $\CWLattlenZdot{}$, we can parametrise $\CWLattlenZ{}$ entirely in terms of $\CWLattdihedral{}$, yielding
\begin{equation}
\CWLattlenZ{} = \frac{4 \sqrt{2} \pi M}{2\pi - 3\, \CWLattdihedral{}} \tan \left ( \frac{1}{2} \, \CWLattdihedral{} \right).
\label{ZerothRegge-param}
\end{equation}

An expression for $\dpsidot{}$ can be deduced by taking the perturbative expansion of $\dot{\psi}$ as given by \eqref{psidot}; we thus obtain
\begin{equation}
\dpsidot{} = \frac{\sqrt{6}}{\CWLattlenZ{}^2 \, \CWLattlenZdot{}} \left[ \CWLattlenZ{} \left( \CWLattdldot{0}{} - \CWLattdldot{1}{} \right) - \CWLattlenZdot{} \left( \CWLattdl{0}{} - \CWLattdl{1}{} \right) \right].
\label{dpsidot}
\end{equation}

The quantities $\CWLattddihedralS{i}{}$ can be deduced from the perturbative expansions of relations \eqref{theta0} to \eqref{theta3}.  The zeroth order terms of these relations are all identical to \eqref{cosq-cont-time}, as expected.  The first order terms yield
\begin{IEEEeqnarray}{rCl}
\CWLattddihedralS{0}{} & = & -\frac{\CWLattdldot{1}{}}{4\sqrt{2}} \, \sqrt{\frac{3 \cos \CWLattdihedral{} - 1}{1 - \cos \CWLattdihedral{}}} \left(1 + \cos \CWLattdihedral{} \right), \label{dtheta0}\\
\CWLattddihedralS{1}{} & = & -\frac{\CWLattdldot{1}{}}{4\sqrt{2}} \, \sqrt{\frac{3 \cos \CWLattdihedral{} - 1}{1 - \cos \CWLattdihedral{}}} \left(1 + \cos \CWLattdihedral{} \right)\nonumber\\
&& {} + \frac{1}{8\sqrt{2}} \frac{\left(2\pi - 3\, \CWLattdihedral{}\right)}{\pi M} \left(1 + \cos \CWLattdihedral{} \right) \left(\CWLattdl{0}{} - \CWLattdl{1}{}\right), \label{dtheta1}\\
\CWLattddihedralS{2}{} & = & \frac{\left(2\pi - 3\, \CWLattdihedral{}\right)}{4\sqrt{2}\, \pi M} \left( \frac{1+\cos\CWLattdihedral{}}{1-\cos\CWLattdihedral{}} \right) \left(2\cos\CWLattdihedral{} - 1 \right) \left( \CWLattdl{0}{} - \CWLattdl{1}{} \right) \nonumber\\
&& {} - \frac{1}{4\sqrt{2}} \, \sqrt{\frac{3 \cos \CWLattdihedral{} - 1}{1 - \cos \CWLattdihedral{}}} \left(1 + \cos \CWLattdihedral{} \right) \left( 2\, \CWLattdldot{0}{} - \CWLattdldot{1}{} \right) \nonumber\\
&& {} + \frac{4}{\sqrt{3}}\, \frac{\pi M}{\left(2\pi - 3\, \CWLattdihedral{}\right)} \left(3 \cos \CWLattdihedral{} - 1 \right) \dpsidot{},\\
\CWLattddihedralS{3}{} & = & -\frac{\left(2\pi - 3\, \CWLattdihedral{}\right)}{4\sqrt{2}\, \pi M} \left( \frac{1+\cos\CWLattdihedral{}}{1-\cos\CWLattdihedral{}} \right) \cos\CWLattdihedral{} \left( \CWLattdl{0}{} - \CWLattdl{1}{} \right) \nonumber\\
&& {} - \frac{\CWLattdldot{1}{}}{4\sqrt{2}} \, \sqrt{\frac{3 \cos \CWLattdihedral{} - 1}{1 - \cos \CWLattdihedral{}}} \left(1 + \cos \CWLattdihedral{} \right),\label{dtheta3}
\end{IEEEeqnarray}
where we have made use of \eqref{ldot} and \eqref{ZerothRegge-param} to express these as functions of $\CWLattdihedral{}$.  We note that only $\CWLattddihedralS{2}{}$ depends upon the boost parameter $\psi$; if we substitute in relation \eqref{dpsidot} for $\dpsidot{}$, then $\CWLattddihedralS{2}{}$ becomes
\begin{equation}
\begin{aligned}
\CWLattddihedralS{2}{} = {} & -\frac{1}{4\sqrt{2}} \left[ \frac{\left(2\pi - 3\, \CWLattdihedral{}\right)}{\pi M} \left( \frac{1+ \cos \CWLattdihedral{}}{1- \cos \CWLattdihedral{}} \right) \cos \CWLattdihedral{} \left( \CWLattdl{0}{} - \CWLattdl{1}{} \right) \right. \\
& \left. \hphantom{-\frac{1}{4\sqrt{2}} [} {} + \sqrt{\frac{3 \cos \CWLattdihedral{} - 1}{1 - \cos \CWLattdihedral{}}} \left( 1+ \cos \CWLattdihedral{} \right) \CWLattdldot{1}{} \right],
\end{aligned}
\label{dtheta2}
\end{equation}
and we note that $\CWLattdldot{0}{}$ has now dropped out of this expression; in fact, now none of the angle perturbations depends on $\CWLattdldot{0}{}$.

From the perturbative expansion of \eqref{Regge1_Comoving}, the first order term yields
\begin{equation}
\begin{split}
& \CWLattdl{0}{} \left(2\pi - 3\, \CWLattdihedral{}\right) - 3 \, \CWLattlenZ{} \, \CWLattddihedralS{2}{} \\
& \qquad {} = -\frac{\pi M}{\sqrt{1 - \frac{1}{8} \CWLattlenZdot{}^2}} \, \frac{1}{2} \, \CWLattlenZdot{} \left[ \frac{\CWLattlenZdot{}}{\CWLattlenZ{}} \left( \CWLattdl{0}{} - \CWLattdl{1}{} \right) + \CWLattdldot{1}{} \right],
\end{split}
\label{Regge1FirstOrder}
\end{equation}
where we have substituted for $\dpsidot{}$ using \eqref{dpsidot}, and from the perturbative expansion of \eqref{Regge2_Comoving}, the first order term yields
\begin{equation}
\begin{split}
& 4 \pi \dM \sqrt{1 - \frac{1}{8} \CWLattlenZdot{}^2} \\
& \qquad {} = \left( \CWLattdl{0}{} + 3 \, \CWLattdl{1}{} \right) \left(2\pi - 3\, \CWLattdihedral{}\right) - 3 \, \CWLattlenZ{} \left( \CWLattddihedralS{0}{} + 2 \CWLattddihedralS{1}{} + \CWLattddihedralS{3}{} \right) \\
& \qquad \hphantom{{} = } {} + \frac{\left(2\pi - 3\, \CWLattdihedral{}\right)}{1 - \frac{1}{8} \CWLattlenZdot{}^2 } \left[ \frac{1}{8} \, \CWLattlenZdot{}^2 \left( \CWLattdl{0}{} - \CWLattdl{1}{} \right) + \frac{1}{2} \, \CWLattlenZ{} \, \CWLattlenZdot{} \, \CWLattdldot{1}{} \right].
\end{split}
\label{Regge2FirstOrder}
\end{equation}
We note that none of these relations depends on $\CWLattdldot{0}{}$ either.

In the unperturbed model, we used the dihedral angle $\CWLattdihedral{}$ to parametrise $\CWLattlenZdot{}$ through relation \eqref{ldot}.  We shall do something similar here and parametrise $\CWLattdldot{1}{}$ with respect to one of the angle perturbations.  It is easiest to do this with relation \eqref{dtheta0}, which then yields
\begin{equation}
\CWLattdldot{1}{} = -4 \sqrt{2} \, \sqrt{ \frac{1-\cos\CWLattdihedral{}}{3\cos\CWLattdihedral{}-1} } \frac{\CWLattddihedralS{0}{}}{\left( 1+\cos\CWLattdihedral{} \right)}.\label{dldot1-param}
\end{equation}
Since none of the angle perturbations depends on $\CWLattdldot{0}{}$, a similar parametrisation is not possible for $\CWLattdldot{0}{}$.  However, the first order terms \eqref{Regge1FirstOrder} and \eqref{Regge2FirstOrder} of the two Regge equations \eqref{Regge1_Comoving} and \eqref{Regge2_Comoving} do not depend on $\CWLattdldot{0}{}$ anyway, so such a parametrisation of $\CWLattdldot{0}{}$ is not necessary.

Using relations \eqref{dtheta0}, \eqref{dtheta1}, \eqref{dtheta2}, and \eqref{dtheta3} to substitute for $\CWLattddihedralS{0}{}$, $\CWLattddihedralS{1}{}$, $\CWLattddihedralS{2}{}$, $\CWLattddihedralS{3}{}$, relation \eqref{dldot1-param} to substitute for $\CWLattdldot{1}{}$, and relations \eqref{ZerothRegge-param} and \eqref{ldot} to substitute for $\CWLattlenZ{}$ and $\CWLattlenZdot{}$, we can now solve \eqref{Regge1FirstOrder} and \eqref{Regge2FirstOrder} for $\CWLattdl{0}{}$ and $\CWLattdl{1}{}$ and express them exclusively in terms of the parameters $\CWLattdihedral{}$ and $\CWLattddihedralS{0}{}$.  We find that
\begin{widetext}
\begin{equation}
\begin{split}
\CWLattdl{0}{} = {} & \frac{4\sqrt{2}\pi M}{3 \left(2\pi - 3\, \CWLattdihedral{}\right)} \left[ 2 \sin \CWLattdihedral{} \left( \frac{1+2 \cos \CWLattdihedral{}}{1+\cos \CWLattdihedral{}} \right) + \left(2\pi - 3\, \CWLattdihedral{}\right) \right]^{-1}\\
& {} \times \left[ \frac{\dM}{M} \left[\frac{1-\cos \CWLattdihedral{}}{1+\cos \CWLattdihedral{}} \right]^\frac{1}{2} \left[ 6 \sin \CWLattdihedral{} \left[ \frac{\cos \CWLattdihedral{}}{1+\cos \CWLattdihedral{}} \right] + \left(2\pi - 3\, \CWLattdihedral{}\right) \left[ \frac{3 \cos \CWLattdihedral{} - 1}{1+\cos \CWLattdihedral{}} \right] \right] \right.\\
& \left. \hphantom{{} \times [} {} + \frac{3 \, \CWLattddihedralS{0}{}}{\left(2\pi - 3\, \CWLattdihedral{}\right)} \frac{1}{1+\cos \CWLattdihedral{}} \left[ 3 \sin \CWLattdihedral{} + \left(2\pi - 3\, \CWLattdihedral{}\right) \vphantom{\left( \frac{1+2 \cos \CWLattdihedral{}}{1+\cos \CWLattdihedral{}} \right)} \right] \left[ 2 \sin \CWLattdihedral{} \left( \frac{1+2 \cos \CWLattdihedral{}}{1+\cos \CWLattdihedral{}} \right) + \left(2\pi - 3\, \CWLattdihedral{}\right) \right] \vphantom{\left[\frac{1-\cos \CWLattdihedral{}}{1+\cos \CWLattdihedral{}} \right]^\frac{1}{2}} \right]
\end{split}
\label{dl0}
\end{equation}
and that
\begin{equation}
\begin{split}
\CWLattdl{1}{} = {} & \frac{4\sqrt{2}\pi M}{3 \left(2\pi - 3\, \CWLattdihedral{}\right)} \left[ 2 \sin \CWLattdihedral{} \left( \frac{1+2 \cos \CWLattdihedral{}}{1+\cos \CWLattdihedral{}} \right) + \left(2\pi - 3\, \CWLattdihedral{}\right) \right]^{-1} \\
& {} \times \left[ \frac{\dM}{M} \left[\frac{1-\cos \CWLattdihedral{}}{1+\cos \CWLattdihedral{}} \right]^\frac{1}{2} \left[ 6 \sin \CWLattdihedral{} \left[ \frac{\cos \CWLattdihedral{}}{1+\cos \CWLattdihedral{}} \right] + \left(2\pi - 3\, \CWLattdihedral{}\right) \right] \right.\\
& \left. \hphantom{{} \times [} {} + \frac{3 \, \CWLattddihedralS{0}{}}{\left(2\pi - 3\, \CWLattdihedral{}\right)} \frac{1}{1+\cos \CWLattdihedral{}} \left[ 3 \sin \CWLattdihedral{} + \left(2\pi - 3\, \CWLattdihedral{}\right) \vphantom{\left( \frac{1+2 \cos \CWLattdihedral{}}{1+\cos \CWLattdihedral{}} \right)} \right] \left[ 2 \sin \CWLattdihedral{} \left( \frac{1+2 \cos \CWLattdihedral{}}{1+\cos \CWLattdihedral{}} \right) + \left(2\pi - 3\, \CWLattdihedral{}\right) \right] \vphantom{\left[\frac{1-\cos \CWLattdihedral{}}{1+\cos \CWLattdihedral{}} \right]^\frac{1}{2}} \right].
\end{split}
\label{dl1}
\end{equation}
\end{widetext}

Finally, we note that our two parameters $\CWLattdihedral{}$ and $\CWLattddihedralS{0}{}$ are not independent of each other; rather both are functions of the underlying time parameter $t$; therefore if one parameter evolves, so must the other.  We can relate the two parameters to $t$ through the system of differential equations
\begin{IEEEeqnarray}{rCl}
\CWLattlenZdot{} &=& \frac{d}{dt} \CWLattlenZ{}, \label{DiffEqn1}\\
\CWLattdldot{1}{} &=& \frac{d}{dt} \CWLattdl{1}{}, \label{DiffEqn2}
\end{IEEEeqnarray}
where the left-hand side denotes the quantities given by \eqref{ldot} and \eqref{dldot1-param}, while the right-hand side denotes the explicit differentiation of \eqref{ZerothRegge-param} and \eqref{dl1} with respect to $t$.  The first equation involves only $\CWLattdihedral{}$; it can be solved on its own to yield $\CWLattdihedral{}(t)$.  This can then be substituted into the second equation to give a differential equation for $\CWLattddihedralS{0}{}$.  We shall solve these equations numerically.  To determine a unique solution though, we must also specify a set of initial conditions: we shall require the perturbed model to obey the initial value equation at its moment of time symmetry.  This equation will be explained in the next section, and it implies specific conditions on $\theta$ and $\CWLattddihedralS{0}{}$ which we shall derive.

The range of the parameter $t$ will be constrained by the requirement that all struts remain time-like.  As this constraint depends on $\dM$, the resulting range of $t$ will also depend on $\dM$.

Before leaving this section, we wish to remark on an advantage that local variation has afforded over global variation.  For each model, local variation has yielded a pair of Regge equations that, when expanded perturbatively, gave three distinct equations, an identical equation from their zeroth order terms and two distinct equations from their first order terms.  Had we directly varied the action globally instead, we would only have obtained one Regge equation corresponding to a linear combination of the two local Regge equations, and the perturbative expansion of this global equation would give just two independent equations.  Thus local variation has provided us with an extra independent equation, allowing us to specify one more of the five quantities, $\CWLattlenZ{}$, $\CWLattlenZdot{}$, $\CWLattdl{0}{}$, $\CWLattdl{1}{}$, $\CWLattdldot{1}{}$, that we needed to solve for.  This has allowed us to pick one unique solution out of many in the solution space of the global Regge equations.

\section{Initial value equation for perturbed models}

We now have a set of equations, in \eqref{DiffEqn1} and \eqref{DiffEqn2}, that should determine the entire Regge skeleton from a set of initial conditions on $\theta$ and $\CWLattddihedralS{0}{}$.  However, the question remains as to what initial conditions would be appropriate.  To answer this, we shall consider the analogous (3+1)-formulation of general relativity wherein the entire space-time is similarly determined by evolving forwards in time from some Cauchy surface $\Cauchyt{0}$ with initial data; naturally, the evolution equation would be derived from the Einstein field equations.  It has been shown \cite{HawkingEllis} that, for such a formulation, the required initial data consists of the first and second fundamental forms, $\Tensb{\firstform}$ and $\Tensb{\secondform}$; the former corresponds to the projection of the metric $\Tensb{\metric}$ into $\Cauchyt{0}$ and effectively determines the 3-dimensional intrinsic curvature of $\Cauchyt{0}$; the latter effectively determines the extrinsic curvature of $\Cauchyt{0}$ within the overall space-time.  

However, for this initial data to be consistent with the Einstein field equations, there is a set of constraint equations that it must satisfy.  Let us express the Einstein field equations in the form
$$\Tensb{\EinsTens} = 8 \pi \Tensb{\StressEnergy},$$
where $\Tensb{\EinsTens}$ is the Einstein tensor and $\Tensb{\StressEnergy}$ the stress-energy tensor.  Let $\Tensb{n}$ denote a field of normalised one-forms everywhere orthogonal to $\Cauchyt{0}$.  By making use of the Gauss equation
\begin{equation}
{}^{(3)}\Riemann_{\mu\nu\sigma\rho} = \Riemann_{\alpha\beta\gamma\delta} \, \firstprojection{\alpha}{\mu} \, \firstprojection{\beta}{\nu} \, \firstprojection{\gamma}{\sigma} \, \firstprojection{\delta}{\rho} - \secondform_{\mu\sigma} \, \secondform_{\nu\rho} + \secondform_{\mu\rho} \, \secondform_{\nu\sigma},
\label{Gauss}
\end{equation}
which relates the 3-dimensional intrinsic curvature ${}^{(3)}\Riemann_{\mu\nu\sigma\rho}$ of $\Cauchyt{0}$ to its extrinsic curvature $\Tensb{\secondform}$ and its 4-dimensional intrinsic curvature $\Riemann_{\alpha\beta\gamma\delta}$, we can express the relation $\Tensb{\EinsTens}\left(\Tensb{n}, \Tensb{n} \right) = 8\pi \Tensb{\StressEnergy}\left(\Tensb{n}, \Tensb{n} \right)$ as
\begin{equation}
{}^{(3)}\RicScal + \left(\secondform^{\mu\nu} \, \firstform_{\mu\nu} \right)^2 - \secondform^{\mu\nu} \, \secondform^{\rho\sigma} \, \firstform_{\mu\rho} \, \firstform_{\nu\sigma}  = 16 \pi \rho,
\label{Hamiltonian}
\end{equation}
where ${}^{(3)}\RicScal$ is the 3-dimensional Ricci scalar of $\Cauchyt{0}$ and $\rho$ is the energy density of the matter source as measured by an observer co-moving with respect to $\Cauchyt{0}$.  Equation \eqref{Hamiltonian} gives the first constraint equation; it is actually the Hamiltonian constraint of the ADM formalism, where it is customarily derived by extremising the ADM action with respect to the lapse function \cite{MTW}.  Let $\left\{\Tensb{u}_i\right\}$, for $i=1,2,3$, denote a set of normalised basis vectors tangent to $\Cauchyt{0}$, and let $\vert$ denote covariant differentiation with respect to the metric connection implied by $\Tensb{\firstform}$.  By making use of the Gauss-Codazzi equation
\begin{equation}
\RicTens_{\sigma\rho} \, n^\sigma \firstprojection{\rho}{\mu} = \secondform^\sigma_{\phantom{\sigma} \mu \vert \sigma} - \secondform^\sigma_{\phantom{\sigma} \sigma \vert \mu},
\label{Gauss-Codazzi}
\end{equation}
which relates the extrinsic curvature $\Tensb{\secondform}$ of $\Cauchyt{0}$ to its 4-dimensional intrinsic curvature in the form of the Ricci tensor $\RicTens_{\sigma\rho}$, we can express the relation $\Tensb{\EinsTens}\left(\Tensb{n}, \Tensb{u}_i \right) = 8\pi \Tensb{\StressEnergy}\left(\Tensb{n}, \Tensb{u}_i \right)$ as
\begin{equation}
\left( \secondform^{\sigma\mu}_{\phantom{\sigma\mu} \vert \mu} \, \firstform_{\sigma\nu} - \secondform^{\sigma\mu}_{\phantom{\sigma\mu} \vert \nu} \, \firstform_{\sigma\mu} \right) u^\nu_{\phantom{\nu}i} = 8 \pi \, \StressEnergy_{\mu\nu} \, n^\mu \, u^\nu_{\phantom{\nu}i},
\label{Momentum}
\end{equation}
which is actually a set of three equations, one for each $i$.  This gives the rest of the constraint equations.  These are the momentum constraints of the ADM formalism, where they are customarily derived by extremising the ADM action with respect to the shift functions \cite{MTW}.

Quite often, the initial surface $\Cauchyt{0}$ is chosen to be the surface at a moment of time symmetry, that is, the moment when the surface's extrinsic curvature, as given by the second fundamental form $\Tensb{\secondform}$, vanishes.  In this case, the momentum constraints would vanish while the Hamiltonian constraint would simplify to
\begin{equation}
^{(3)}\RicScal = 16 \pi \rho;
\label{GenInitVal}
\end{equation}
this is known as the initial value equation at the moment of time symmetry.

To determine the appropriate initial conditions for our Regge model, we shall require our model to satisfy this equation as well at its moment of time symmetry.  However, we note that this equation and the Einstein field equations, from which it is derived, will only be satisfied in an average manner on a Regge Cauchy surface.  Curvature in the surface is concentrated only at the hinges, yet matter can be distributed away from the hinges where the skeleton is flat, which is indeed the case for our Regge model; thus the two sides of the equation will not agree in a point-wise manner.  This contradiction arises because the Einstein field equations actually apply to smooth manifolds rather than Regge skeletons; they come about by varying the Einstein-Hilbert action when the underlying manifold is smooth rather than discrete.  Thus by using the Einstein equations in this manner, we are effectively varying the Einstein-Hilbert action on a smooth manifold first and then applying the resulting field equations on a discrete manifold afterwards.  The standard approach in Regge calculus is to use a discrete manifold from the very beginning, with the field equations obtained being different as a result.  Clearly, the two approaches are not equivalent.

We shall now apply \eqref{GenInitVal} to our Regge model and thereby deduce a set of initial conditions on $\theta$ and $\CWLattddihedralS{0}{}$.  The perturbed models attain a moment of time symmetry when all lengths cease expanding or contracting, that is, when $\CWLattlendot{0}{} = \CWLattlendot{1}{} = 0$.  We shall perform an averaging of \eqref{GenInitVal} by integrating it over such a time-symmetric CW Cauchy surface $\Cauchyt{0}$.  The left-hand side of \eqref{GenInitVal} then becomes \cite{Regge}
$$
\int_{\Cauchyt{0}} {}^{(3)}\RicScal \; d^3 x = 2 \mkern-20mu \sum_{i \,\in\, \left\{ \text{hinges}\right\}} \mkern-20mu \ell_i \CWdeficit{i},
$$
where the integration measure is unity because the Regge tetrahedra are flat, the summation is over all edges in $\Cauchyt{0}$ because these would be the hinges of a 3-dimensional skeleton, $\ell_i$ is the length of an edge, and $\CWdeficit{i}$ is its corresponding 3-dimensional deficit angle; the right-hand side evaluates to
$$
\displaystyle{\int_{\Sigma_0} \rho \; d^3 x} = 5M + \dM.
$$
Therefore, the averaged initial value equation for the perturbed model can be expressed as
\begin{equation}
\sum_{i \in \{\text{edges}\}} \CWLattlenZ{i} \CWdeficit{i} = 8 \pi \left( 5M + \dM \right).
\label{InitVal}
\end{equation}
By requiring our model to satisfy this form of the initial value equation, we shall deduce the necessary initial conditions on $\theta$ and $\CWLattddihedralS{0}{}$.

The only quantities in \eqref{InitVal} that have yet to be determined are the deficit angles.  A Cauchy surface of the perturbed universe has only two distinct types of hinges, the edges of length $\CWLattlen{0}{i}$ and the edges of length $\CWLattlen{1}{i}$.  Each edge is connected to three faces separating three tetrahedra, so each tetrahedron at the edge will contribute one dihedral angle to the edge's deficit angle.  A Cauchy surface of the perturbed universe also has only two distinct types of tetrahedra, the Type I and Type II tetrahedra.  As the Type I tetrahedron is equilateral, it will contribute the same dihedral angle to each of its six edges, and we denote this angle by $\CWLattdihedralPhi{0}{i}$.  In the Type II tetrahedron, all edges of $\CWLattlen{0}{i}$ are identical to each other, as are all edges of length $\CWLattlen{1}{i}$.  Thus this tetrahedron will contribute the same dihedral angle $\CWLattdihedralPhi{1}{i}$ to each of its $\CWLattlen{0}{i}$ edges and the same dihedral angle $\CWLattdihedralPhi{2}{i}$ to each of its $\CWLattlen{1}{i}$ edges.  To determine the dihedral angle $\CWLattdihedralPhi{i}{i}$ between any pair of faces, we shall again take the scalar product of the unit normals to the two faces, as this product will yield $\cos \CWLattdihedralPhi{i}{i}$.

Let us first consider the dihedral angles in the Type I tetrahedron.  We can use co-ordinate system \eqref{vertices} for this tetrahedron, dropping the time co-ordinate so that we work in a purely 3-dimensional spatial co-ordinate system and replacing lengths $\CWLattlenZ{i}$ with $\CWLattlen{1}{i}$, as this tetrahedron has edges of length $\CWLattlen{1}{i}$; we can then use this co-ordinate system to assign co-ordinates to any of the normal vectors.  We can calculate $\CWLattdihedralPhi{0}{i}$ using the faces meeting at edge $AB$; these faces are $ABC$ and $ABD$, and the scalar product of their unit normals yields
\begin{equation}
\cos \CWLattdihedralPhi{0}{i} = \frac{1}{3}.
\end{equation}

For the Type II tetrahedron, we can work with co-ordinate system \eqref{SLvertices}, again dropping the time co-ordinate to obtain a purely 3-dimensional spatial system.  We can use edge $AE$ to calculate the dihedral angle $\CWLattdihedralPhi{1}{i}$ at an edge of length $\CWLattlen{0}{i}$; the faces meeting at $AE$ are $ABE$ and $ACE$, and the scalar product of their unit normals yields
\begin{equation}
\cos \CWLattdihedralPhi{1}{i} = \frac{\frac{1}{2}\left(\CWLattlen{0}{i}\right)^2 - \frac{1}{4}\left(\CWLattlen{1}{i}\right)^2}{\left(\CWLattlen{0}{i}\right)^2 - \frac{1}{4}\left(\CWLattlen{1}{i}\right)^2}.
\end{equation}
Similarly, we can use edge $AB$ to calculate the dihedral angle $\CWLattdihedralPhi{2}{i}$ of an $\CWLattlen{1}{i}$ edge; the faces meeting at $AB$ are $ABC$ and $ABE$, and the scalar product of their unit normals yields
\begin{equation}
\cos \CWLattdihedralPhi{2}{i} = \frac{\CWLattlen{1}{i}}{2\sqrt{3}} \frac{1}{\sqrt{\left( \CWLattlen{0}{i} \right)^2 - \frac{1}{4} \left( \CWLattlen{1}{i} \right)^2 }}.
\end{equation}
To obtain the continuum time limit of these expressions to leading order in $dt$, we simply need to drop all subscripts $i$.  If we next take the perturbative expansion of these continuum time expressions, such that
\begin{align}
\CWLattdihedralPhi{1}{} \approx{}& \phi + \CWLattddihedralPhi{1}{},\nonumber\\
\CWLattdihedralPhi{2}{} \approx{}& \phi + \CWLattddihedralPhi{2}{},\nonumber\\
\intertext{we find that}
\phi ={}& \frac{1}{3},\\
\intertext{and that}
\CWLattddihedralPhi{1}{} = -\CWLattddihedralPhi{2}{} ={}& -\frac{\sqrt{2}}{3 \, \CWLattlenZ{}} \left( \CWLattdl{0}{} - \CWLattdl{1}{} \right).
\end{align}

To calculate the deficit angles at an edge, we simply subtract the three relevant dihedral angles from $2\pi$.  Since only Type II tetrahedra have $\CWLattlen{0}{}$ edges, an $\CWLattlen{0}{}$ edge must be connected exclusively to Type II tetrahedra, and its deficit angle $\CWLattdeficit{0}{}$ must therefore be
\begin{equation}
\begin{aligned}
\CWLattdeficit{0}{} &= 2\pi - 3\, \CWLattdihedralPhi{1}{}\\
&\approx 2\pi - 3 \,\arccos \frac{1}{3} + \frac{\sqrt{2}}{\CWLattlenZ{}} \left( \CWLattdl{0}{} - \CWLattdl{1}{} \right).
\end{aligned}
\end{equation}
As there is only one Type I tetrahedron on the entire Cauchy surface, an $\CWLattlen{1}{}$ edge can only be connected to one Type I tetrahedron, and its other two tetrahedra must be Type II; its deficit angle $\CWLattdeficit{1}{}$ must therefore be
\begin{equation}
\begin{aligned}
\CWLattdeficit{0}{} &= 2\pi - \CWLattdihedralPhi{0}{} - 2\, \CWLattdihedralPhi{2}{}\\
&\approx 2\pi - 3 \,\arccos \frac{1}{3} - \frac{2\sqrt{2}}{3 \, \CWLattlenZ{}} \left( \CWLattdl{0}{} - \CWLattdl{1}{} \right).
\end{aligned}
\end{equation}

Having now determined the deficit angles, we can now substitute all relevant geometric quantities into \eqref{InitVal}.  As mentioned previously, a 5-tetrahedra Cauchy surface will have a total of 10 edges; six must come from the Type I tetrahedron and must therefore be of length $\CWLattlen{1}{}$; the remaining four must therefore be of length $\CWLattlen{0}{}$.  Thus \eqref{InitVal} can be expressed as
\begin{align}
&8\pi \left( 5M + \dM \right)\nonumber\\
&\qquad {} = 4\, \CWLattlen{0}{} \left[ 2\pi - 3 \,\arccos \frac{1}{3} + \frac{\sqrt{2}}{\CWLattlenZ{}} \left( \CWLattdl{0}{} - \CWLattdl{1}{} \right) \right] \nonumber\\
&\qquad \hphantom{{} = } {} + 6\, \CWLattlen{1}{} \left[ 2\pi - 3 \,\arccos \frac{1}{3} - \frac{2\sqrt{2}}{3 \, \CWLattlenZ{}} \left( \CWLattdl{0}{} - \CWLattdl{1}{} \right) \right] \nonumber\\
&\qquad {} \approx \left[10\, \CWLattlenZ{} + 2 \left( 2\, \CWLattdl{0}{} + 3\, \CWLattdl{1}{} \right) \right] \left( 2\pi - 3\, \arccos \frac{1}{3} \right).\label{InitValuePerturbed}
\end{align}

The zeroth order term corresponds to the initial value equation for the unperturbed model.  At the moment of time symmetry in the unperturbed model, we have that $\CWLattlenZdot{} = 0$, and from \eqref{cosq-cont-time}, it follows that $\CWLattdihedral{}$ is
\begin{equation}
\CWLattdihedral{} = \arccos \frac{1}{3}.\label{theta_condition}
\end{equation}
Substituting this into the unperturbed model's Regge equation \eqref{ZerothRegge}, we deduce that
\begin{equation}
4 \pi M = \CWLattlenZ{} \left( 2\pi - 3\, \arccos \frac{1}{3} \right),
\end{equation}
which is identical to the initial value equation for the unperturbed model.  Thus the Regge equation of the unperturbed model satisfies its initial value equation at the moment of time symmetry.  Therefore, for the zeroth order component of the perturbed models' initial value equation to be satisfied, we must also require that $\CWLattdihedral{}$ satisfy condition \eqref{theta_condition}.  This is the initial condition on $\CWLattdihedral{}$.

We can next deduce the condition on $\CWLattddihedralS{0}{}$ by solving for it from the first order term of \eqref{InitValuePerturbed}.  After using \eqref{theta_condition} to substitute for $\CWLattdihedral{}$ as well as \eqref{dl0} and \eqref{dl1} to substitute for $\CWLattdl{0}{}$ and $\CWLattdl{1}{}$, we find that $\CWLattddihedralS{0}{}$ must satisfy
\begin{equation}
\CWLattddihedralS{0}{} = 0.\label{dq0_condition}
\end{equation}

We note that the behaviour of $\CWLattdldot{1}{}$ near time symmetry cannot be determined from \eqref{dldot1-param}.  On the one hand, condition \eqref{dq0_condition} suggests it may approach zero, while on the other, condition \eqref{theta_condition} suggests it may diverge.  As we shall see below however, $\CWLattdldot{1}{}$ is indeed well-behaved and tends towards zero as time symmetry is approached.

Finally, even after imposing conditions \eqref{theta_condition} and \eqref{dq0_condition} at some moment $t=T_\text{max}$, we must still ensure that $\CWLattdldot{0}{} = \CWLattdldot{1}{} = 0$ at that moment; otherwise, $t=T_\text{max}$ would not be a moment of time symmetry.  We have just mentioned that $\CWLattdldot{1}{}$ does tend toward zero as time symmetry is approached.  Given this information, it can also be shown that $\CWLattdldot{0}{}$ will be zero as well.  First, by comparing \eqref{dl0} and \eqref{dl1}, we note that $\CWLattdl{0}{}$ can be expressed in terms of $\CWLattdl{1}{}$ to give
\begin{equation}
\begin{split}
\CWLattdl{0}{} = {}& \CWLattdl{1}{} - \frac{8\sqrt{2}\pi \dM}{3} \left[\frac{1-\cos \CWLattdihedral{}}{1+\cos \CWLattdihedral{}} \right]^\frac{3}{2} \\
& \hphantom{\CWLattdl{1}{} - } \; {} \times \left[ 2 \sin \CWLattdihedral{} \left( \frac{1+2 \cos \CWLattdihedral{}}{1+\cos \CWLattdihedral{}} \right) + \left(2\pi - 3\, \CWLattdihedral{}\right) \right]^{-1} \mkern-5mu.
\end{split}
\label{dl0_dl1_relation}
\end{equation}
If we differentiate this with respect to $t$, we obtain an expression of the form
$$
\CWLattdldot{0}{} = \CWLattdldot{1}{} - \frac{8\sqrt{2}\pi \dM}{3} F_1 \left( \CWLattdihedral{} \right) \CWLattdihedraldot{}.
$$
At the moment of time symmetry, we have said that $\CWLattdldot{1}{}$ will vanish and that condition \eqref{theta_condition} must be satisfied; then it can be shown that $F_1 \left( \CWLattdihedral{} = \arccos \frac{1}{3} \right)$ will not vanish, and therefore $\CWLattdldot{0}{}$ will vanish if and only if $\CWLattdihedraldot{}$ vanishes.  To see that $\CWLattdihedraldot{}$ does indeed vanish, we next differentiate $\CWLattlenZ{}$, as given by \eqref{ZerothRegge-param}, with respect to $t$ to obtain an expression of the form
$$
\CWLattlenZdot{} = F_2 \left( \CWLattdihedral{} \right) \CWLattdihedraldot{}.
$$
The left-hand side is given by \eqref{ldot} and is zero at time symmetry.  It can also be shown that $F_2 \left( \CWLattdihedral{} = \arccos \frac{1}{3} \right)$ will not vanish; thus for the two sides of the expression to equal, it follows that $\CWLattdihedraldot{}$ must vanish at time symmetry.  Therefore, we deduce that $\CWLattdldot{0}{}$ vanishes at time symmetry provided $\CWLattdldot{1}{}$ vanishes as well.

\section{Discussion of the models}
\begin{figure}[t]
\includegraphics{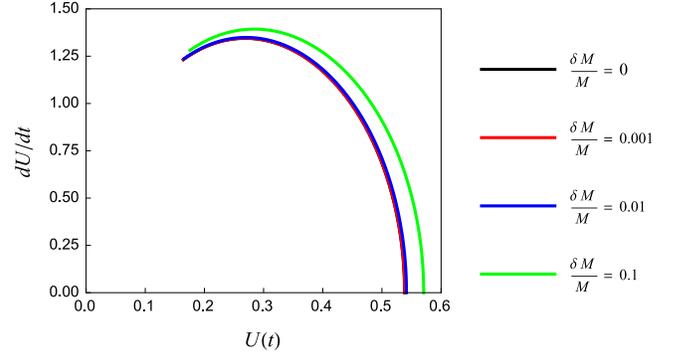}
\caption{\label{fig:PertVolGraphs}The expansion rate of the universe's volume $dU / dt$ versus the volume $U$ itself.  Graphs corresponding to four different values of $\dM / M$ are shown, with $\dM / M = 0$ corresponding to the unperturbed model.  Owing to $\dM / M = 0.001$ and $\dM / M = 0.01$ being very small perturbations, the graph for $\dM / M = 0.01$ has almost completely covered the graphs for $\dM / M = 0$ and $\dM / M = 0.001$.  The left of the graphs have been truncated at the moment the struts turn null.  In all four models, the mass $M$ has been fixed to be $M= 1/5$.}
\end{figure}

We shall now examine the behaviour of the Regge model just obtained, comparing the behaviour for various mass perturbations against each other and against the behaviour of the unperturbed 5-tetrahedra model.  We begin by examining the expansion rate of the universe's volume $dU / dt$ against the volume $U$ itself, a relation which has been plotted in \figref{fig:PertVolGraphs}.
\begin{figure}[t]
\includegraphics{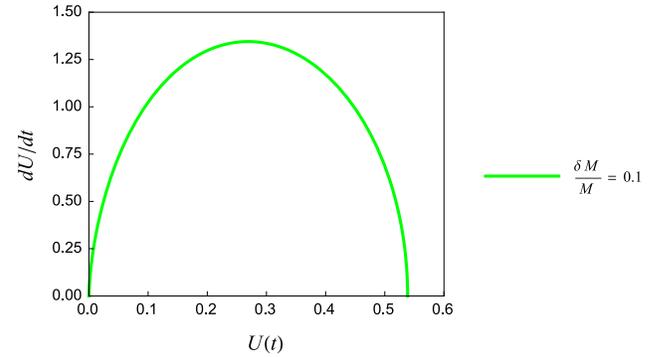}
\caption{\label{fig:ExtendedPertVolGraph}The same graph as in \figref{fig:PertVolGraphs} for the $\dM / M = 0.1$ model, but with the graph extended all the way back to $t=0$.}
\end{figure}
The volume of the unperturbed universe is given by \eqref{ParentU} and its expansion rate by \eqref{ParentdUdt}.  To first order in the perturbative expansion, the volume of a perturbed universe is given by
\begin{equation}
U = \frac{5}{6\sqrt{2}}\, \CWLattlenZ{}^3 + \frac{1}{2\sqrt{2}}\, \CWLattlenZ{}^2 \left( 2\, \CWLattdl{0}{} + 3\, \CWLattdl{1}{} \right),
\label{PerturbedVol}
\end{equation}
and the expansion rate by
\begin{equation}
\begin{split}
\frac{dU}{dt} ={} & \frac{5}{2\sqrt{2}}\, \CWLattlenZ{}^2\, \CWLattlenZdot{} + \frac{1}{\sqrt{2}}\, \CWLattlenZ{} \CWLattlenZdot{} \left( 2\, \CWLattdl{0}{} + 3\, \CWLattdl{1}{} \right)\\
& {} + \frac{1}{2\sqrt{2}}\, l^2 \left( 2\, \CWLattdldot{0}{} + 3\, \CWLattdldot{1}{} \right),
\end{split}
\end{equation}
where $\CWLattdldot{0}{}$ would be given by the explicit time-derivative of $\CWLattdl{0}{}$.
Across all perturbations in mass, the evolution of the universe's volume is very stable, indeed closely resembling the evolution of the unperturbed universe.  The effect of increasing the perturbation $\dM / M$ is for the universe to attain larger volumes and faster expansion rates.  All graphs have been truncated on the left at the moment the struts turn null.  However, we note that these models actually remain well-behaved past this point all the way back to $t=0$, as \figref{fig:ExtendedPertVolGraph} shows for the $\dM / M = 0.1$ model.

\begin{figure}[t]
\subfloat{\includegraphics{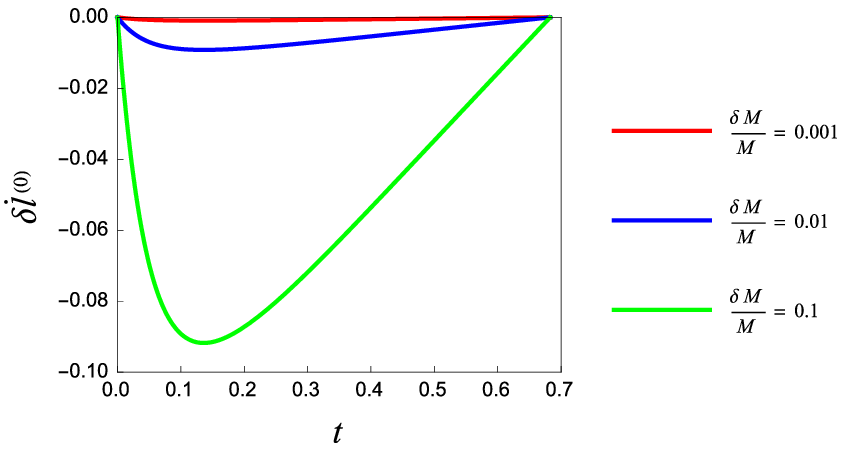}}\\
\subfloat{\includegraphics{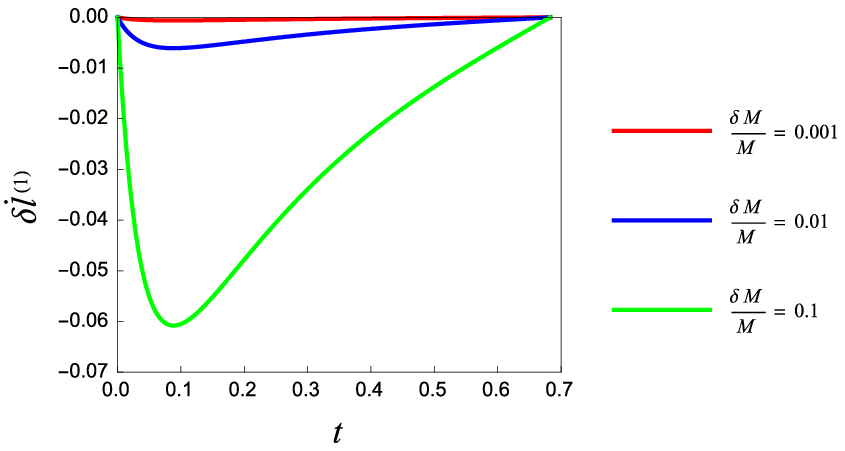}}
\caption{\label{fig:dldot_graphs}Plots of $\CWLattdldot{0}{}(t)$ (top) and $\CWLattdldot{1}{}(t)$ (bottom) against $t$.  The graphs have been extended all the way back to $t=0$.}
\end{figure}

\figref{fig:dldot_graphs} shows the behaviour of $\CWLattdldot{0}{}$ and $\CWLattdldot{1}{}$ as functions of time; these graphs have also been extended all the way back to $t=0$.  They reveal that $\CWLattdldot{0}{}$ and $\CWLattdldot{1}{}$ are indeed well-behaved near $t = T_\text{max}$ and approach zero as $t \to T_\text{max}$, which is required for a moment of time symmetry.  They also reveal that $\CWLattdldot{0}{}$ and $\CWLattdldot{1}{}$ start from zero as well at $t=0$.  Finally, they show that the absolute magnitudes of the graphs increase with $\dM / M$.

\begin{table}[t]
\renewcommand{\arraystretch}{1.2}
\caption{\label{tab:TimeSymmVals}The numerical values for $\CWLattlenZdot{}$, $\CWLattdldot{0}{}$, and $\CWLattdldot{1}{}$ at $t=T_\text{max}$.}
\begin{tabular}{>{\centering\arraybackslash}m{1.25cm} >{\centering\arraybackslash}m{2.75cm} >{\centering\arraybackslash}m{2.75cm} >{\centering\arraybackslash}m{1.25cm}}
\hline
$\frac{\dM}{M} \vphantom{\frac{\dM}{M}_\frac{1}{2}^\frac{1}{2}}$ & $\CWLattlenZdot{}$ & ${}^{\phantom{(0)}} \CWLattdldot{0}{}$ & ${}^{\phantom{(1)}} \CWLattdldot{1}{}$ \\
\hline\hline
0.001 & $5.16191 \times 10^{-8}$ & $-1.5465 \times 10^{-11}$ & 0\\
0.01 & $5.16191 \times 10^{-8}$ & $-1.5465 \times 10^{-10}$ & 0\\
0.1 & $5.16191 \times 10^{-8}$ & $-1.5465 \times 10^{-9}$ & 0\\
\hline
\end{tabular}
\end{table}

We have computed the numerical values for $\CWLattlenZdot{}$, $\CWLattdldot{0}{}$, and $\CWLattdldot{1}{}$ at $t=T_\text{max}$ to verify that they are indeed zero; the results are given in \tabref{tab:TimeSymmVals}.  As far as the computer is concerned, $\CWLattdldot{1}{}$ is exactly zero for all $\dM / M$.  All other quantities are infinitesimally small, and the difference from zero can be attributed to numerical error.  Thus, we can conclude that $\CWLattlenZdot{}$, $\CWLattdldot{0}{}$, and $\CWLattdldot{1}{}$ are indeed all zero at $t=T_\text{max}$, and hence $t=T_\text{max}$ is a moment of time symmetry, as required for our choice of initial conditions to be valid.

\begin{figure}[t]
\subfloat{\includegraphics{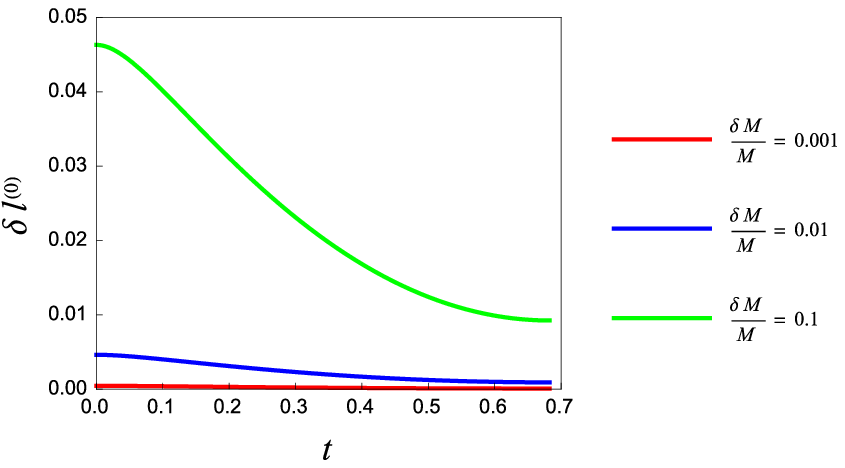}}\\
\subfloat{\includegraphics{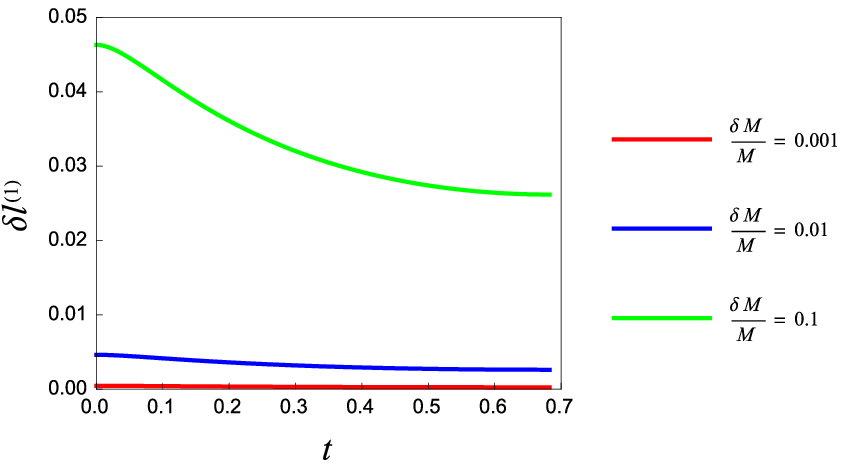}}
\caption{\label{fig:dl_graphs}Plots of $\CWLattdl{0}{}(t)$ (top) and $\CWLattdl{1}{}(t)$ (bottom) against $t$.  The graphs have been extended all the way back to $t=0$.}
\end{figure}

\figref{fig:dl_graphs} shows the behaviour of the length perturbations $\CWLattdl{0}{}$ and $\CWLattdl{1}{}$ as functions of time; these graphs have also been extended back to $t=0$.  We see that the magnitudes of these graphs also increase with $\dM / M$, which is consistent with the models' attaining larger volumes in \figref{fig:PertVolGraphs} as the perturbation is increased.  We also see that $\CWLattdl{0}{}$ and $\CWLattdl{1}{}$ are always well-behaved and non-zero.  The latter fact implies that there is never a moment when the tetrahedra have edges of zero length.  Rather, there is actually a moment when the tetrahedra are equilateral; this happens at $t = 0$, because $\CWLattdihedral{}$ is then zero, and from \eqref{dl0_dl1_relation}, it therefore follows $\CWLattdl{0}{}$ and $\CWLattdl{1}{}$ are equal.  This suggests that our choice of $\alpha = \beta = \frac{1}{4}$ for co-moving particle trajectories at the end of \secref{ParticleTrajSec} was appropriate.  Finally, we see that the length perturbations decrease with time, consistent with the fact that both $\CWLattdldot{0}{}$ and $\CWLattdldot{1}{}$ are always negative, as \figref{fig:dldot_graphs} shows.

Before closing, we wish to comment on our choice of approach to the perturbed Regge models.  These models were based on a specific triangulation of the CW skeleton where all diagonals on Type II quadrilateral hinges terminated at $E^\prime$.  This triangulation had the virtue of simplicity, but there were other equally valid but inequivalent triangulations we could have worked with.  In terms of the triangulation algorithm described in \secref{GlobLocSec}, we used a labelling of the Cauchy surface's five vertices such that the common apex to the Type II tetrahedra was ordered last.  By shifting the order of that vertex's label, one could generate an alternative triangulation that would certainly not be equivalent to the one we used.  For instance, if the orders of vertices $A$ and $E$ were swapped, then the diagonals on all hinges attached to strut $AA^\prime$ would now terminate on $A^\prime$ instead; none would terminate on $A$.  Thus, compared to the original triangulation, some of the diagonals on Type II quadrilateral hinges would get swapped, and the new pair of triangular hinges that result would be geometrically different from the original pair.  It remains to be seen how many distinct triangulations there are, although we know there must be at least five, as there are five possible orderings of the label for the Type II tetrahedra's common apex; given a specific ordering for this vertex, it remains to be seen whether permutations of the other vertices' ordering would generate inequivalent triangulations or not.  More importantly, it remains to be seen whether these alternative triangulations would lead to the same Regge equations or to something new.  For example, it is possible that in the continuum time limit, the alternative triangulations would still reduce to the same Regge equations at leading order in $dt$.  Even if we had a new set of Regge equations, the solutions of both this new set and the original set would be equally valid as both of them would satisfy the global Regge equation, as discussed in \secref{GlobLocSec}.  We shall leave a more thorough investigation of these alternative models to future study.

\begin{acknowledgments}
The authors would like to thank Leo Brewin for much helpful discussion as well as Ulrich Sperhake and Tim Clifton for greatly appreciated comments, including pointing out an error in an earlier version of \figref{fig:UnperturbedGraphs}.  RGL acknowledges partial financial support from the Cambridge Commonwealth Trust.
\end{acknowledgments}

\appendix

\setcounter{section}{0}
\renewcommand{\thesection}{\Alph{section}}
\titleformat{\section}[block]
  {\normalfont\bfseries}{APPENDIX \thesection:}{1em}{\centering \MakeUppercase{#1}}

\section{Regular lattices in 3-spaces of constant curvature}
\label{App_Latt}
In this appendix, we shall list all possible lattices that cover 3-spaces of constant curvature with a single regular polyhedral cell.  The cell is tiled to completely cover the 3-space without any gaps or overlaps.  This tessellation problem has been thoroughly studied by Coxeter \cite{Coxeter}.  Clifton and Ferreira \cite{CF, *CF-err} have succinctly summarised Coxeter's results relevant to our discussion, and we have presented their summary in \tabref{tab:Tab_Latt}.
\begin{table} [t]
\caption{\label{tab:Tab_Latt}All possible lattices obtained by tessellating 3-spaces of constant curvature with a single regular polyhedron.  We note that the second column, which indicates how many cells meet at any lattice edge, effectively determines the lattice's structure.}
\begin{tabular*}{8.6cm}{@{\extracolsep{\fill} } S N B T }
\hline\hline
\textbf{Elementary cell shape} & \textbf{Number of cells at a lattice edge} & \textbf{Background curvature} & \textbf{Total cells in lattice} \\
\hline
tetrahedron & 3 & + & 5 \\
cube & 3 & + & 8 \\
tetrahedron & 4 & + & 16 \\
octahedron & 3 & + & 24 \\
dodecahedron & 3 & + & 120 \\
tetrahedron & 5 & + & 600 \\
cube & 4 & 0 & $\infty$ \\
cube & 5 & - & $\infty$ \\
dodecahedron & 4 & - & $\infty$ \\
dodecahedron & 5 & - & $\infty$ \\
icosahedron & 3 & - & $\infty$ \\
\hline\hline
\end{tabular*}
\end{table}

\begin{figure}[htb]
{\fontsize{8pt}{9.6pt}\input{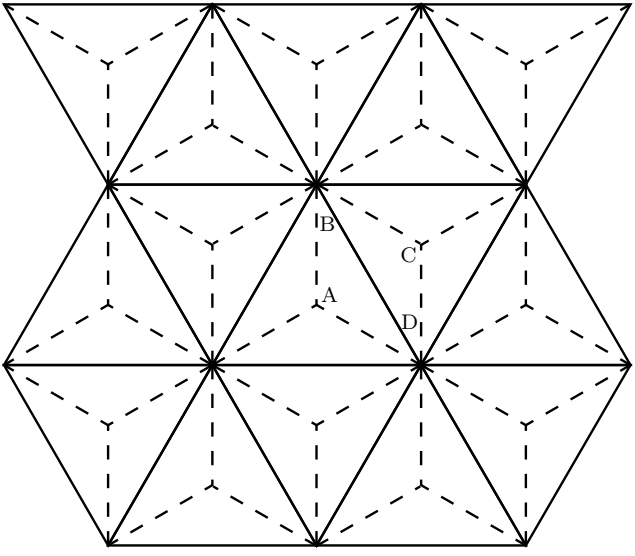}}
\caption{\label{fig:new_lattice}A lattice of equilateral triangles tessellating flat 2-dimensional space with a new lattice, derived from the original, centred on the mid-points of the triangular edges; the triangular lattice has been drawn in solid lines and the new lattice in dashed lines.  The new lattice cells do not correspond to regular polytopes but are instead rhombi; although their edge-lengths are equal, their interior angles are not.  For the new cell marked ABCD, the angles at the centres of triangles, marked A and C, are $120\degree$, while the angles at vertices of triangles, marked B and D, are $60\degree$.  These angles imply that three dual cells would meet at A and C, and six at B and D; thus the cell vertices are not identical either.}
\end{figure}

The Coxeter lattices are the only possible lattices that use a single regular polytope as its elementary cell.  However, if we allow for elementary cells that are not regular polytopes, then further regular lattices are possible.  For instance, one can obtain a new lattice from the closed 600-tetrahedra lattice by using the centres of the original lattice's triangles as the new cell centres; one would then partition out new cells by erecting new boundaries between pairs of nearest-neighbouring triangular centres.  Each tetrahedra has four triangles, but each triangle is shared between two tetrahedra, so the 600-tetrahedra lattice has a total of 1200 triangles; thus this new lattice has a total of 1200 cells.  Clearly, this cannot be a Coxeter lattice, and its cell therefore cannot be a regular polytope; yet this new lattice is regular because the triangles are distributed in a regular manner.  For an analogous situation, consider a lattice tessellating flat 2-dimensional space with equilateral triangles, as shown in \figref{fig:new_lattice}, and a new lattice with cells centred on the mid-points of the original edges.
This new lattice is still regular, but its cells do not correspond to regular polytopes: their internal angles are not equal, even though their edges are; thus the cells are actually rhombi.  Since the internal angles are not equal, the vertices themselves are not identical either, with three new cells meeting at vertices like the one marked A and six at vertices like the one marked B.  Therefore, such cells would not have as high a rotational symmetry as its regular counterpart, the square; the rhombus has only an order 2 rotational symmetry, while the square has order 4.  For such reasons, the Coxeter lattices are the lattices with the highest symmetry.

For the closed lattices, if one constructs new lattices using cells centred on the original lattice vertices, it appears these new lattices form Coxeter lattices as well.  We shall refer to these new lattices as dual lattices.  The 5-tetrahedra lattice is dual to itself, as is the 24-octahedra lattice.  The 8-cube and 16-tetrahedra lattices are duals of each other, as are the 120-dodecahedra and 600-tetrahedra lattices.  However, new lattices using cells centred on the mid-points of edges or the centres of faces are not Coxeter lattices in general.

\section{Variation of particle path-lengths with respect to the struts}
\label{PathVariation}
In this appendix, we shall explain the derivation of results \eqref{Type2ComovingTrajAA}, \eqref{Type2ComovingTrajEE}, and \eqref{Type1TrajCont}, which give the local variation of the particle path-lengths with respect to each strut-length.  Let us consider the length $\RegLinEl{i}$ of an arbitrary particle through a 4-block; we shall work with the 4-block of a Type II tetrahedron because it is more general; it can easily be reduced to the Type I case by setting $\psi_i = 0$, $\CWLattlen{1}{i} = \CWLattlen{0}{i}$, and $\CWLattlen{1}{i+1} = \CWLattlen{0}{i+1}$.  In order to vary $\RegLinEl{i}$ with respect to each strut locally, we need to express it first in terms of the lengths of all four struts.  We denote the four strut-lengths by $\CWLattstrutv{A}{i}$, $\CWLattstrutv{B}{i}$, $\CWLattstrutv{C}{i}$, and $\CWLattstrutv{E}{i}$, with the superscript labelling the lower vertex to which the strut is attached.  Our approach will be to use a new co-ordinate system for the 4-block such that the co-ordinates of the vertices are given in terms of the lengths of the edges, including the tetrahedral edges, the diagonals, and the struts.  As we are only interested in varying the struts, we can greatly simplify our co-ordinate system if we first impose the symmetries on all other edges, constraining them to be $\CWLattlen{0}{i}$, $\CWLattlen{0}{i+1}$, $\CWLattlen{1}{i}$, $\CWLattlen{1}{i+1}$, $\CWLattdiag{i}$, $\CWLattdiag{i+1}$, $\CWLattdiagp{i}$, $\CWLattdiagp{i+1}$ accordingly; this is permissible because when we locally vary with respect to one edge, all other edges must be held constant.  We then calculate $\RegLinEl{i}$ in this new co-ordinate system, differentiate it with respect to each of $\CWLattstrutv{A}{i}$, $\CWLattstrutv{B}{i}$, $\CWLattstrutv{C}{i}$, $\CWLattstrutv{E}{i}$, and then impose the relevant strut-length constraints on $\CWLattstrutv{A}{i}$, $\CWLattstrutv{B}{i}$, $\CWLattstrutv{C}{i}$, $\CWLattstrutv{E}{i}$.

Once again, the 4-block has been triangulated in the manner described in \secsref{CWRegge}{GlobLocSec}, and this introduces the diagonals $AE^\prime$, $BE^\prime$, $CE^\prime$, which have length $\CWLattdiag{i}$, and the diagonals $AB^\prime$, $AC^\prime$, and $BC^\prime$, which have length $\CWLattdiagp{i}$.

We shall now construct our new co-ordinate system.  We can freely fix the co-ordinates of the upper tetrahedron's vertices to be
\begin{equation}
\renewcommand{\arraystretch}{2}
\begin{aligned}
A^\prime &= \displaystyle{\left(-\frac{\CWLattlen{1}{i+1}}{2}, -\frac{\CWLattlen{1}{i+1}}{2\sqrt{3}}, 0, \, 0 \right)},\\
B^\prime &=\displaystyle{\left(\frac{\CWLattlen{1}{i+1}}{2}, -\frac{\CWLattlen{1}{i+1}}{2\sqrt{3}}, 0, \, 0 \right)},\\
C^\prime &= \displaystyle{\left(0, \frac{\CWLattlen{1}{i+1}}{\sqrt{3}}, 0, \, 0 \right)},\\
E^\prime &= \displaystyle{\left(0, 0, \axislen{i+1}, 0 \right)}.
\end{aligned}
\end{equation}
Since the upper tetrahedron's co-ordinates are fixed, the dependence on the strut-lengths must appear in the lower tetrahedron's co-ordinates.

Vertex $A$ is constrained by the lengths
\begin{align*}
\distance{AA^\prime} &= \CWLattstrutv{A}{i}, & \distance{AB^\prime} = \distance{AC^\prime}  &= \CWLattdiagp{i}, & \distance{AE^\prime}  &= \CWLattdiag{i}.
\end{align*}
As $A$ is equidistant to $B^\prime$ and $C^\prime$, its co-ordinates will have the form
$$
A = \left(-\frac{1}{2} a_A,\, -\frac{1}{2\sqrt{3}} a_A,\, c_A,\, - \imath d_A \right).
$$
Using our new co-ordinates for the vertices, we can calculate the edge-lengths above in terms of $a_A$, $c_A$, and $d_A$.  This leads to the equations
\begin{IEEEeqnarray*}{rCl}
\left(\CWLattstrutv{A}{i}\right)^2 \! &=& \frac{1}{4}\left(\CWLattlen{1}{i+1} - a_A \right)^2 \!\!  + \frac{1}{12}\left(\CWLattlen{1}{i+1} - a_A \right)^2 \!\!  + c_A^2 - d_A^2,\\
\left(\CWLattdiagp{i}\right)^2 \! &=& \frac{1}{4}\left(\CWLattlen{1}{i+1} + a_A \right)^2 \!\!  + \frac{1}{12}\left(\CWLattlen{1}{i+1} - a_A \right)^2 \!\!  + c_A^2 - d_A^2,\\
\left(\CWLattdiag{i}\right)^2 \! &=& \frac{1}{3} a_A^2 + \left( \axislen{i+1} - c_A \right)^2 \! - d_A^2.
\end{IEEEeqnarray*}
Since we are only interested in the first derivative of $\RegLinEl{i}$ with respect to the strut-lengths, we need only determine $a_A$, $c_A$, and $d_A$ and similar quantities to first order in $\CWLattdstrutv{A}{i}$, $\CWLattdstrutv{B}{i}$, $\CWLattdstrutv{C}{i}$, and $\CWLattdstrutv{E}{i}$.  So by expressing $\CWLattstrutv{A}{i}$ as $\CWLattstrutv{A}{i}\approx \CWLattstrutlenp{i} + \CWLattdstrutv{A}{i}$ and making use of \eqref{strutlenp} to \eqref{diagp}, we can solve the above system of equations to first order in $\CWLattdstrutv{A}{i}$, obtaining
\begin{widetext}
\begin{IEEEeqnarray*}{rCl}
a_A &\approx& \CWLattlen{1}{i} - 2\, \frac{\CWLattstrutlenp{i}}{\CWLattlen{1}{i+1}}\, \CWLattdstrutv{A}{i},\\
c_A &\approx& \frac{\delta \CWLattlen{1}{i}}{2\sqrt{6}} \cosh \psi_i + \sinh \psi_i\, \delta \RegTime{i} + \frac{1}{3} \frac{\CWLattstrutlenp{i}}{\axislen{i+1}}\, \CWLattdstrutv{A}{i},\\
d_A &\approx& \frac{\delta \CWLattlen{1}{i}}{2\sqrt{6}} \sinh \psi_i + \cosh \psi_i\, \delta \RegTime{i} + \frac{1}{3} \frac{\CWLattstrutlenp{i}}{\axislen{i+1}} \frac{\frac{\delta \CWLattlen{1}{i}}{2\sqrt{6}} \cosh \psi_i + \sinh \psi_i\, \delta \RegTime{i} - \left[2\frac{\CWLattlen{1}{i}}{\CWLattlen{1}{i+1}} +1 \right]\axislen{i+1}}{\frac{\delta \CWLattlen{1}{i}}{2\sqrt{6}} \sinh \psi_i + \cosh \psi_i\, \delta \RegTime{i}}\, \CWLattdstrutv{A}{i}.
\end{IEEEeqnarray*}
\end{widetext}

Vertex $B$ is constrained by the lengths
\begin{align*}
\distance{BB^\prime} &= \CWLattstrutv{B}{i}, & \mkern-11mu \distance{BC^\prime} &= \CWLattdiagp{i}, & \mkern-11mu \distance{BE^\prime}  &= \CWLattdiag{i}, & \mkern-11mu \distance{AB}  &= \CWLattlen{1}{i},
\end{align*}
and we shall express its co-ordinates in the form
$$
B = \left(\frac{1}{2} a_B,\, -\frac{1}{2\sqrt{3}} b_B,\, c_B,\, -\imath d_B\right).
$$
Furthermore, based on symmetries and the co-ordinates just obtained for $A$, we know what the co-ordinates of $B$ should be when $\CWLattstrutv{B}{i} = \CWLattstrutlenp{i}$, so we can express $a_B$, $b_B$, $c_B$, and $d_B$ as
\begin{IEEEeqnarray*}{rCl}
a_B &\approx& \CWLattlen{1}{i} + \delta a_B,\\
b_B &\approx& \CWLattlen{1}{i} + \delta b_B,\\
c_B &\approx& \frac{\delta \CWLattlen{1}{i}}{2\sqrt{6}} \cosh \psi_i + \sinh \psi_i\, \delta \RegTime{i} + \delta c_B,\\
d_B &\approx& \frac{\delta \CWLattlen{1}{i}}{2\sqrt{6}} \sinh \psi_i + \cosh \psi_i\, \delta \RegTime{i} + \delta d_B,
\end{IEEEeqnarray*}
where $\delta a_B$, $\delta b_B$, $\delta c_B$, $\delta d_B$ are linear in $\CWLattdstrutv{A}{i}$ and $\CWLattdstrutv{B}{i}$.  We can use the new co-ordinates to express the lengths above in terms of $a_B$, $b_B$, $c_B$, and $d_B$, yielding
\begin{IEEEeqnarray*}{rCl}
\left(\CWLattstrutv{B}{i}\right)^2 &=& \frac{1}{4} \left( \CWLattlen{1}{i+1} - a_B \right)^2 \! + \frac{1}{12} \left( \CWLattlen{1}{i+1} - b_B \right)^2 \! + c_B^2 - d_B^2,\\
\left(\CWLattdiagp{i}\right)^2 &=& \frac{1}{4} a_B^2 + \frac{1}{3} \left( \CWLattlen{1}{i+1} + \frac{1}{2} b_B \right)^2 \! + c_B^2 - d_B^2,\\
\left(\CWLattdiag{i}\right)^2 &=& \frac{1}{4} a_B^2 + \frac{1}{12} b_B^2 + \left(\axislen{i+1}-c_B\right)^2 - d_B^2,\\
\left(\CWLattlen{1}{i}\right)^2 &=& \frac{1}{4} \left(a_A + a_B \right)^2 + \frac{1}{12} \left(b_B - a_A\right)^2 \\
&& {} + \left(c_B - c_A\right)^2 - \left(d_B - d_A\right)^2.
\end{IEEEeqnarray*}
Then by taking $\CWLattstrutv{B}{i}\approx \CWLattstrutlenp{i} + \CWLattdstrutv{B}{i}$ and making use of \eqref{strutlenp} to \eqref{diagp}, we match the first order terms and solve to obtain
\begin{equation}
\begin{aligned}
\delta a_B \approx{}& 2\, \frac{\CWLattstrutlenp{i}}{\CWLattlen{1}{i+1}}\, \CWLattdstrutv{A}{i},\\
\delta b_B \approx{}& - 2\, \frac{\CWLattstrutlenp{i}}{\CWLattlen{1}{i+1}}\, \left(\CWLattdstrutv{A}{i} + 2\, \CWLattdstrutv{B}{i} \right),\\
\delta c_B \approx{}& \frac{1}{3} \frac{\CWLattstrutlenp{i}}{\axislen{i+1}}\, \left(\CWLattdstrutv{A}{i} + 2\, \CWLattdstrutv{B}{i} \right),\\
\delta d_B \approx{}& 
- \frac{1}{3} \CWLattstrutlenp{i} \left[\frac{\delta \CWLattlen{1}{i}}{2\sqrt{6}} \sinh \psi_i + \cosh \psi_i\, \delta \RegTime{i}\right]^{-1}\\
& {} \times \! \left[ \left[ 1 \! - \! \frac{\frac{\delta \CWLattlen{1}{i}}{2\sqrt{6}} \cosh \psi_i \! + \! \sinh \psi_i\, \delta \RegTime{i}}{\axislen{i+1}} \right] \mkern-5mu \left(\CWLattdstrutv{A}{i} \! + \! 2 \, \CWLattdstrutv{B}{i} \right) \right. \\
&\hphantom{{} \times  \! [} {} - \left. \frac{\CWLattlen{1}{i}}{\CWLattlen{1}{i+1}} \left(\CWLattdstrutv{A}{i} - \CWLattdstrutv{B}{i} \right) \vphantom{\frac{\frac{\delta \CWLattlen{1}{i}}{2\sqrt{6}}}{\axislen{i+1}}} \right].
\end{aligned}
\end{equation}

Vertex $C$ is constrained by the lengths
\begin{align*}
\distance{CC^\prime} &= \CWLattstrutv{C}{i}, & \distance{AC} = \distance{BC}  &= \CWLattlen{1}{i}, & \distance{CE^\prime}  &= \CWLattdiag{i},
\end{align*}
and we shall express its co-ordinates in the form
$$
C = \left(a_C,\, \frac{1}{\sqrt{3}} b_C,\, c_C,\, -\imath d_C\right).
$$
We also know what the co-ordinates of $C$ should be when $\CWLattstrutv{C}{i} = \CWLattstrutlenp{i}$, so we can express $a_C$, $b_C$, $c_C$, and $d_C$ as
\begin{IEEEeqnarray*}{rCl}
a_C &\approx& \delta a_C,\\
b_C &\approx& \CWLattlen{1}{i} + \delta b_C,\\
c_C &\approx& \frac{\delta \CWLattlen{1}{i}}{2\sqrt{6}} \cosh \psi_i + \sinh \psi_i\, \delta \RegTime{i} + \delta c_C,\\
d_C &\approx& \frac{\delta \CWLattlen{1}{i}}{2\sqrt{6}} \sinh \psi_i + \cosh \psi_i\, \delta \RegTime{i} + \delta d_C,
\end{IEEEeqnarray*}
where $\delta a_C$, $\delta b_C$, $\delta c_C$, $\delta d_C$ are linear in $\CWLattdstrutv{A}{i}$, $\CWLattdstrutv{B}{i}$, and $\CWLattdstrutv{C}{i}$.  Following a similar method to that of the previous two vertices, we find that
\begin{IEEEeqnarray*}{rCl}
\delta a_C &\approx& \frac{\CWLattstrutlenp{i}}{\CWLattlen{1}{i+1}}\, \left(\CWLattdstrutv{A}{i} - \CWLattdstrutv{B}{i}\right),\\
\delta b_C &\approx& \frac{\CWLattstrutlenp{i}}{\CWLattlen{1}{i+1}}\, \left(\CWLattdstrutv{A}{i} + \CWLattdstrutv{B}{i} \right),\\
\delta c_C &\approx& \frac{1}{3} \frac{\CWLattstrutlenp{i}}{\axislen{i+1}}\, \left(\CWLattdstrutv{A}{i} + \CWLattdstrutv{B}{i} + 3\, \CWLattdstrutv{C}{i} \right),\\
\delta d_C &\approx& \frac{1}{3} \CWLattstrutlenp{i} \left[\frac{\delta \CWLattlen{1}{i}}{2\sqrt{6}} \sinh \psi_i + \cosh \psi_i\, \delta \RegTime{i}\right]^{-1}\\
&& {} \times \left[ \left[ \frac{\frac{\delta \CWLattlen{1}{i}}{2\sqrt{6}} \cosh \psi_i + \sinh \psi_i\, \delta \RegTime{i}}{\axislen{i+1}} - 1 \right] \right.\\
&& \hphantom{ {} \times [} {} \times \left(\CWLattdstrutv{A}{i} + \CWLattdstrutv{B}{i} + 3 \, \CWLattdstrutv{C}{i} \right)\\
&& \hphantom{ {} \times [} {} + \left. \frac{\CWLattlen{1}{i}}{\CWLattlen{1}{i+1}} \left(\CWLattdstrutv{A}{i} + \CWLattdstrutv{B}{i} \right)\vphantom{^{2^2}} \vphantom{\frac{\frac{\delta \CWLattlen{1}{i}}{2\sqrt{6}}}{\axislen{i+1}} } \right].
\end{IEEEeqnarray*}

Finally, vertex $E$ is constrained by the lengths
\begin{align*}
\distance{EE^\prime} &= \CWLattstrutv{E}{i}, & \distance{AE} = \distance{BE} = \distance{CE}  &= \CWLattlen{0}{i},
\end{align*}
and we shall express its co-ordinates in the form
$$
E = \left(a_E,\, b_E,\, c_E,\, -\imath d_E\right).
$$
We also know what the co-ordinates of $E$ should be when $\CWLattstrutv{E}{i} = \CWLattstrutlen{i}$, so we can express $a_E$, $b_E$, $c_E$, and $d_E$ as
\begin{IEEEeqnarray*}{rCl}
a_E &\approx& \delta a_E,\\
b_E &\approx& \delta b_E,\\
c_E &\approx& \frac{\delta \CWLattlen{1}{i}}{2\sqrt{6}} \cosh \psi_i + \sinh \psi_i\, \delta \RegTime{i} + \axislen{i} \cosh \alpha_i + \delta c_E,\\
d_E &\approx& \frac{\delta \CWLattlen{1}{i}}{2\sqrt{6}} \sinh \psi_i + \cosh \psi_i\, \delta \RegTime{i} + \axislen{i} \sinh \alpha_i + \delta d_E,
\end{IEEEeqnarray*}
where $\alpha_i$ is a yet to be determined boost parameter and $\delta a_E$, $\delta b_E$, $\delta c_E$, $\delta d_E$ are linear in $\CWLattdstrutv{A}{i}$, $\CWLattdstrutv{B}{i}$, $\CWLattdstrutv{C}{i}$, and $\CWLattdstrutv{E}{i}$.  Unlike the other struts, $\CWLattstrutv{E}{i}$ has the perturbative expansion $\CWLattstrutv{E}{i} \approx \CWLattstrutlen{i} + \CWLattdstrutv{E}{i}$.  We can follow a similar method to that of the previous vertices to solve the equations above for $\alpha_i$, $\delta a_E$, $\delta b_E$, $\delta c_E$, $\delta d_E$.  By matching the zeroth order terms, we deduce that
\begin{equation}
\alpha_i = \psi_i.
\end{equation}
Next, by matching the first order terms and then solving, we find that
\begin{widetext}
\begin{IEEEeqnarray*}{rCl}
\delta a_E & \approx & \frac{\CWLattstrutlenp{i}}{\CWLattlen{1}{i+1}}\, \CWLattdstrutv{A}{i} - \frac{1}{3}\frac{\CWLattstrutlenp{i}}{\CWLattlen{1}{i}}\left[\frac{\CWLattlen{1}{i}}{\CWLattlen{1}{i+1}} + 2 \frac{\axislen{i}}{\axislen{i+1}}\cosh \psi_i \right] \CWLattdstrutv{B}{i}\\
&& {} + \frac{1}{3} \frac{\CWLattstrutlenp{i}}{\CWLattlen{1}{i}} \frac{\axislen{i} \sinh \psi_i}{\frac{\delta \CWLattlen{1}{i}}{2\sqrt{6}} \sinh \psi_i + \cosh \psi_i\, \delta \RegTime{i}} \left[ \frac{\CWLattlen{1}{i}}{\CWLattlen{1}{i+1}} \left(3\, \CWLattdstrutv{A}{i} - \CWLattdstrutv{B}{i} \right) - 2 \left(1 - \frac{ \frac{\delta \CWLattlen{1}{i}}{2\sqrt{6}} \cosh \psi_i + \sinh \psi_i\, \delta \RegTime{i} }{\axislen{i+1}} \right) \CWLattdstrutv{B}{i} \right], \\
\delta b_E & \approx & \frac{1}{\sqrt{3}} \frac{\CWLattstrutlenp{i}}{\CWLattlen{1}{i+1}}\left[1 + \frac{\axislen{i} \sinh \psi_i}{\frac{\delta \CWLattlen{1}{i}}{2\sqrt{6}} \sinh \psi_i + \cosh \psi_i\, \delta \RegTime{i}}\right]\left(\CWLattdstrutv{A}{i} + \CWLattdstrutv{B}{i}\right) - \frac{2}{\sqrt{3}} \frac{\CWLattstrutlenp{i}}{\CWLattlen{1}{i}} \frac{\axislen{i}}{\axislen{i+1}}\frac{\delta \RegTime{i} + \axislen{i+1} \sinh \psi_i}{\frac{\delta \CWLattlen{1}{i}}{2\sqrt{6}} \sinh \psi_i + \cosh \psi_i\, \delta \RegTime{i}} \, \CWLattdstrutv{C}{i}, \\
\delta c_E & \approx & -\frac{\CWLattstrutlen{i} \sinh \psi_i}{\delta \RegTime{i} + \axislen{i+1}\sinh \psi_i}\, \CWLattdstrutv{E}{i} + \frac{1}{3} \frac{\CWLattstrutlenp{i}}{\axislen{i+1}} \left[ 1 + \frac{\axislen{i} \sinh \psi_i}{\frac{\delta \CWLattlen{1}{i}}{2\sqrt{6}} \sinh \psi_i + \cosh \psi_i\, \delta \RegTime{i}} \right] \left( \CWLattdstrutv{A}{i} + \CWLattdstrutv{B}{i} + \CWLattdstrutv{C}{i} \right), \\
\delta d_E & \approx & -\frac{\CWLattstrutlen{i} \cosh \psi_i}{\delta \RegTime{i} + \axislen{i+1}\sinh \psi_i}\, \CWLattdstrutv{E}{i} + \frac{1}{3} \frac{\CWLattstrutlenp{i}}{\axislen{i+1}} \left( \CWLattdstrutv{A}{i} + \CWLattdstrutv{B}{i} + \CWLattdstrutv{C}{i} \right) \frac{\frac{\delta \CWLattlen{1}{i}}{2\sqrt{6}} \cosh \psi_i + \sinh \psi_i\, \delta \RegTime{i} + \axislen{i} \cosh \psi_i - \axislen{i+1}}{\frac{\delta \CWLattlen{1}{i}}{2\sqrt{6}} \sinh \psi_i + \cosh \psi_i\, \delta \RegTime{i}}.
\end{IEEEeqnarray*}
\end{widetext}

Using this new co-ordinate system, we shall now vary the particle's path-length with respect to each of the struts.  Based on the particle's position given by \eqref{comoving_particle} and its counterpart on Cauchy surface $\Cauchyt{i+1}$, the particle should follow a trajectory $\Tensb{\LinEl}_i$ given by
$$
\Tensb{\LinEl}_i = \frac{1}{4} \left( \overrightarrow{AA^\prime} + \overrightarrow{BB^\prime} + \overrightarrow{CC^\prime} + \overrightarrow{EE^\prime} \right).
$$
We note that since $\Tensb{\LinEl}_i$ is just a linear combination of the four strut vectors, it will always be time-like if all four strut vectors are time-like.  For each vertex $X=A,B,C,E$, let us express the perturbative expansion of the corresponding strut vector $\overrightarrow{XX^\prime}$ as
$$
\overrightarrow{XX^\prime} \approx \Tensb{\CWstrut{}}_X + \delta \Tensb{\CWstrut{}}_X + \cdots,
$$
where $\Tensb{\CWstrut{}}_X$ denotes a vector corresponding to the zeroth order component of $\overrightarrow{XX^\prime}$ and $\delta \Tensb{\CWstrut{}}_X$ denotes a vector corresponding to the component of $\overrightarrow{XX^\prime}$ that is first order in $\CWLattdstrutv{A}{i}$, $\CWLattdstrutv{B}{i}$, $\CWLattdstrutv{C}{i}$, and $\CWLattdstrutv{E}{i}$.  Then the trajectory length $\RegLinEl{i}$ can be expressed as the perturbative expansion
\begin{widetext}
\begin{equation*}
\begin{aligned}
\RegLinEl{i} = {} & \lvert \Tensb{\RegLinEl{0}} \rvert + \frac{\Tensb{\RegLinEl{0}}}{\lvert \Tensb{\RegLinEl{0}} \rvert} \cdot \delta \Tensb{\LinEl} + \cdots\\
=&  \lvert \Tensb{\RegLinEl{0}} \rvert + \frac{\partial \RegLinEl{i}}{\partial \CWLattstrutv{A}{i}} \CWLattdstrutv{A}{i} + \frac{\partial \RegLinEl{i}}{\partial \CWLattstrutv{B}{i}} \CWLattdstrutv{B}{i} + \frac{\partial \RegLinEl{i}}{\partial \CWLattstrutv{C}{i}} \CWLattdstrutv{C}{i} + \frac{\partial \RegLinEl{i}}{\partial \CWLattstrutv{E}{i}} \CWLattdstrutv{E}{i} + \cdots,
\end{aligned}
\end{equation*}
\end{widetext}
where $\Tensb{\RegLinEl{0}}$ denotes the vector
$$
\Tensb{\RegLinEl{0}} = \frac{1}{4} \left( \Tensb{\CWstrut{}}_A + \Tensb{\CWstrut{}}_B + \Tensb{\CWstrut{}}_C + \Tensb{\CWstrut{}}_E \right),
$$
and $\delta \Tensb{\LinEl}$ the vector
$$
\delta \Tensb{\LinEl} = \frac{1}{4} \left( \delta \Tensb{\CWstrut{}}_A + \delta \Tensb{\CWstrut{}}_B + \delta \Tensb{\CWstrut{}}_C + \delta \Tensb{\CWstrut{}}_E \right).
$$
In the new co-ordinate system, it can be shown that $\Tensb{\RegLinEl{0}}$ has co-ordinates
\begin{widetext}
$$
\Tensb{\RegLinEl{0}} = \left(0, 0, \frac{1}{4} \left(\axislen{i+1} - \axislen{i}\cosh\psi_i\right) - \left(\frac{\delta \CWLattlen{1}{i}}{2\sqrt{6}} \cosh\psi_i + \sinh\psi_i \, \delta \RegTime{i}\right), \frac{\imath}{4}\, \axislen{i}\sinh\psi_i + \imath \left(\frac{\delta \CWLattlen{1}{i}}{2\sqrt{6}} \sinh\psi_i + \cosh\psi_i \, \delta \RegTime{i}\right) \right).
$$
\end{widetext}
Since the first two co-ordinates are zero, then to calculate $\frac{\Tensb{\RegLinEl{0}}}{\lvert \Tensb{\RegLinEl{0}} \rvert} \cdot \delta \Tensb{\LinEl}$ and hence $\partial \RegLinEl{i} / \partial \CWLattstrutv{X}{i}$, we need only the third and fourth co-ordinates of $\delta \Tensb{\LinEl}$.  The third co-ordinate is
\begin{widetext}
$$
-\frac{1}{3}\frac{\CWLattstrutlenp{i}}{\axislen{i+1}}\left( 1 + \frac{\frac{1}{4}\, \axislen{i} \sinh\psi_i}{\frac{\delta \CWLattlen{1}{i}}{2\sqrt{6}} \sinh\psi_i + \cosh\psi_i \, \delta \RegTime{i}} \right) \left( \CWLattdstrutv{A}{i} + \CWLattdstrutv{B}{i} + \CWLattdstrutv{C}{i} \right) + \frac{\frac{1}{4}\, \CWLattstrutlen{i}\sinh\psi_i}{\delta \RegTime{i} + \axislen{i+1} \sinh\psi_i}\, \CWLattdstrutv{E}{i},
$$
and the fourth is
$$
\frac{\imath}{3}\frac{\CWLattstrutlenp{i}}{\axislen{i+1}} \frac{\frac{\delta \CWLattlen{1}{i}}{2\sqrt{6}} \cosh\psi_i + \sinh\psi_i \, \delta \RegTime{i} - \axislen{i+1} + \frac{1}{4}\, \axislen{i} \cosh\psi_i}{\frac{\delta \CWLattlen{1}{i}}{2\sqrt{6}} \sinh\psi_i + \cosh\psi_i \, \delta \RegTime{i}}\left( \CWLattdstrutv{A}{i} + \CWLattdstrutv{B}{i} + \CWLattdstrutv{C}{i} \right) - \frac{\frac{\imath}{4}\, \CWLattstrutlen{i}\cosh\psi_i}{\delta \RegTime{i} + \axislen{i+1} \sinh\psi_i}\, \CWLattdstrutv{E}{i}.
$$
\end{widetext}
We can then obtain $\partial \RegLinEl{i} / \partial \CWLattstrutv{X}{i}$ for each $X$ by reading off the factor multiplying $\CWLattdstrutv{X}{i}$ in $\frac{\Tensb{\RegLinEl{0}}}{\lvert \Tensb{\RegLinEl{0}} \rvert} \cdot \delta \Tensb{\LinEl}$, and we find that
\begin{widetext}
\begin{align}
\frac{\partial \RegLinEl{i}}{\partial \CWLattstrutv{A}{i}} = \frac{\partial \RegLinEl{i}}{\partial \CWLattstrutv{B}{i}} = \frac{\partial \RegLinEl{i}}{\partial \CWLattstrutv{C}{i}} &= \frac{1}{4}\, \frac{\CWLattstrutlenp{i}}{\lvert \Tensb{\RegLinEl{0}} \rvert} \left(1+ \frac{\frac{1}{4}\, \axislen{i} \sinh\psi_i}{\frac{\delta \CWLattlen{1}{i}}{2\sqrt{6}} \sinh\psi_i + \cosh\psi_i \, \delta \RegTime{i}} \right),\\
\frac{\partial \RegLinEl{i}}{\partial \CWLattstrutv{E}{i}} &= \frac{1}{4}\, \frac{\CWLattstrutlen{i}}{\lvert \Tensb{\RegLinEl{0}} \rvert} \frac{\frac{1}{4} \, \axislen{i+1} \sinh\psi_i + \delta \RegTime{i}}{ \axislen{i+1} \sinh\psi_i + \delta \RegTime{i}},
\end{align}
where
\begin{equation}
\lvert \Tensb{\RegLinEl{0}} \rvert = \left[ \left[\frac{1}{4} \left(\axislen{i+1} - \axislen{i} \cosh\psi_i\right) - \left( \frac{\delta \CWLattlen{1}{i}}{2\sqrt{6}} \cosh\psi_i + \sinh\psi_i \, \delta \RegTime{i} \right)\right]^2 - \left[ \frac{1}{4}\, \axislen{i} \sinh\psi_i + \frac{\delta \CWLattlen{1}{i}}{2\sqrt{6}} \sinh\psi_i + \cosh\psi_i \, \delta \RegTime{i}  \right]^2 \right]^\frac{1}{2}.
\end{equation}
\end{widetext}
Taking the continuum time limit, we have that
\begin{equation}
\begin{aligned}
\lvert \Tensb{\RegLinEl{0}} \rvert &\to \RegLinEldot{0}\, dt + \Odt{2}\\
&= \left[ \left(\frac{1}{4}\, \axislendot{} - \frac{\CWLattlendot{1}{}}{2\sqrt{6}}\right)^2 - \left( \frac{1}{4} \axislen{} \dot{\psi} + 1 \right)^2 \right]^\frac{1}{2} dt + \Odt{2},
\end{aligned}
\end{equation}
and in this limit, it follows that
\begin{IEEEeqnarray*}{rCl}
\frac{\partial \RegLinEl{}}{\partial \CWLattstrutv{A}{}} = \frac{\partial \RegLinEl{}}{\partial \CWLattstrutv{B}{}} = \frac{\partial \RegLinEl{}}{\partial \CWLattstrutv{C}{}} &=& \frac{1}{4}\, \frac{\CWLattstrutlenpdot{}}{\RegLinEldot{0}} \left( \frac{1}{4}\, \axislen{} \dot{\psi} + 1 \right) + \Odtone,\\
\frac{\partial \RegLinEl{}}{\partial \CWLattstrutv{E}{}} &=& \frac{1}{4} \frac{\CWLattstrutlendot{}}{\RegLinEldot{0}} \frac{\frac{1}{4} \axislen{} \dot{\psi} + 1}{\axislen{} \dot{\psi} + 1} + \Odtone,
\end{IEEEeqnarray*}
which are relations \eqref{Type2ComovingTrajAA} and \eqref{Type2ComovingTrajEE} as required.

Finally, to obtain the equivalent results for the particle in the Type I tetrahedron, we simply set $\psi_i$ to be zero, replace vertex $E$ with vertex $D$, and replace the lengths $\CWLattlen{0}{i}$, $\CWLattlen{0}{i+1}$, and $\CWLattstrutlen{i}$ with $\CWLattlen{1}{i}$, $\CWLattlen{1}{i+1}$, and $\CWLattstrutlenp{i}$, respectively.  Since the mass in this tetrahedron is the perturbed mass, the path-length is now $\RegLinEl{i}^\prime$.  We then find that
\begin{equation}
\frac{\partial \RegLinEl{i}^\prime}{\partial \CWLattstrutv{A}{i}} = \frac{\partial \RegLinEl{i}^\prime}{\partial \CWLattstrutv{B}{i}} = \frac{\partial \RegLinEl{i}^\prime}{\partial \CWLattstrutv{C}{i}} = \frac{\partial \RegLinEl{i}^\prime}{\partial \CWLattstrutv{D}{i}}  = -\frac{\imath}{4}\, \frac{\CWLattstrutlenp{i}}{\delta \RegTime{i}},
\label{Type1TrajDiscrete}
\end{equation}
and in the continuum time limit, this becomes
$$
\frac{\partial \RegLinEl{}^\prime}{\partial \CWLattstrutv{A}{}} = \frac{\partial \RegLinEl{}^\prime}{\partial \CWLattstrutv{B}{}} = \frac{\partial \RegLinEl{}^\prime}{\partial \CWLattstrutv{C}{}} = \frac{\partial \RegLinEl{}^\prime}{\partial \CWLattstrutv{D}{}} = -\frac{\imath}{4}\, \CWLattstrutlenpdot{},
$$
which is relation \eqref{Type1TrajCont} as required.

\bibliographystyle{apsrev4-1}
\bibliography{Regge_lattice_universe}

\end{document}